\documentclass[fleqn,usenatbib]{aa}  

\usepackage{graphicx}
\usepackage{txfonts}
\begin{document}

\title
{The quenching of galaxies, bulges, and disks since cosmic noon:}

\subtitle
{A machine learning approach for identifying causality in astronomical data}

\titlerunning
{Bulge \& Disk Quenching}

\author
{Asa F. L. Bluck\inst{1,2,3}, Roberto Maiolino\inst{1,2}, Simcha Brownson\inst{1,2}, Christopher J. Conselice\inst{4}, Sara L. Ellison\inst{5},\\ Joanna M. Piotrowska\inst{1,2} \& Mallory D. Thorp\inst{5}}

\authorrunning
{Asa F. L. Bluck et al.}

\institute
{Kavli Institute for Cosmology, University of Cambridge, Madingley Road, Cambridge, CB3 0HA, UK
\and
Cavendish Laboratory Astrophysics Group, University of Cambridge, 19 JJ Thomson Avenue, Cambridge, CB3 0HE, UK
\and
Department of Physics, Florida International University, 11200 SW 8th Street, Miami, FL, USA
\and
Jodrell Bank Centre for Astrophysics, University of Manchester, Oxford Road, Manchester UK
\and
Department of Physics \& Astronomy, University of Victoria, Finnerty Road, Victoria, British Columbia, V8P 1A1, Canada
\\ \email{ab2531@cantab.ac.uk}}


\abstract
{We present an analysis of the quenching of star formation in galaxies, bulges, and disks throughout the bulk of cosmic history, from $z=2-0$. We utilise observations from the Sloan Digital Sky Survey (SDSS) and the Mapping Nearby Galaxies at Apache Point Observatory survey (MaNGA) at low redshifts. We complement these data with observations from the Cosmic Assembly Near-Infrared Deep Extragalactic Legacy Survey (CANDELS) at high redshifts. Additionally, we compare the observations to detailed predictions from the LGalaxies semi-analytic model. To analyse the data, we developed a machine learning approach utilising a Random Forest classifier. We first demonstrate that this technique is extremely effective at extracting causal insight from highly complex and inter-correlated model data, before applying it to various observational surveys. Our primary observational results are as follows: At all redshifts studied in this work, we find bulge mass to be the most predictive parameter of quenching, out of the photometric parameter set (incorporating bulge mass, disk mass, total stellar mass, and $B/T$ structure). Moreover, we also find bulge mass to be the most predictive parameter of quenching in both bulge and disk structures, treated separately. Hence, intrinsic galaxy quenching must be due to a stable mechanism operating over cosmic time, and the same quenching mechanism must be effective in both bulge and disk regions. Despite the success of bulge mass in predicting quenching, we find that central velocity dispersion is even more predictive (when available in spectroscopic data sets). In comparison to the LGalaxies model, we find that all of these observational results may be consistently explained through quenching via preventative `radio-mode' active galactic nucleus (AGN) feedback. Furthermore, many alternative quenching mechanisms (including virial shocks, supernova feedback, and morphological stabilisation) are found to be inconsistent with our observational results and those from the literature.}



 \maketitle


\section{Introduction}

In many ways, classification is the first step towards knowledge. Or, as Linnaeus put it, `classification and name-giving will be the foundation of our science'. The process of classifying objects in the physical Universe inevitably leads to an understanding of the differences between them. As such, even today, classification is a major branch of science. The overarching goal of this work is to develop a robust classifier to separate physical classes of galaxies, and in the process reveal what is fundamentally different about them. More specifically, we utilise classification as a tool to understand the physical origin of galaxy quenching (i.e. why some galaxies cease to form stars). In so doing, we build on a large history of similar approaches. However, our present method is distinguished by its success at extracting causality from complex astronomical data, via systematically controlling for nuisance parameters.

The population of galaxies in the local Universe is observed to exhibit strong bimodality in a variety of fundamental properties. More specifically, bimodality is observed in rest-frame optical colours, specific star formation rates, and stellar ages (e.g. \citealt{Strateva2001, Brinchmann2004, Peng2010, Bluck2014}). In this work we refer to actively star forming, blue, and young galaxies as `star forming' and quiescent, red, and old galaxies as `quenched'. Additionally, strong bimodality is observed in morphological, structural, and kinematic parameters (e.g. \citealt{Driver2006, Cameron2009, Cameron2009a, Cappellari2011, Simard2011, Mendel2014, Brownson2022}). In this work we generally refer to these two classes of galaxies as `disk-dominated' (predominantly rotationally supported) and `bulge-dominated' (predominantly pressure supported), where pure spheroids are taken to be the natural extremum of bulge-dominated systems. 

As a result of the observed bimodalities in galaxy properties, it is natural to divide galaxies in the local Universe into classes on the basis of their level of star formation and structural, or  kinematic, type.  Furthermore, there is now much evidence for a deep connection between these two forms of bimodality (see e.g. \citealt{Bell2008, Bell2012, Cameron2009a, Cameron2009, Cheung2012, Fang2013, Omand2014, Lang2014, Bluck2014, Bluck2016}). Hence, galaxy evolution models must explain the nature of bimodality in both star formation and structure, as well as their observed connection. Indeed, this avenue of research is leading to powerful new constraints on the physics of feedback in modern simulations (e.g. \citealt{Lang2014, Bluck2016, Bluck2019, Bluck2020a, Terrazas2016, Terrazas2020, Brennan2017, Piotrowska2021, Brownson2022}).

Since the turn of the century, enormous progress has been made on extending the detailed observations of galaxy populations in the local Universe to high redshifts (e.g. \citealt{Giavalisco2004, Scoville2007, Lilly2007, Grogin2011, Conselice2011}). Bimodality in colour, and in sSFR, is found up to $z \sim 2 - 3$ (e.g. \citealt{Williams2009, Santini2009, Whitaker2011, Bauer2011, Brammer2011}). Moreover, it is now well established that the comoving star formation rate density in the Universe has declined by over an order of magnitude from $z \sim 2$ to the present (see \citealt{Lilly1996, Madau1996}; and \citealt{Madau2014} for a review). Additionally, galaxies are observed to exhibit morphological and kinematic diversity, including bimodality, in the early Universe as well (see \citealt{ForsterSchreiber2004, ForsterSchreiber2006, ForsterSchreiber2009, Bluck2012, Mortlock2013, Buitrago2013, Lang2014}; and \citealt{Conselice2014} for a review). Hence, bimodality in both star formation rate and in structure has been evident in the galaxy population since at least `cosmic noon' (the peak epoch of star formation, and consequently optical brightness, in the history of the Universe).

The past decade has seen a revolution in optical astronomy, whereby the conventional techniques of photometry and spectroscopy have been unified in the spectacular application of integral field unit (IFU) spectroscopy (see \citealt{Sanchez2020} for a review). This revolution has enabled the study of sub-galactic regions within galaxies, resulting in unprecedentedly detailed measurements revealing the physics operating on $\sim$kpc scales in local, and higher redshift, galaxies (see \citealt{Cappellari2011, Sanchez2012, Bryant2015, Bundy2015}). One remarkable observation from these studies is that sub-galactic regions also exhibit strong bimodality in star formation (see \citealt{Bluck2020a, Bluck2020b}). Thus, individual regions within galaxies may be classified as star forming or quenched, as well as the galaxy as a whole. Generally, this categorisation is achieved by considering the level of star formation in a spaxel relative to the resolved main sequence (i.e. the $\Sigma_{\rm SFR} - \Sigma_*$ relation, see \citealt{Sanchez2013, Wuyts2013, CanoDiaz2016, GonzalezDelgado2016, Ellison2018}). Of course, it is also of great interest that there exists a resolved analogue to the global main sequence of \cite{Brinchmann2004} in and of itself (see \citealt{Ellison2021a} for a detailed discussion).

Utilising spatially resolved measurements of star formation, many studies have found evidence for `inside-out' quenching of star formation, whereby the inner regions within galaxies first exhibit reduced levels of star formation followed by the outskirts (e.g. \citealt{Tacchella2015, GonzalezDelgado2016, Belfiore2017, Belfiore2018, Ellison2018, Medling2018, Bluck2020b}). However, in \cite{Bluck2020b} we identify a number of important caveats to this observation. First, although `green valley' systems (galaxies with intermediate levels of star formation) do exhibit a clear signature of inside-out quenching, the vast majority of local galaxies are either star forming or quenched everywhere in radial extent. Thus, most galaxies exhibit sub-galactic conformity, whereby all regions within the galaxy have the same star forming state as the galaxy as a whole. Second, we note that inside-out quenching is primarily a feature of central galaxies, particularly those with high stellar masses. Conversely, satellite (and low stellar mass) systems show no evidence of inside-out quenching and even a hint of the opposite: outside-in quenching (see \citealt{Bluck2020b}). 

From a theoretical perspective, bimodality in star formation poses a serious challenge to models of galaxy formation and evolution (see \citealt{Somerville2015} for a review). In the absence of energetic baryonic feedback, the vast majority of baryons are expected to have collapsed into stars by the present epoch (e.g. \citealt{Cole2000, Bower2008, Henriques2015, Henriques2019}). Yet, observations reveal that only $\sim$10\% of baryons currently reside in stars (e.g. \citealt{Fukugita2004, Shull2012}). Thus, star formation must be significantly less efficient\footnote{We note that here we are referring to the global star formation efficiency: $\epsilon_{\rm SF} \equiv M_*/M_{\rm Halo}$ (\citealt{Moster2010}), as opposed to the efficiency of gas conversion to stars within galaxies: $\mathrm{SFE \equiv SFR/M_{\rm gas}}$ (\citealt{Kennicutt1998}).} than naively expected from simple models of cooling and gravitational collapse. Moreover, the suppression in star formation efficiency is not evenly distributed across dark matter halo mass scales. This manifests in such a way that star formation is suppressed in both low and high mass haloes, reaching a peak at $M_{\rm Halo} \sim 10^{12} M_{\odot}$, roughly equivalent to $M_* \sim 10^{10.5} M_{\odot}$ (see \citealt{Guo2010, Moster2010, Moster2013}). There are no known explanations for the scale-dependence on star formation efficiency utilising gravitational physics and gas cooling alone. Hence, theoretical attention has turned to complex feedback processes as a result of active galactic nuclei (AGN), supernovae, and virial shocks. 

In \cite{Bluck2020a} we derive in a general analytical manner how the energy released from preventative AGN feedback, supernova feedback, and virial shock heating depends on galaxy, and halo, observables. Naturally enough, we find that quenching from AGN feedback must scale primarily with black hole mass (i.e. the time integral of accretion rate; see also \citealt{Soltan1982, Silk1998}); quenching via supernova feedback must scale primarily with total stellar mass (i.e. the time integral of SFR; see also \citealt{Henriques2015}); and quenching from virial shocks must scale primarily with halo mass (i.e. the gravitational potential; see also \citealt{Dekel2006}). Utilising star formation measurements on a global scale (\citealt{Bluck2016, Teimoorinia2016}), and on a spatially resolved scale (\citealt{Bluck2020a, Bluck2020b}), we directly test which out of stellar, halo and black hole mass is more constraining of quenching. We find that black hole mass (estimated via the $M_{BH} - \sigma_\star$ relation) is far more predictive of quenching (on both scales) than stellar mass or halo mass (estimated from abundance matching in Yang et al. 2007, 2009). Additionally, for a much smaller sample of galaxies with dynamically measured black hole masses, \cite{Terrazas2016, Terrazas2017} confirm that black hole mass is superior to stellar mass for parameterising quenching in central galaxies. Taken together, these results clearly favour quenching via AGN feedback over virial shock heating or supernova feedback.  

In this paper, we draw on the phenomenal statistical power of the Sloan Digital Sky Survey (SDSS; \citealt{Abazajian2009}) to study the global star forming properties of galaxies, bulges, and disks in the local Universe. Furthermore, we expand on this data set with both high redshift photometric observations from the Cosmic Assembly Near-Infrared Deep Extragalactic Legacy Survey (CANDELS; \citealt{Grogin2011}), and spatially resolved spectroscopic observations from the Mapping Nearby Galaxies at Apache Point Observatory survey (MaNGA; \citealt{Bundy2015}), to yield a comprehensive view of star formation and quenching from cosmic noon to the present era. We consider both star formation and structural bimodality in the galaxy population by analysing state-of-the-art structural measurements and bulge - disk decompositions from \cite{Mendel2014} and \cite{Simard2011}) in the SDSS, and from \cite{Dimauro2018} in CANDELS. Moreover, we developed powerful new machine learning methods to accurately predict quenching within a multi-dimensional parameter space, incorporating morphological, structural, mass, and (when available) kinematic and environmental parameters. We establish that our machine learning technique is extremely effective at isolating causality within highly inter-correlated model data, before applying it to our observational data sets.

This paper is novel in three key aspects. First, we utilise a Random Forest classifier to determine the most important variables governing quenching across cosmic time, which we demonstrate is far more robust and effective than conventional correlation-based approaches. This technique has never been applied to the SDSS or CANDELS, although we have previously applied it in a simpler mode to MaNGA (see \citealt{Bluck2020a, Bluck2020b}). Second, we analyse the quenching of bulge and disk structures independently from galaxies as a whole for the first time in all of these observational surveys. This is particularly important because it yields a partially spatially resolved analysis in photometric data, helping to bridge the gap between photometry and IFU spectroscopy. The clear advantage of this approach is that photometric data still dwarfs resolved spectroscopic data in terms of the number of galaxies observed (by many orders of magnitude). Third, we compare our observational results in a fully self-consistent manner to a leading cosmological model (LGalaxies, \citealt{Henriques2015}). We utilise the model both to rigorously test our machine learning approach, and to aid in the interpretation of our observational results.

Additionally, we demonstrate that our machine learning method can recover known results on galaxy-wide quenching across cosmic time (e.g. from \citealt{Bluck2014, Bluck2016}; and \citealt{Lang2014}), which first adds further credence to our novel method, and second places these prior results on a more statistically robust footing. We also note that our CANDELS bulge - disk sample is a factor of three times larger than that analysed in \cite{Lang2014}, which is highly advantageous for our statistical approach. Finally, we present a detailed discussion considering a wide range of theoretically proposed quenching mechanisms from the literature, and critically assess these against our observational findings. 

The scope of this paper is quite large - we aim to explore quenching across cosmic time, in observations and simulations, globally, and on spatially resolved scales. We believe that it is only by considering star formation quenching in all of its relevant aspects that we can make a significant advance in our overall understanding of galaxy evolution. Hence, a key motivation for this work is to place the analysis of all of the various relevant data on the same methodological footing. We do this by analysing everything with a Random Forest classifier, which we demonstrate is a highly effective tool for revealing hidden causality within complex inter-correlated astronomical data.

The paper is structured as follows. In Section 2 we describe our data sources, explain the measurements used throughout the paper, and discuss sample selection, measurement uncertainty, and additional checks we have made on the reliability of the data. In Section 3 we present our methods, including a discussion on categorising star forming and quenched galaxies, bulges, disks, and spaxels in the various data sources. Additionally in Section 3, we introduce our Random Forest method (with further tests and details provided in Appendix B), and extract several testable quenching predictions from the LGalaxies semi-analytic model. In Section 4, we present our results for the SDSS and MaNGA surveys at low redshifts, analysing bulges and disks separately, as well as galaxies as a whole. In Section 5, we extend our analysis to high redshifts using the CANDELS data set, also providing separate analyses for bulges and disks, as well as for galaxies as a whole. In Section 6, we discuss our results within the context of the literature, and attempt a causal explanation for intrinsic galaxy quenching. We summarise the major contributions of this work in Section 7. Finally, in the Appendix we present a detailed mathematical description of the LGalaxies quenching model, including a novel re-parameterisation (Appendix A); and provide a detailed technical description of our Random Forest technique (Appendix B).   

Throughout the paper we assume a spatially flat $\Lambda$CDM cosmology, with the following parameters: $\{ \Omega_M, \Omega_{\Lambda}, H_0 \}$ = $\{ 0.3, 0.7, 70\,{\rm km\,s^{-1}/Mpc} \}$. Additionally, we adopt the AB magnitude system, unless otherwise specified.


\section{Data}

In this paper we draw on various observational and simulated public data sources, in order to analyse the star formation quenching of galaxies, bulges and disks across cosmic time. In this section we list the sources of our data, and briefly explain the key measurements utilised throughout the paper. 

catalogue 

\subsection{SDSS}

We utilise the Sloan Digital Sky Survey Data Release 7 (SDSS DR7) as our primary low redshift data source (see \citealt{Abazajian2009}). The SDSS provides $u$, $g$, $r$, $i$ \& $z$-band photometry and imaging for over a million galaxies, along with single aperture spectroscopy for $\sim$0.65 million galaxies, at $z \lesssim 0.2$. Additionally, we employ a host of public value added catalogues for the SDSS in this work. More specifically, we take star formation rates (SFR) from \cite{Brinchmann2004}; bulge - disk decompositions and stellar masses from \cite{Mendel2014}; additional morphological and structural parameters from \cite{Simard2011}; stellar velocity dispersions from \cite{Bernardi2007}; and environmental parameters, including halo masses and central - satellite classifications, from \cite{Yang2007, Yang2009}. All of these datasets are public, and may be accessed via links given in the papers, or direct from the SDSS data server\footnote{SDSS: \href{https://classic.sdss.org/dr7/access/}{https://classic.sdss.org/dr7/access/}}.

Star formation rates are inferred from a two step procedure in \cite{Brinchmann2004} whereby galaxies with strong emission lines without the signature of AGN have SFRs determined through dust corrected emission lines, whereas galaxies without strong emission lines (or else that exhibit AGN contamination) have their SFRs determined via an empirical relationship between specific SFR (sSFR) and the strength of the 4000\AA \, break (D4000). Additionally, a fibre-to-global correction is made, based on colours outside of the fibre region (see \citealt{Brinchmann2004}). As a consequence of the two stage approach, low SFRs from D4000 estimates must be interpreted as upper limits on star formation, given that they constrain only the absence of young stellar features in a given spectrum. Some care must be taken to correctly interpret these low values in SFR. In this work, we mostly consider two broad classes: star forming and quenched objects. As such, the upper-limits are not seriously problematic for our scientific approach.

Bulge, disk, and total stellar masses are derived in \cite{Mendel2014} via SED fitting of multi-waveband bulge - disk decompositions by light, utilising the {\small GIM2D} package (see \citealt{Simard2002, Simard2011}). In \cite{Mendel2014} and \cite{Simard2011}, bulge - disk decompositions are performed assuming an exponential ($n = 1$) disk plus De Vaucouleurs ($n=4$) bulge model. Extensive testing of the accuracy and reliability of these measurements is provided in \cite{Mendel2014} and \cite{Simard2011}. Additionally, further tests are presented in the appendices of \cite{Bluck2014, Bluck2019}. Typical statistical uncertainties on the bulge, disk, and total stellar masses are found to be $\sim$0.1 - 0.2\,dex, with additional systematic uncertainty engendered from the choice of IMF, stellar population synthesis code, star formation history parameterisation, and dust model. The total uncertainties on these parameters are estimated in \cite{Mendel2014} to be $\sim$0.2 - 0.3\,dex.

In this work we consider two values of total stellar mass: the combined bulge + disk mass ($M_* = M_B + M_D$); and the total stellar mass derived from a S\'{e}rsic model for the light profiles ($M_{\rm Sers}$). Additionally, we consider the ratio of bulge-to-total stellar mass, defining $B/T \equiv M_{B} / (M_B + M_D)$. This parameter gives a useful measurement of the morphological type of a galaxy, whereby $B/T = 0$ indicates a pure disk structure (with no bulge component) and $B/T = 1$ indicates a pure spheroidal structure (with no disk component). The majority of galaxies in the SDSS are found to host both a bulge and disk structure. However, for the subset of galaxies where there is not strong statistical evidence in favour of dual components (evaluated from the F-test, see \citealt{Simard2011}), we treat these systems as pure S\'{e}rsic galaxies for some parts of the analysis. 

More specifically, if the probability of a galaxy being a single component S\'{e}rsic system ($P_{pS}$) is higher than 0.32, and the preliminary measured $B/T$ is within 0.3 of an extremal value (i.e. zero or one), we relocate the $B/T$ value to its closest extremum. In these cases, we also utilise the S\'{e}rsic stellar mass (deemed to be more appropriate in this case) and set the value of the other (negligible) component to 1/100 times the value of the total stellar mass. The reason we do not set this to zero is simply because we usually work in logarithmic units. This approach is a simplified application of the scheme developed in \cite{Bluck2014} and \cite{Thanjavur2016}, which is designed to remove `false disks', although we apply it here also to the case of low $B/T$ systems without strong evidence of a bulge. Again, we emphasise that this approach is used for some, but not all, analyses in this paper. Moreover, our primary conclusions are completely stable to the choice of applying a pure S\'{e}rsic correction to these measurements or not. 

Group halo masses are estimated from an abundance matching approach applied to the total stellar mass of groups and clusters in \cite{Yang2007}. Additional halo masses for lower mass systems are derived in \cite{Yang2009}. Briefly, an iterative linking length algorithm is used to assign galaxies to groups, whereupon total dark matter masses are estimated by rank ordering the halo and stellar mass functions. This enables the estimation of a virial radius, and then the procedure may be iterated to exclude galaxies well beyond the virial radius from the group, or include new galaxies not included in the first run. In this work, we take central galaxies to be the most massive galaxy in their group, with satellite galaxies being any other group member. 

We also utilise stellar velocity dispersions ($\sigma_\star$) from \cite{Bernardi2007}, derived from continuum and absorption line fitting of emission-line subtracted observed galaxy spectra to broadened model templates. We adopt the Princeton measurement as opposed to the standard SDSS pipeline here, since the latter does not provide velocity dispersions for late-type systems. We restrict to continuum measurements with S/N $>$ 3.5 for all analyses (and test against a higher threshold of S/N $>$ 10). Due to the instrumental resolution of the SDSS, $\sigma_\star$ values of $\lesssim70$\,km/s are not accurately constrained. We take this into account when analysing these data. Additionally, we apply an aperture correction from Jorgensen et al. (1995) to place all measurements of $\sigma_\star$ on approximately the same spatial footing. See \cite{Bluck2016} for more details on these measurements and our tests on their reliability.

We adopt the same sample selection of SDSS galaxies as in \cite{Bluck2016}, selecting systems with $0.02 < z < 0.20$ and $8 < \log(M_*/M_{\odot}) < 12$, although for some analyses we additionally exclude $M_* < 10^{9} M_{\odot}$ systems. We also require that all galaxies are identified in each of the catalogues discussed above, and additionally have a good measurement (i.e. not null, not NaN, not infinite and so forth). This yields a sample of 538,000 galaxies (423,000 centrals and 115,000 satellites). For some analyses, we additionally weight statistics by the inverse of the co-moving volume in which any given galaxy may be observed in the survey ($1/V_{\rm max}$). This statistic is derived from the absolute magnitudes of galaxies (from SED fitting) and the apparent magnitude limit of the SDSS. See \cite{Thanjavur2016} for tests on this approach, and an application to deriving robust stellar mass functions. See \cite{Bluck2014, Bluck2016, Bluck2019} for further discussion on these data, including detailed consideration of sample biases and measurement uncertainties.

\subsection{MaNGA}

To analyse a sub-set of the low redshift SDSS galaxies in more detail, we additionally utilise the Mapping Nearby Galaxies at Apache Point Observatory (MaNGA) survey, see \citealt{Bundy2015} for an introduction. More specifically, we access these data via the SDSS DR15 release (see \citealt{Aguado2019}). All of the data used in this paper are publicly available from the SDSS data archive\footnote{MaNGA: \href{https://www.sdss.org/dr15/manga/}{https://www.sdss.org/dr15/manga/}}. Additionally, we utilise derived data products from full spectrum fitting utilising the {\small PIPE3D} package (see \citealt{Sanchez2016, Sanchez2016a}). These data are also publicly available from the SDSS archive\footnote{MaNGA Pipe3D: \href{https://www.sdss.org/dr15/manga/manga-data/manga-pipe3d-value-added-catalog/}{https://www.sdss.org/dr15/manga/manga-data/manga-pipe3d-value-added-catalog/}}. Full details on the survey, sample selection, scientific motivation and possible applications are given in \cite{Bundy2015} and \cite{Law2015}. This is the same data and sample as analysed previously in \cite{Bluck2020a, Bluck2020b}, and hence we direct interested readers to our prior publications for full details on these data. Here we give only a brief review of the most important features of the MaNGA data set for this paper.

MaNGA will provide spatially resolved integral field unit (IFU) spectroscopy for $\sim$10,000 local galaxies ($z \leq 0.1$) as part of SDSS IV. Currently, 4600 galaxies are publicly available. The MaNGA survey selects galaxies from the SDSS legacy survey (discussed above), and imposes an approximately flat mass distribution from $M_* = 10^9 - 10^{11.5} M_{\odot}$. Additionally, galaxies are carefully selected to span a wide range of morphological, star forming, and environmental types (from clusters to the field). As such, MaNGA offers a very diverse sample of galaxies in the nearby Universe in which one can analyse the spatially resolved physics within galaxies. More specifically, information on kinematics, stellar ages and metallicities, ionised gas properties, mass and light distributions may be inferred, enabling a high-resolution view on galaxy evolution at late cosmic times.

From the {\small PIPE3D} value added MaNGA catalogs, we utilise: spaxel emission line fluxes and errors, the D4000 index, stellar and gas phase metallicities, stellar mass surface densities ($\Sigma_*$), and central stellar velocity dispersions ($\sigma_c$). We adopt star formation rate surface densities ($\Sigma_{\rm SFR}$) derived in \cite{Bluck2020a} from dust corrected H$\alpha$ luminosity (where possible) and from an analogous method to \cite{Brinchmann2004} utilising here the resolved empirical sSFR - D4000 relation. Emission lines are utilised only when there are secure detections (S/N $>$ 3 in all relevant lines), and additionally the emission line regions are free of AGN contamination, as assessed by the [NII] - BPT diagram (see \citealt{Baldwin1981}). We have extensively tested these measurements against alternative approaches, for example via photometric $\Sigma_{\rm SFR}$ measurements from SED fitting (see appendix A of \citealt{Bluck2020a}). As with the SFRs in the SDSS, as a result of our calibration approach, low values of $\Sigma_{\rm SFR}$ estimated via D4000 must be treated as effective upper limits. We sidestep this issue in this paper by utilising a classification technique of spaxels, in exact analogy to the global classification of galaxies (see Section 3).

In our analysis of MaNGA galaxies, we also incorporate several measurements of galaxy properties taken from the SDSS. This includes the global SFR, total stellar mass, and morphological parameters (particularly from the bulge - disk decompositions). As such, to select MaNGA galaxies to analyse in this paper we first match to the SDSS parent catalog, require measurements of good quality in all of the relevant SDSS catalogs, and impose a 2 arcsecond threshold for a secure match. Application of these cuts yields a final sample of 3523 galaxies (2550 centrals and 973 satellites), containing over 5 million spaxels.

\subsection{CANDELS}

To extend our low redshift analysis to high redshifts, we utilise public data from the CANDELS survey (\citealt{Grogin2011, Koekemoer2011}). More specifically, we utilise measurements of bulge, disk and galaxy stellar masses, along with rest-frame colours and photometric SFRs, from \cite{Dimauro2018}\footnote{CANDELS: \href{https://vm-weblerma.obspm.fr/huertas/form\_CANDELS}{https://vm-weblerma.obspm.fr/huertas/form\_CANDELS}}. The basic survey data, used as an input for the bulge - disk decomposition in \cite{Dimauro2018}, originates in H-band selected galaxy catalogs from 5 sub-surveys: i) UDS (\citealt{Galametz2013}); ii) GOODS-S (\citealt{Guo2013}); iii) GOODS-N (\citealt{Barro2017}); iv) COSMOS (\citealt{Stefanon2017}); and v) AEGIS (\citealt{Stefanon2017}). 

\cite{Dimauro2018} present a bulge - disk decomposition of 17,600 CANDELS galaxies, selected to have F160W $<$ 23 and reside at $0 < z < 2$ (with magnitudes and redshifts given by the official CANDELS release). Dimauro et al. utilise a deep learning approach with convolutional neural networks (CNN), expanding on earlier machine learning approaches by \cite{HuertasCompany2015, HuertasCompany2016}). Two component light distribution models are extracted for 4 - 7 filters (in the 430 - 1600\,nm range), depending on the sub-survey region. Once light is assigned to bulge and disk structures, SED fitting for the entire galaxy, and bulge and disk regions treated individually, is performed enabling the estimation of the stellar masses of galaxies, bulges and disks in CANDELS. Single S\'{e}rsic fits (from \citealt{Wel2012}), deep learning visual morphologies (from \citealt{HuertasCompany2015}), and total stellar masses (from \citealt{HuertasCompany2016}) are utilised to test the deep learning approach. Final statistical uncertainties on the stellar masses of components are estimated to be $\sim$20\% of the value, comparable to the statistical uncertainties on the SDSS measurements (see above). Additional systematic uncertainties are engendered from the choice of IMF, fitting code, SSP library and dust extinction parameterization. 

In slightly more detail, \cite{Dimauro2018} apply a three stage methodology for deriving bulge, disk and total stellar masses in CANDELS. First, they utilise a novel CNN method to classify galaxies as one of four categories (pure disks, pure spheroids, disk + classical bulge, disk + pseudo bulge). Second, once the preferred model is ascertained via the machine learning classification, \cite{Dimauro2018} apply {\small GALFIT-M} in the {GALAPAGOS-2} wrapper from the {\small MEGAMORPH} project (\citealt{Haussler2013, Barden2012, Vika2014}). This software enables multi-waveband simultaneous multi-component fitting of light profiles, based on the {\small GALFIT} fitting package (\citealt{Peng2002}). The output of this step is a bulge - disk decomposition in light, that is giving magnitudes in multiple wavebands for bulge and disk structures. Finally, standard SED fitting techniques are applied to the entire galaxy, and to the bulge and disk components separately, utilising the {\small FAST} fitting code (\citealt{Kriek2009}). Stellar population models are taken from \cite{Bruzual2003}, and a \cite{Chabrier2003} IMF and \cite{Calzetti2000} extinction law are assumed. In addition to the masses of galaxies, bulges, and disks, rest-frame dust-corrected UVJ magnitudes (and hence colours) are provided direct from the best fitting model (see \citealt{Dimauro2018}).

As a result of the above approach, galaxies which are determined to be pure S\'{e}rsic systems via the CNN step are treated as such in the bulge - disk decomposition, and consequently have $B/T$ set equal to zero or one (as appropriate). Hence, the CANDELS bulge - disk decompositions are more in line with the pure S\'{e}rsic corrected bulge - disk decompositions in the SDSS, rather than the raw \cite{Mendel2014} and \cite{Simard2011} catalogue values. As such, when comparing between these data sets we are careful to use pure S\'{e}rsic corrected values in the SDSS to enable fair comparison with the CANDELS data. However, this makes no difference to our primary results or conclusions.

\subsection{LGalaxies model}

\begin{figure*}
\includegraphics[width=0.49\textwidth]{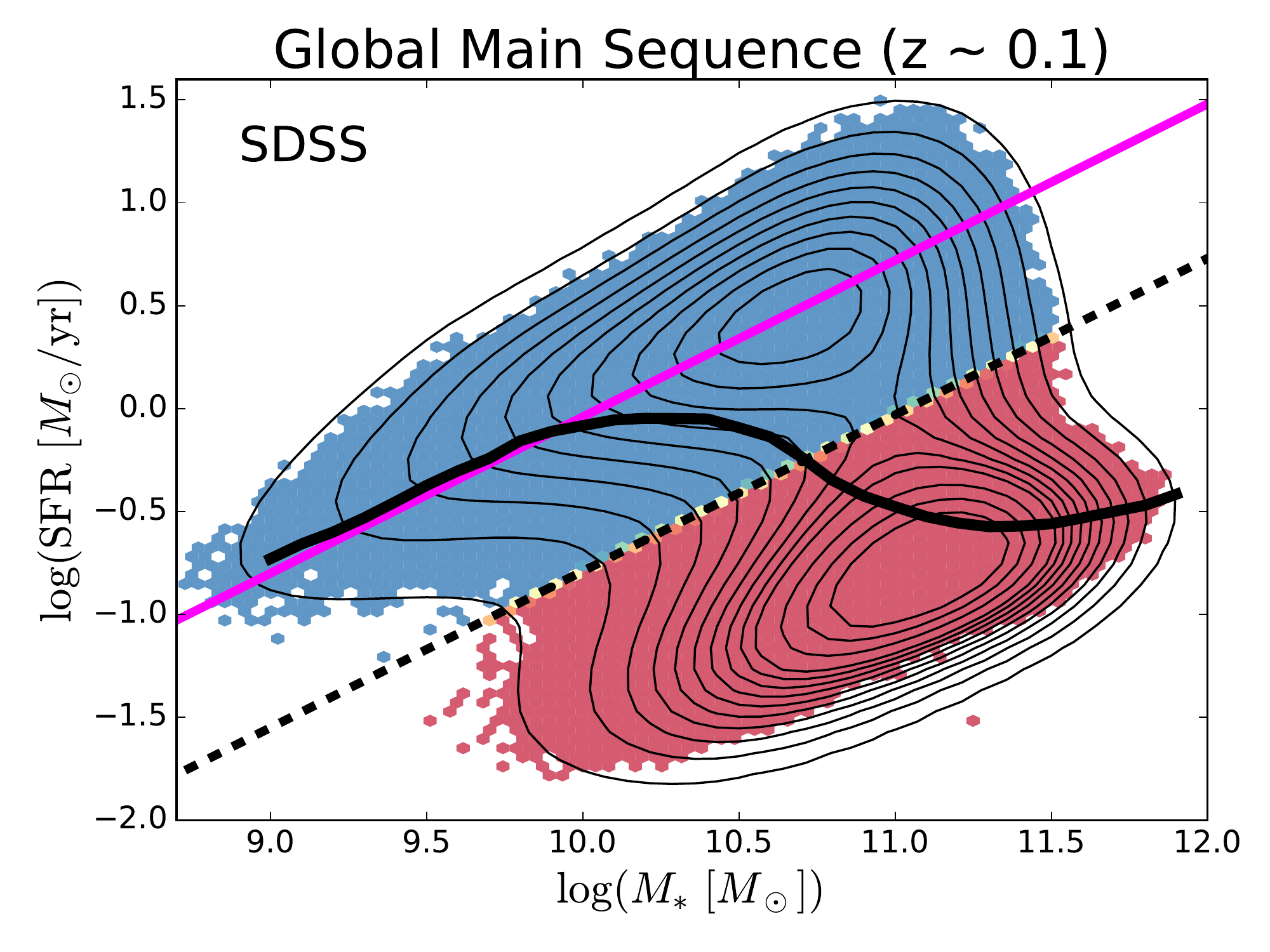}
\includegraphics[width=0.49\textwidth]{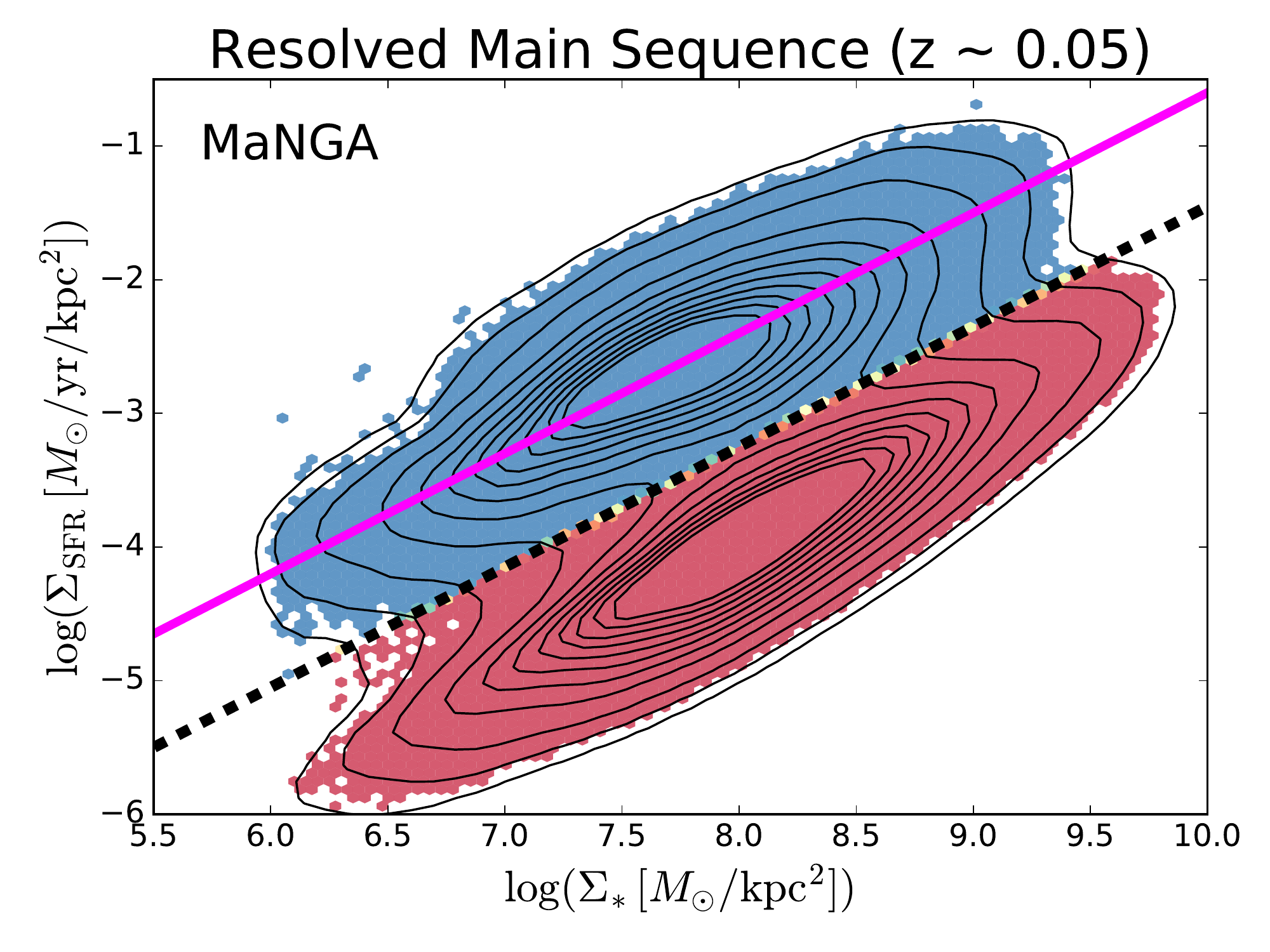}
\caption{Illustration of methods to classify star forming and quenched objects in the local Universe. {\it Left panel: } Global star forming main sequence (${\rm SFR} - M_*$ relation) for SDSS galaxies (at z $\sim$ 0.1). {\it Right panel: } Spatially resolved star forming main sequence ($\Sigma_{\rm SFR} - \Sigma_*$ relation) for MaNGA spaxels (at z $\sim$ 0.05). In each panel, density contours are shown as light black lines, revealing clear bimodality in both the global and resolved realisations of the star forming main sequence. However, it must be stressed that the quiescent population in both panels are only upper limits on their true SFRs and hence in reality form a sequence to arbitrarily low values. Our definition of star forming and quenched classes for both data sets are indicated by the colour shading (blue for star forming; red for quenched). Additionally, the quenching threshold is highlighted by a dashed black line and the best fit relation for the star forming subset is indicated by a solid magenta line, shown on each panel. For the global main sequence, the median SFR - $M_*$ relation is displayed by a thick black line, which shows a transition from star forming to quenched at $M_* \sim 10^{10.5} M_{\odot}$.}
\end{figure*}

In order to compare our multi-epoch observational results to a cosmological framework for galaxy evolution, we make use of the public release of the LGalaxies semi-analytic model (see \citealt{Henriques2015})\footnote{LGalaxies: \href{http://galformod.mpa-garching.mpg.de/public/LGalaxies/}{http://galformod.mpa-garching.mpg.de/public/LGalaxies/}}. LGalaxies (often referred to as the `Munich model') has a long history, with prior publications including: \cite{Kauffmann1993, DeLucia2006, DeLucia2007, Guo2011, Henriques2013}. In its current form, the model is applied to galaxy merger trees extracted from the Millennium Simulation (\citealt{Springel2005}) and Millennium-II (\citealt{BoylanKolchin2009}). The halo clustering is updated from these simulations utilising the approach of \cite{Angulo2010} to approximate the correct dark matter structural evolution for the Planck cosmological parameters (\citealt{Planck2014}). The detailed physical prescriptions for modelling the non-dark matter (i.e. baryonic) processes within galaxies have been developed throughout the model's history (see the supplemental material in \cite{Henriques2015} for a detailed account).

In this paper, we analyse two snapshots of the LGalaxies model (at z = 0.1 and z = 1), to compare to our low and high redshift observational data sets. More specifically, we utilise the application of the LGalaxies model to the full Millennium Simulation, with a comoving volume of $480.3^{3} [({\rm Mpc}/h)^3]$. In Appendix A we present a detailed account of the quenching and bulge formation models in LGalaxies (the aspects most relevant for our present study) and derive a novel mathematical feature of the model which is particularly useful for explaining our observational results. In Section 3 we present a statistical analysis of quenching in the LGalaxies model, which we test against multi-epoch observational data in Sections 4 \& 5.


\section{Method}

\subsection{Classifying galaxies \& spaxels}

In this paper we analyse the quenching of star formation in three observational galaxy surveys, with very different types of data at both a quantitative and qualitative level. In this section we explain precisely how we identify star forming and quenched galaxies, bulges, and disks, as well as sub-galactic regions within galaxies. Although the method is necessarily somewhat heterogeneous, in order to account for the available data, we make considerable effort to place observations on the same footing as much as possible.

\subsubsection{Global bimodality}

At $z \sim 0.1$ we utilise the SDSS to study the star formation quenching of local galaxies. Following \cite{Bluck2014, Bluck2016}, we utilise the global star forming main sequence (${\rm SFR} - M_*$ relation) to identify galaxies which are forming stars well below the expected level for actively star forming systems at that epoch. In Fig. 1 (left panel) we present the star forming main sequence for SDSS galaxies, with linearly spaced density contours shown in black. Clearly, there is pronounced bimodality in the distribution of galaxies in this 2D plane. Although it is important to appreciate that SFRs in the lower density peak are essentially just upper limits. Roughly speaking, we take the lower density peak to be quenched and the upper density peak to be star forming (as illustrated by the colour shading: red for quenched, blue for star forming). Additionally, we display the median ${\rm SFR} - M_*$ relationship on the left panel of Fig. 1 (solid black line), which shows a sharp transition from the star forming to quenched density peak at $M_{*} \sim 10^{10.5} M_{\odot}$.

In slightly more detail, we utilise the linear best fit for the main sequence ridge line from \cite{Renzini2015} to quantify the location of the star forming main sequence (which is shown on Fig. 1 left panel as a solid magenta line). We then define the distance (in logarithmic units) each galaxy resides at from the main sequence ridge line as:

\begin{equation}
\Delta {\rm SFR} \equiv \log({\rm SFR}) - \log({\rm SFR}_{\rm MS}(M_*)) ,
\end{equation}

\noindent where ${\rm SFR}_{\rm MS}(M_*)$ indicates the expected rate of star formation for a given stellar mass, if the system were forming stars exactly on the main sequence ridge line. Hence, galaxies with a value of $\Delta {\rm SFR} = 0$ are forming stars on the main sequence, galaxies with $\Delta {\rm SFR} > 0$ have enhanced levels of star formation relative to the main sequence, and galaxies with $\Delta {\rm SFR} < 0$ have star formation levels which are suppressed relative to the main sequence. It is important to appreciate that the SFRs of quenched galaxies are essentially upper-limits, whereas the SFRs for star forming galaxies are actual numerical estimates (see \citealt{Brinchmann2004}). As such, the exact value of $\Delta {\rm SFR}$ is often less informative than the class in which a galaxy is situated (defined below).

We identify the minimum in the distribution of $\Delta {\rm SFR}$ between the star forming and quiescent peak, and utilise this threshold to classify galaxies. The minimum occurs at $\Delta {\rm SFR} = -0.75$\, dex. Systems with $\Delta {\rm SFR} > -0.75$\, dex are deemed to be `star forming', and systems with $\Delta {\rm SFR} < -0.75$\, dex are deemed to be `quenched'. This transition is indicated by a dashed black line in the left panel of Fig. 1. Additionally, we define a buffer zone between star forming and quenched classes for some analyses, restricting the conditions to $\Delta {\rm SFR} > -0.5$\, dex for star forming and $\Delta {\rm SFR} < -1$\, dex for quenched (see Bluck et al. 2020a for further justification). None of our results are sensitive to the exact location of these thresholds. Furthermore, it should be noted that these values are somewhat arbitrary in nature, and are ultimately a function of the approximation method used for the low SFR systems (see e.g. \citealt{Brinchmann2004}).

\subsubsection{Local bimodality}

Also at low redshifts, we utilise the MaNGA IFU survey to incorporate spatially resolved information into our quenching analysis (as in Bluck et al. 2020a,b) for a sub-set of the SDSS legacy sample. In MaNGA, we have access to spatially resolved star formation rates, and hence it is important to construct an approach to classify spaxels (spectroscopic pixels) into star forming classes, as well as galaxies as a whole. To achieve this we follow precisely our methodology in \cite{Bluck2020b}. 

In the right panel of Fig. 1 we present the resolved star forming main sequence ($\Sigma_{\rm SFR} - \Sigma_*$ relation) for spaxels within MaNGA galaxies. As with the global main sequence (left panel of Fig. 1), we see a profound bimodality in the distribution of star formation at the local (i.e. spatially resolved) level as well (see \citealt{Bluck2020a} for a discussion on this interesting fact). However, it is important to appreciate that the lower density peak represents upper limits (as in the analogous region of the SDSS global main sequence). As such, we adopt an analogous method for classifying spaxels in MaNGA to the SDSS. More specifically, we define the logarithmic distance to the resolved star forming main sequence as:

\begin{equation}
\Delta \Sigma_{\rm SFR} \equiv \log(\Sigma_{\rm SFR}) - \log(\Sigma_{\rm SFR, MS}(\Sigma_*)),
\end{equation}

\noindent where we take $\Sigma_{\rm SFR, MS}(\Sigma_*)$ as the least squares linear fit to the resolved main sequence from \cite{Bluck2020a}, which is shown as a solid magenta line in the middle panel of Fig. 1. The minimum in the distribution of $\Delta \Sigma_{\rm SFR}$ occurs at -0.85\,dex, which we show as a dashed magenta line in the right panel of Fig. 1. We use this threshold to separate star forming and quenched spaxels. Additionally, we define a buffer zone of $-1.1 \, {\rm dex} < \Delta \Sigma_{\rm SFR} < -0.6 \, {\rm dex}$ for some analyses, i.e. removing spaxels with ambiguous levels of star formation located in the resolved `green valley'. As with the SDSS, our results are highly stable to the exact location of these thresholds.

It is important to acknowledge that the two primary methods used to quantify star formation rates in this work (i.e. dust corrected H$\alpha$ luminosity and the strength of the 4000\AA\hspace{0.05cm} break) do not precisely trace the same physical process, and consequently have (slightly) different timescale sensitivities. More specifically, emission lines trace the HII-regions around highly ionising stars, whereas the D4000 index is most sensitive to the presence or absence of O- and B-class stars in the integrated stellar continuum. Consequently, there is a timescale difference of $\sim$ 20-30 Myr in sensitivity between the methods. This issue is discussed at length for the SDSS sample in \cite{Brinchmann2004}. For the MaNGA sample, in \cite{Bluck2020a} we performed a variety of tests on our heterogeneous SFR method. Briefly, we found that there is excellent agreement between the various methods for classifying galaxies, even if the precise SFRs recovered differ to a modest degree. Hence, for the classification approach adopted in this work, we are not concerned about small ($\leq$100 Myr) differences in timescale sensitivity. Readers interested in this issue are encouraged to read the appendices of \cite{Bluck2020a} for full details.

\begin{figure}
\includegraphics[width=0.49\textwidth]{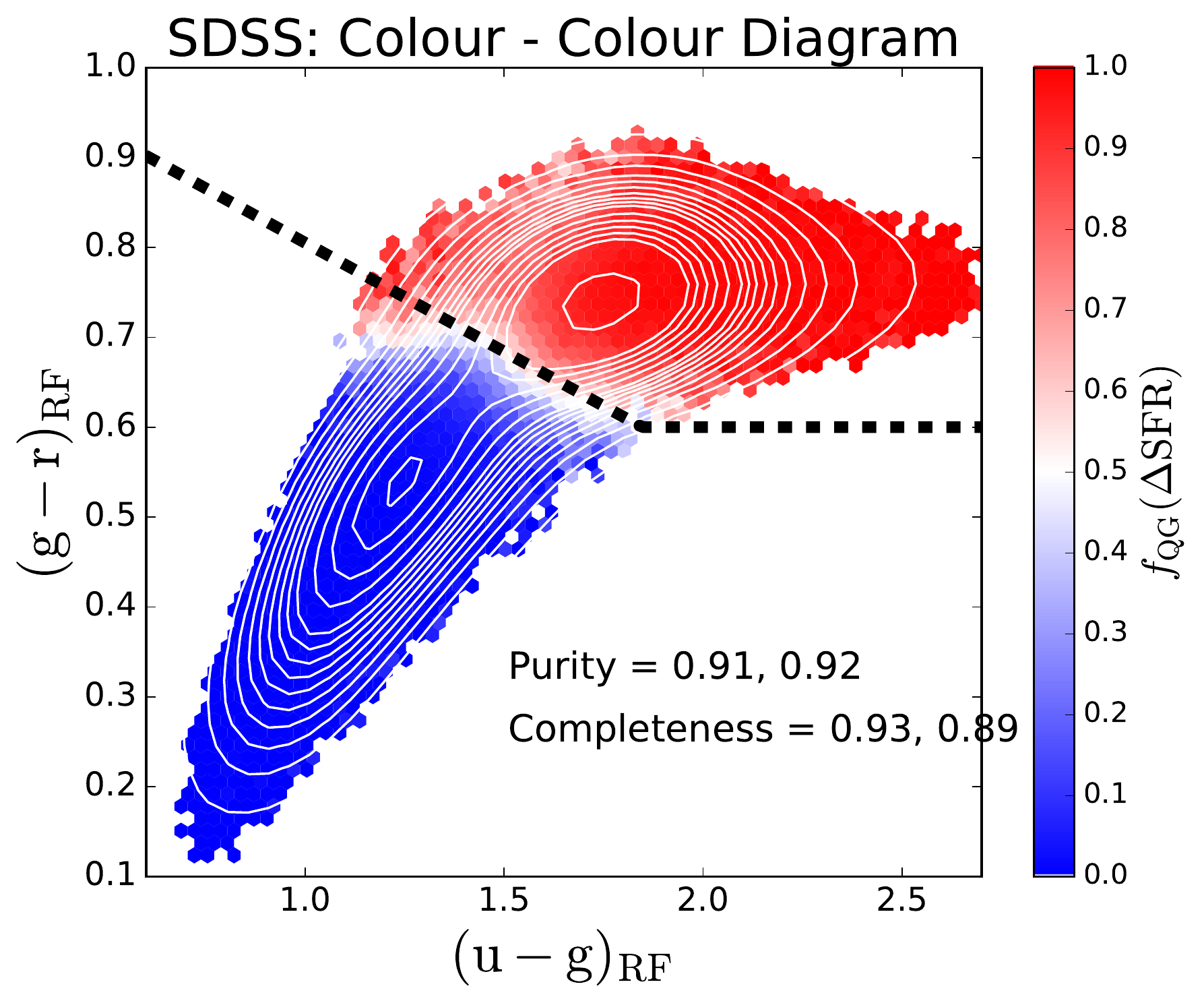}
\caption{Location of SDSS galaxies in rest-frame (g-r) -- (u-g) colour space. Only galaxies with a solid star forming vs. quenched classification are included in this figure. Density contours are shown as light white lines, indicating that bimodality in $\Delta$SFR is accompanied by bimodality in colour. Additionally, the (g-r) -- (u-g)  plane is subdivided into small hexagonal regions, each colour coded by the fraction of quenched galaxies (as indicated by the colour bar). The optimal linear decision boundary in colour space to classify star forming and quenched systems is shown by a dashed black line. Overlaid on the figure is the purity and completeness obtained by this cut (shown respectively for red, blue objects).}
\end{figure}

\subsubsection{Colour-based Classification}

At high redshifts, we utilise the CANDELS galaxy survey which is (to leading order) a photometric-only survey. As such, we do not have access to spectroscopic star formation rates unlike with the low-$z$ data. As an alternative, we utilise a rest-frame, dust corrected UVJ colour - colour classification of galaxies into star forming and quenched classes for the CANDELS dataset (see \citealt{Williams2009, Dimauro2018}). We defer a detailed discussion of our method for classifying CANDELS galaxies to Section 5.1.

Additionally, we leverage the power of resolved spectroscopy in Section 4.3 to study quenching within bulge and disk structures, utilising data from the MaNGA survey. This part of the method is explained in detail in Section 4.3.1. However, the SDSS has a 100-fold increase in the number of galaxies over MaNGA, and hence it is also highly interesting to attempt to explore resolved quenching in this data set as well. 

Utilising the photometric bulge - disk decompositions of \cite{Simard2011} and \cite{Mendel2014}, we have access to rest-frame magnitudes and colours for SDSS galaxies, and their component bulge and disk structures. Hence, if we can find a reliable method to classify star forming and quenched objects based on colour we can additionally explore quenching separately in bulges and disks. To this end, we have explored a variety of extant and novel methods for classifying galaxies into quenched and star forming categories via their rest-frame colours. In the remainder of this sub-section we present one highly effective method, which we utilise in Section 4.3 to probe bulge and disk quenching in the SDSS.

In Fig. 2 we show the distribution of $\Delta$SFR selected star forming and quenched galaxies in rest-frame (g-r) - (u-g) colour space. There is a remarkable separation of star forming and quenched objects in this colour diagnostic diagram (as indicated by the colour of each hexagonal region). We utilise a non-linear optimisation algorithm, based on {\small LMFIT}\footnote{Website: lmfit.github.io/lmfit-py/},  to find the optimal decision boundary between star forming and quenched objects in the (g-r) - (u-g) plane. To achieve this, we simultaneously optimise for red purity, blue purity, red completeness, and blue completeness. Red purity is defined as the fraction of red galaxies (at the colour - colour decision boundary) which are quenched, and red completeness is defined as the fraction of quenched galaxies which are red (at the same decision boundary). The blue purity and blue completeness are defined in exact analogy to the red cases.

The optimal linear colour - colour boundary for selecting quenched systems is at:

\begin{equation}
(g-r)_{\mathrm{RF}} > -0.234 \, (u-g)_{\mathrm{RF}} + 1.03,
\end{equation}

\noindent where the subscript, RF, indicates rest-frame quantities. The typical uncertainties on the coefficients are $\sim$5\% of the values, as determined from bootstrapped random sampling. We additionally require that $(g-r)_{\rm RF} > 0.6$ (see e.g. \citealt{Bluck2014}). 

Using the above cuts in colour space we achieve $\sim$90\% purity and completeness for both the red and blue populations. Hence, we can use colour as a proxy for $\Delta$SFR to classify galaxies into star forming and quenched classes with a very high level of accuracy, and an equally high level of completeness (even though the colours are not explicitly dust corrected). The power of this approach is that it can then be applied to bulge and disk colours to probe the quenching of these sub-galactic structures for the full SDSS data set. Note that we could select slightly higher still purity (or completeness), but this always comes at the expense of the other statistic. As such, the above strategy is the optimal compromise between purity and completeness in this colour diagnostic. It is also worth noting that a slight improvement may be achieved using a polynomial decision boundary. However, more complex decision boundaries are more likely to pick up pathological features in the data, and hence are less likely to be universal in application.

Given that the colours we utilise for classification in the SDSS are not dust corrected (unlike for CANDELS), this may cause an additional bias for disk structures (which are more prone to dust extinction than bulge structures for both geometric and galaxy evolution reasons). For galaxies as a whole, we know from the above analysis that colours yield a very good proxy for $\Delta$SFR (despite the lack of explicit dust correction). Yet, for disks, the accuracy of this method is likely to be reduced. To combat this issue, in Section 4.3 we additionally impose axial ratio cuts to minimise the impact of dust extinction on our results. We note in advance that our results are stable to even the most severe cuts in $b/a$ (yielding only face-on disks in the sample). Moreover, we also perform a complementary analysis of bulge and disk quenching in MaNGA where the impact of dust is fully removed, from explicit dust correction of emission lines (via the Balmer decrement) or the use of the D4000 break. See \cite{Bluck2020a} for full details on the MaNGA SFRs, including detailed tests with alternative methods (e.g. from SED fitting).


\subsection{Random forest classification}

\subsubsection{Overview}

Throughout this paper we utilise a Random Forest classifier to predict the star forming state of galaxies, bulges, disks and spaxels. The Random Forest approach is a powerful generalisation of the decision tree, enabling the identification of highly non-linear classification boundaries in complex multi-dimensional data. Moreover, the Random Forest method also provides a natural and intuitive definition of the importance of any given feature (i.e. variable made available to the classifier) for the task of predicting a given class. We utilise the prior classification of galaxies and spaxels into star forming and quenched categories from the preceding section as the training (and testing) sets for our Random Forest. The goal of this approach is not to predict unknown star forming states (although it could be used for this purpose), but rather to ascertain how connected various parameters are to the process of quenching.  

For all variants of Random Forest classification used in this paper, the immediate task for the classifier is to learn to predict the star forming state (i.e. quenched or star forming) of certain data, based on the available features. For the SDSS and CANDELS the target class is for galaxies (and bulges and disks treated separately), whereas for MaNGA the target class pertains to spaxels instead (in various groupings). The input features used by the Random Forest to predict the star forming class are separated in this work into two sets: a Photometric Bulge - Disk Parameter set, which consists of bulge mass ($M_B$), disk mass ($M_D$), total stellar mass ($M_*$), and $B/T$ structure; and an Expanded Parameter set, which incorporates the photometric bulge - disk parameters (above) plus dark matter halo mass ($M_{\rm Halo}$), central stellar velocity dispersion ($\sigma_\star$), and local galaxy over-density evaluated at the 5th nearest neighbour ($\delta_5$). 

We then extract quantitatively how effective each parameter is at accurately separating the classes (see Appendix B.1 for details). Note that the first set of parameters are all measurable in photometric data, and hence do not explicitly rely on spectroscopy; whereas the second set of parameters can only be measured reliably (or at all) in spectroscopic data. The point of the separation is to enable fair and consistent comparison between different galaxy surveys, with different available parameters. Before using the above data for training, we first ensure that all parameters are given in logarithmic units, subtract the median value, and normalise by the interquartile range. This approach ensures that all data are treated equally by the Random Forest classifier, ensuring no bias in the calculation of relative importance.

\subsubsection{Practical implementation \& avoiding over-fitting}

In Appendix B.1 we present a detailed mathematical description of the Random Forest architecture (which we recommend anyone unfamiliar with this form of machine learning to read before proceeding). Additionally, in Appendix B.2 we present several detailed tests on the Random Forest approach for our purpose of extracting causal insights from complex, inter-correlated data. Importantly, we find that in the `All Parameter' mode, the underlying causality within mock data can be clearly identified in realistic (i.e. noisy, non-linear and probabilistic) data for levels of inter-correlation up to $\rho \sim 0.99$. This remarkable property of the Random Forest technique is ideal for the scientific goals of this paper, i.e. to utilise classification as a tool to reveal causality in galaxy evolution.

As in \cite{Bluck2020a, Bluck2020b}, for the practical application of Random Forest classification in this paper we utilised {\small RANDOMFORESTCLASSIFIER} from the powerful {\small SCIKIT-LEARN} python package\footnote{Scikit-Learn website: \href{https://www.scikit-learn.org}{https://www.scikit-learn.org}} (\citealt{Pedregosa2011}). We separated each sample into a training and testing set, each containing half of the available data. In training, we allowed each tree to develop to a maximum of 250 nodes (an arbitrary high number, never actually needed) and utilised 250 decision trees per forest, with differences ensured by bootstrapped random sampling. We adjusted the {\it min-samples-leaf} hyper-parameter ($\chi$) so as to avoid over-fitting the data. This specifies the number of objects required in a node in order to split further.

Specifically, we systematically reduced the $\chi$ hyper-parameter from an initial high guess limit in order to increase the overall performance on the Random Forest classifier in training. We quantified the overall performance of the classification by the area under the receiver operator true positive - false positive curve (AUC; see \citealt{Teimoorinia2016, Bluck2019} for details). This statistic is highly correlated with the more intuitive `fraction of correct classifications', but has several mathematical properties which are advantageous for technical reasons (see \citealt{Teimoorinia2016}). Without any further constraint we could reduce $\chi$ all the way to a value of two (the logical minimum needed to split), which will naturally lead to the most accurate performance in training. However, this does not necessarily lead to the most accurate performance in unseen data (i.e. in the testing phase). This is the crucial problem of over-fitting. To combat this problem, we additionally required that the AUC achieved on the unseen test data (not used in training) is similar in value to the AUC achieved in training. For galaxy-level objects, we required a very small difference in AUC of $|\Delta {\rm AUC}| \leq 0.01$; whereas in spaxel-level data we relax this slightly to $|\Delta {\rm AUC}| \leq 0.02$, to allow slightly more flexibility with this much more complex data. As a result, we ensured that our Random Forest classifier does not learn pathological features of the training data, but only those features which extrapolate well to unseen data.

\begin{figure*}
\includegraphics[width=0.49\textwidth]{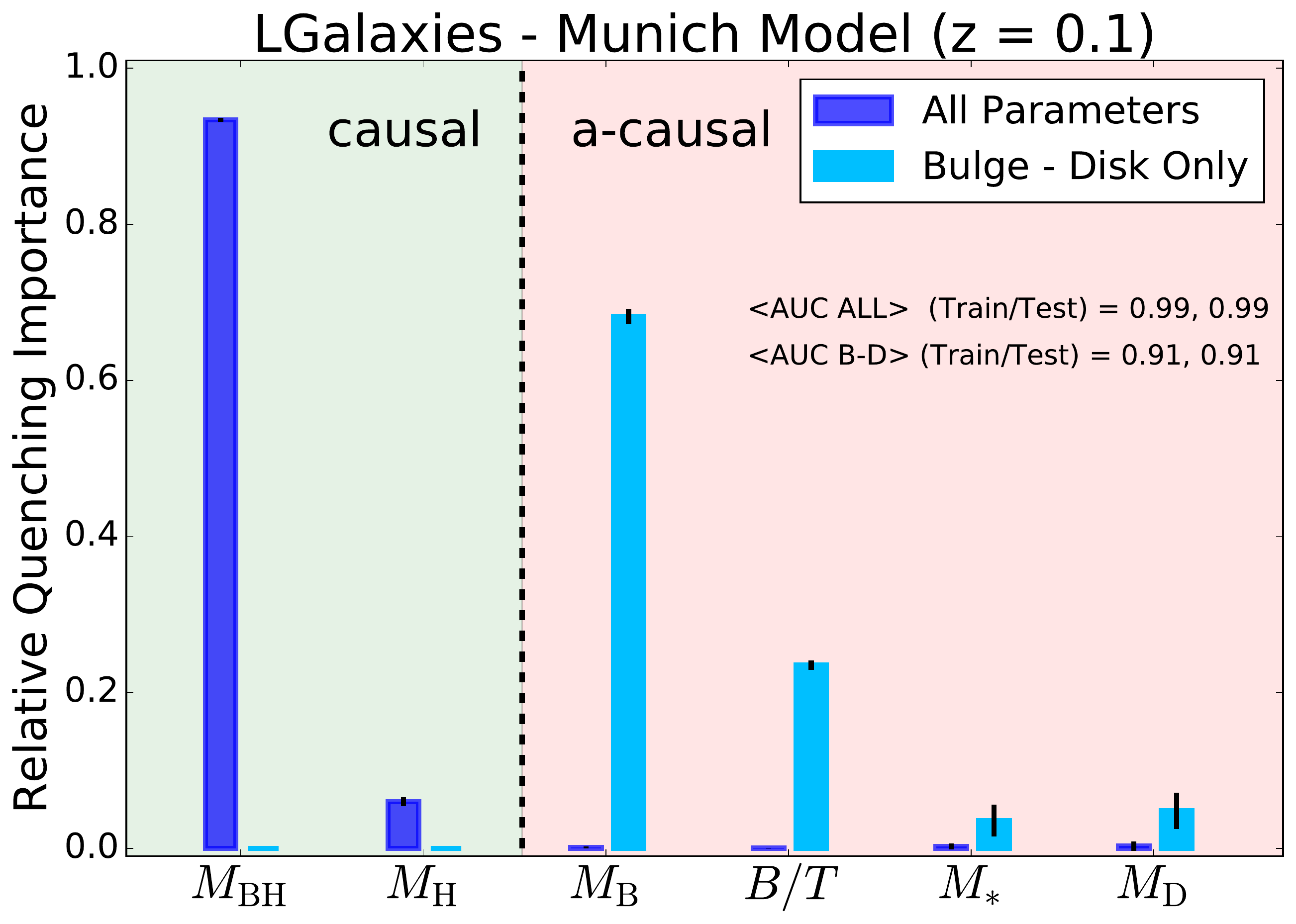}
\includegraphics[width=0.49\textwidth]{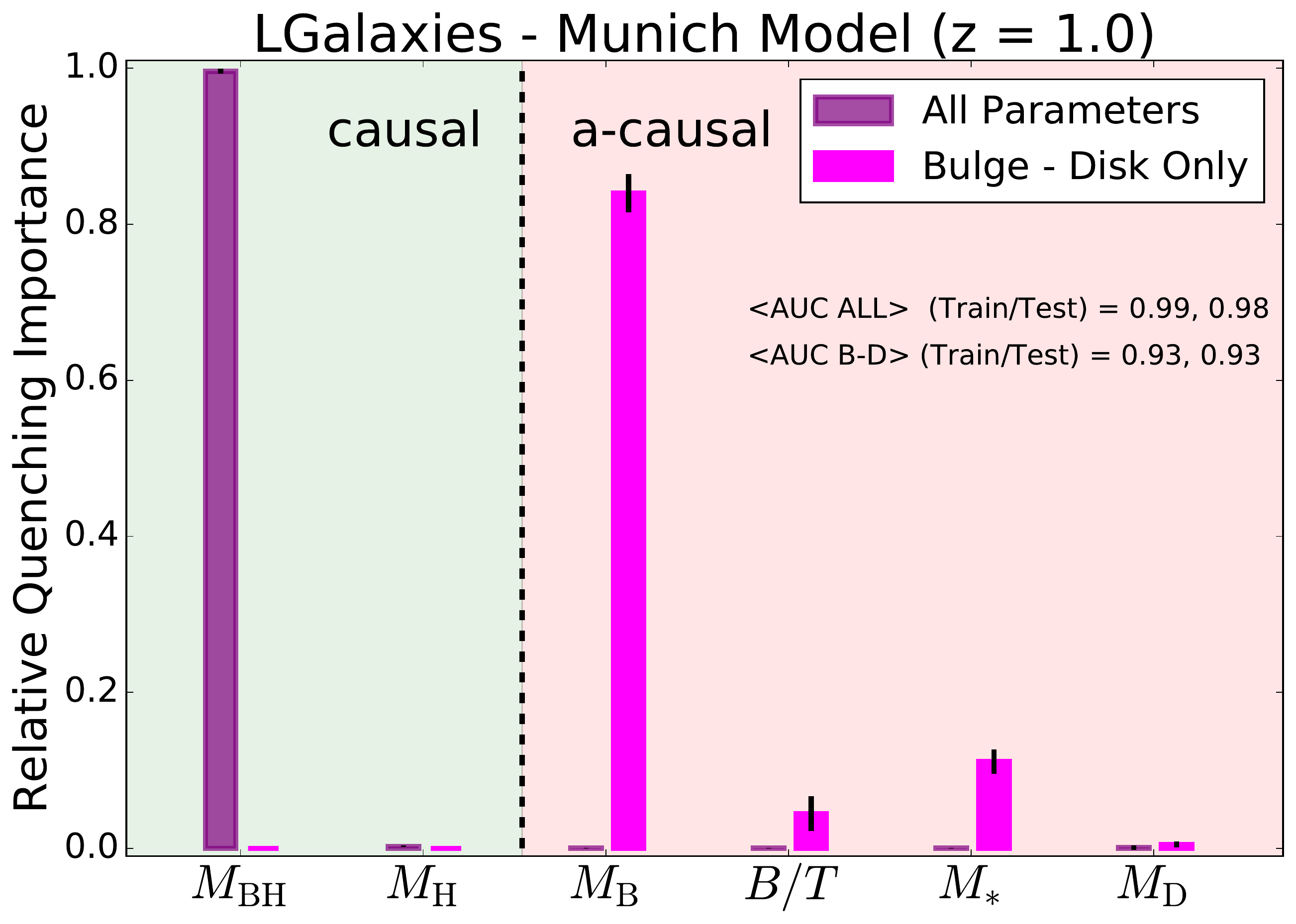}
\caption{Random Forest classification analysis to predict the quenching of central galaxies in the LGalaxies semi-analytic model. The left panel shows results from the z = 0.1 snapshot (shown in blue colours) and the right panel shows results from the z = 1.0 snapshot (shown in purple colours). Results are shown separately for the bulge - disk parameter set (i.e. $M_B$, $M_D$, $M_*$ and $B/T$) and for this set plus black hole mass ($M_{BH}$) and dark matter halo mass ($M_{H}$), as indicated on the legends. Quenched galaxies are defined in training to be forming stars at least an order of magnitude below the main sequence. The AUC for training and testing is shown on each panel for each data set. Errors on the relative importances are given as 5 $\times$ the dispersion across ten independent training and testing runs. Coloured shading on the plots indicates which parameters are known mathematically to be causal or a causal for quenching. For the full parameter set, it is absolutely clear that $M_{BH}$ is ultimately the most important parameter, and essentially no importance is given to any other parameter (except at a very low level with $M_H$ at low redshifts). This is expected in the model since LGalaxies quenches centrals exclusively through preventative AGN feedback. Hence, our RF classifier correctly extracts the causation in the LGalaxies model. In the absence of $M_{BH}$ \& $M_{H}$, bulge mass is clearly found to be the most important parameter governing quenching at both redshifts, which may be interpreted as a key prediction of the model in the a causal bulge - disk parameter space.}
\end{figure*}

As a final test, for each Random Forest classification, we iterated ten times over the training and testing phase, randomly selecting different galaxies or spaxels for training and testing in each case. We took the final performance of the Random Forest classification to be the mean of the ten AUC values, and extracted the average feature importance for each variable across the ten independent sets. Thus, in total the final feature importance quoted is based on the usefulness of that parameter to constrain the class of objects in 2500 independent decision tress, containing up to 250 decision forks in each. Consequently, our Random Forest method is capable of accurately modelling highly non-linear and complex boundaries between classes in the multi-dimensional parameter space, whilst still being conceptually simple enough to accurately interpret the meaning of the results. Finally, we took the variance across the feature importances for each parameter across the ten independent Random Forest training and testing runs as the statistical uncertainty on the mean feature importance.

Ultimately, the power of the Random Forest classification technique lies in its competitive nature. At each fork in each decision tree, features `compete' for the opportunity to be used to separate the data. Only the variable which achieves the greatest reduction in Gini impurity is used. This process is repeated at all forks throughout all trees in the Random Forest to achieve the final classification prediction. Hence, for two highly correlated parameters, one of which is taken to be fundamental for the classification at hand and one of which is merely incidental, the slight difference in their performance (guaranteed by their high correlation) will be amplified by the choices made by the classifier at each node. Consequently, the Random Forest is able to pick up on small differences in the data and break subtle degeneracies in the relationships of the parameters to the classification task. We demonstrate the power of this characteristic of the Random Forest in the following sub-section (and in further detail in Appendix B).

\subsection{Semi-analytic model test \& predictions}

The purpose of this sub-section is two-fold. First, we test the ability of our Random Forest classifier to extract the known causal dependence of quenching in a semi-analytic model. This tests the capability of our statistical machine learning method in a much more complex (and directly relevant) application than considered in Appendix B.2. Second, we extract detailed quantitative predictions from the semi-analytic model, in both the full parameter space (incorporating all causal parameters) and in a useful sub-space restricted to bulge - disk parameters (which are relatively straightforward to measure in photometric data). Throughout the results sections of this paper we rigorously test these predictions against data from a variety of observational galaxy surveys.

Here we analyse the LGalaxies semi-analytic model (often referred to as the `Munich Model', see \citealt{Henriques2013, Henriques2015}). The advantage of utilising a semi-analytic model (as opposed to a hydrodynamical simulation) for this test is that the physical processes at work in quenching are much clearer in the semi-analytic model. Indeed, a large amount of post-processing is required in a hydrodynamical simulation to ascertain what is cause and what is effect in any given process (see \citealt{Piotrowska2021}). Alternatively, in semi-analytics, galaxy evolution is modelled as a coupled set of partial differential equations, and hence it is much more straightforward to ascertain precisely which process is responsible for any given observable. 

To this end, in Appendix A we give a thorough description of the quenching model in LGalaxies, and present a novel reconceptualisation of the quenching criterion, which is particularly instructive for comparison to our observational results\footnote{We encourage readers interested in the full mathematical description of quenching in LGalaxies (and all of the assumptions involved) to read Appendix A before continuing to the statistical analyses presented here.}. Briefly, intrinsic galaxy quenching in LGalaxies occurs solely as a result of preventative `radio-mode' AGN feedback, which leads to a dominant causal dependence of central galaxy quenching on black hole mass. Additionally, there is a strong correlation between bulge mass and supermassive black hole mass in the model, which emerges as a consequence of both components growing primarily in merger events. These two features of the model are the most important aspects for understanding the results of this sub-section.

In Fig. 3 we present results from several Random Forest classification analyses of the LGalaxies semi-analytic model. On the left-hand panel, we analyse the z = 0.1 snapshot, which may be compared to our observational analysis of the SDSS \& MaNGA at the same epoch (see Section 4). On the right-hand panel, we analyse the z = 1 snapshot of the LGalaxies model, which may be compared to the full redshift range CANDELS analysis (see Section 5). For both snapshots, we restrict to central galaxies to minimise the impact of environment, and we consider two groupings of the data: (i) the photometric bulge - disk parameters (i.e. $M_B$, $B/T$, $M_*$, $M_D$); and (ii) the bulge - disk parameters plus supermassive black hole mass ($M_{BH}$) and dark matter halo mass ($M_{\rm Halo}$). Parameters are listed along the $x$-axis and the relative importance for quenching is displayed on the $y$-axis. Uncertainties on the relative importances are given as the dispersion across ten independent training and testing runs, with bootstrapped random sampling of the data.

When $M_{BH}$ is made available to the Random Forest classifier, it is clearly identified as the most important parameter governing quenching in the simulation (with $>10\sigma$ confidence at z = 0.1 and z = 1). Furthermore, $M_{BH}$ has a relative importance greater than a factor of ten times higher than the second most predictive variable ($M_{\rm Halo}$) at both epochs. The success of black hole mass is no surprise in the LGalaxies model, since galaxies quench exclusively through radio-mode AGN feedback, which we have shown essentially reduces to a simple threshold on $M_{BH}$ at each epoch (see Appendix A). Additionally, the much weaker secondary dependence on halo mass is also well understood analytically in the model, which is a result of halo cooling via bremsstrahlung (see Appendix A). Hence, our Random Forest architecture is capable of unambiguously identifying the most important quenching parameter in the model at both epochs. This is very reassuring indeed, and clearly demonstrates the value of the Random Forest approach for extracting the underlying causal dependence in complex multi-dimensional and inter-correlated data. Of course, it is not possible to perform a similar test on observational data (since the underlying causation is fundamentally unknown). Hence, this justifies our use of a model with known causation for this test on our method.

Additionally in Fig. 3, we consider the Random Forest classification in the absence of black hole and halo mass, i.e. restricting to the bulge - disk parameters. In essence, this explores the causal projection of AGN quenching into the a causal bulge - disk parameter space. In the absence of $M_{BH}$ and $M_{\rm Halo}$, bulge mass is found to be the most predictive parameter governing quenching in the LGalaxies model, at both epochs considered. This is a direct consequence of the preventative AGN feedback model, in conjunction with the bulge and black hole formation models, implicated in LGalaxies (see Appendix A for full details). Importantly, this can be understood as a key prediction of the model, which can be tested in multi-epoch observational galaxy surveys. Ultimately, the value of looking at this projection is due to the bulge - disk parameters being much easier to accurately measure in observational data than black hole or halo masses. One other highly advantageous feature of this projection is that the bulge - disk parameters can be measured with very similar levels of measurement uncertainty, typically 0.2-0.3 dex, and hence the potential concerns of differential measurement uncertainty are largely removed (see the discussion in Appendix B.2).

In addition to the explicit predictions from the LGalaxies AGN quenching model outlined above, there is also a very important implicit prediction. Since the quenching of central galaxies in the model depends only on black hole mass, redshift and (weakly) halo mass, there is no room for significant sub-galactic variation in star forming state within galaxies. Note that this is a modelling choice in LGalaxies, and not a requirement of SAMs in general. This is the case because the SFR in bulge and disk regions are regulated separately in SAMs, and hence it is possible to have, for example, quenched bulges and star forming disks in principle. Yet, once the black hole mass exceeds a given threshold (determined by the mass of the dark matter halo and epoch), gas accretion into the system will be permanently terminated. Hence, no replenishment of gas (used as fuel for star formation) will enter the system. Consequently, stars will cease to form throughout the galaxy. Therefore, the model clearly predicts that the quenching of all components within a galaxy (i.e. bulge and disk) will ultimately depend on the same fundamental parameters.

It is important to appreciate that very similar results are also found in leading cosmological hydrodynamical simulations as well (see \citealt{Piotrowska2021}) and hence these predictions are quite general to the paradigm of AGN feedback quenching. Moreover, this is the dominant theoretical mechanism for quenching central galaxies in the literature (e.g. \citealt{Vogelsberger2014, Vogelsberger2014a, Schaye2015, Zinger2020, Terrazas2020}). In the Discussion, we additionally consider whether alternatives to the AGN feedback paradigm of intrinsic galaxy quenching (e.g. virial shocks, supernova feedback and morphological stabilisation) may offer viable explanations to our observational results.


\section{Quenching in the local universe}

In \cite{Bluck2014} we declared that `Bulge mass is King (or Queen)', in the sense that bulge mass exhibits a tighter relationship with quenched fraction than disk mass, total stellar mass, or bulge-to-total stellar mass ratio ($B/T$). As such, this is in qualitative agreement with the prediction from preventative AGN feedback models (see Section 3.3). To determine this result, we adopted the conventional technique of assessing how much variation is exhibited in the relationship between each parameter and the quenched fraction by varying each other parameter in turn. The parameter with the least variation in its relationship with quenched fraction (i.e. the tightest relation) was deemed to be the most fundamental parameter governing quenching. Although reasonable (and indeed common in the astronomical literature), this approach is both highly inefficient and sub-optimal for extracting causal insight.

In this section, we revisit the SDSS dataset, and the bulge - disk decompositions of \cite{Mendel2014}, to provide an updated analysis with the Random Forest technique. In Section 3 (and in Appendix B) we have shown that this technique is extremely effective at isolating causal relationships in model and mock data. Here we find that our Random Forest classifier recovers the known observational results of global (galaxy-wide) quenching as in \cite{Bluck2014}. Moreover, we place these results on a much firmer statistical footing, by rigorously testing the potential impact of sample variation, volume completeness, and the technical limitations of the bulge - disk decompositions on our prior results. Having established the effectiveness of the new method, and the robustness of our prior conclusions, we then consider the quenching of bulge and disk structures treated separately within the SDSS for the first time. Additionally, we compare these results to a full spatially resolved analysis of a sub-set of SDSS galaxies as observed in the MaNGA survey. Taken as a whole, this section provides a robust z $\sim$ 0 baseline of quenching dependence in galaxies, bulges and disks. In the next Section we extend this analysis to higher redshifts, utilising CANDELS.


\begin{figure*}
\includegraphics[width=0.49\textwidth]{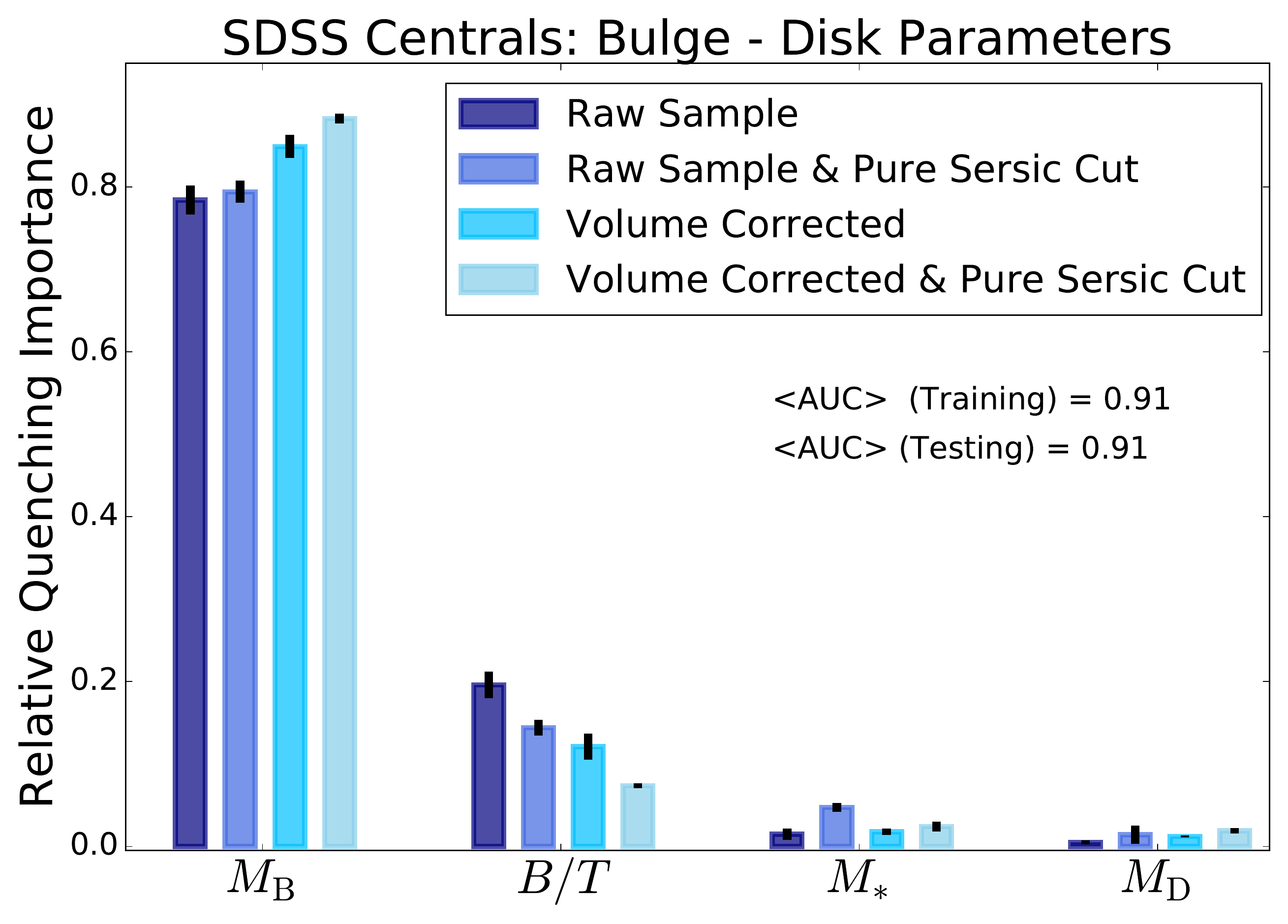}
\includegraphics[width=0.49\textwidth]{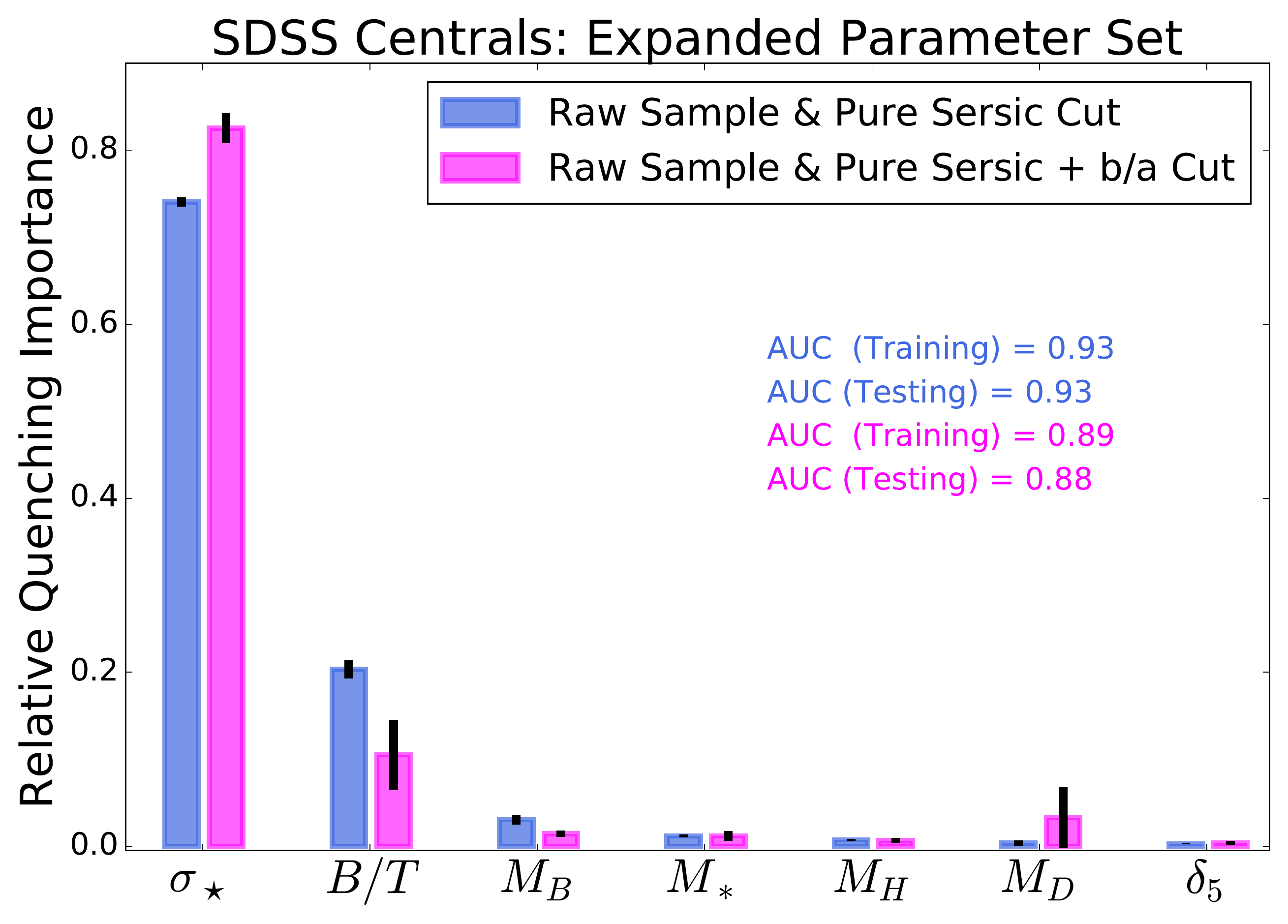}
\caption{Random Forest classification analysis to predict the quenching of central galaxies in the SDSS. {\it Left panel:} Random Forest analysis for the bulge-disk parameters (i.e. $M_B$, $M_D$, $M_*$ and $B/T$). The $y$-axis indicates the relative importance of each parameter for predicting whether central galaxies are star forming or quenched, and the $x$-axis indicates each parameter used to train the Random Forest in turn (ordered from most to least predictive of quenching).  The Random Forest classification analysis is repeated for four different samples: i) raw data (as in \citealt{Mendel2014});  ii) raw data with a pure S\'{e}rsic cut (defined and motivated in Section 2); iii) $1/V_{\rm max}$ volume weighted sample; iv) volume weighted sample with pure S\'{e}rsic cut. The error on each relative importance for each sample is taken as the dispersion across ten independent training and testing runs. It is clear, and highly significant, that bulge mass is the most predictive parameter governing quenching in all sample variants. {\it Right panel: } Random Forest analysis for a broader set of parameters, shown for comparison. Here stellar velocity dispersion ($\sigma_{\star}$), dark matter halo mass ($M_H$), and local galaxy over-density ($\delta_5$) are added to the bulge - disk parameters. This analysis is performed with the raw (unweighted) sample with S\'{e}rsic cuts (shown in blue), and with a sample with an additional cut in axial ratio (shown in magenta). It is clear that $\sigma_\star$ is overwhelmingly the most predictive parameter of central galaxy quenching in the expanded data set.}
\end{figure*}

\subsection{SDSS: random forest classification}

Ultimately, the purpose of this section is to establish which parameters are most predictive of quenching in central galaxies in the SDSS. We utilised a Random Forest to classify central galaxies into star forming and quenched categories on the basis of their bulge - disk parameters. The Random Forest method is discussed in Section 3.2 and in more detail in Appendix B. Briefly, for training, we utilised our $\Delta$SFR cut to pre-classify systems as either star forming or quenched. In this process we removed $\sim$10\% of the sample from the green valley with ambiguous levels of star formation. We trained the Random Forest to make the classification between star forming and quenched objects on the basis of their bulge - disk parameters. The trained Random Forest was then applied to novel data (unseen by the Random Forest in training). 

In both training and testing a `balanced' sample, containing an equal number of star forming and quenched systems, was selected. Throughout the multiple runs of the Random Forest, we explored the full parent data set, but always trained and tested on the same number of star forming and quenched objects. Hence, one class was deliberately under-sampled in each iteration. This is important to avoid trivial biases associated with one class being more frequent than another (see \citealt{Pedregosa2011, Teimoorinia2016}). Nonetheless, we have also tested that our results are stable to modest departure from even sampling.

We optimised the Random Forest to maximise the performance on the test sample (50\% of the full dataset), whilst yielding a good agreement in performance in the training sample (50\% of the full dataset) to avoid over-fitting. We extracted the feature importance for each parameter (see Appendix B.1). Finally, we repeated the entire analysis ten times over, for different randomly chosen training and testing samples. We took our final relative importance result as the mean of the ten runs, and the statistical uncertainty on the importance as the variance across the ten runs. Following the results of extensive testing (see Appendix B.2), we utilised the All Parameter mode\footnote{We note that in the SciKit-Learn application this is achieved by setting {\it max-features} = None.} for the Random Forest throughout this section (and indeed throughout the rest of the paper).

\subsubsection{The bulge - disk parameters}

In Fig. 4 (left panel) we show the relative importance for predicting whether galaxies are star forming or quenched from the photometric bulge - disk parameters. Parameters are arranged from most to least predictive of quenching along the $x$-axis. We repeat the analysis for four different representations of the SDSS bulge - disk parameter set (shown in different shades of blue, as labelled by the legend). Specifically,  we consider the following samples: (i) the raw \cite{Mendel2014} catalog, applying only the essential minimum data quality cuts (as discussed in Section 2); (ii) the application of a pure S\'{e}rsic cut on the basis of the $F$-statistic, allowing systems to appear as pure disk or pure spheroid in the sample if they are fit better (or as well) with a single S\'{e}rsic model; (iii) the raw data weighted by $1/V_{\rm max}$ to yield the statistical appearance of a volume complete sample; and (iv) the pure S\'{e}rsic corrected sample also volume corrected\footnote{It is important to appreciate that whilst the parent samples will in general have different numbers of star forming and quenched systems, we always select a `balanced' sample for both training and testing. The full parent sample is eventually probed through numerous iterations, equally sampling star forming and quenched systems in each case.}. 

All four sample variants yield essentially identical results, with bulge mass being consistently found to be overwhelmingly the most predictive parameter governing central galaxy quenching. Bulge mass is followed by $B/T$, total stellar mass, and finally disk mass as the next most important variables. It is striking how much more predictive power the bulge structure has over central galaxy quenching than information about any other bulge - disk parameter. It is also very interesting to note that bulge mass is much more predictive than either $B/T$ morphology or the total stellar mass of the galaxy. This result agrees with the much simpler statistical analysis of Bluck et al. (2014), therefore we confirm that `bulge mass is king (or queen)'. Additionally these results are in close accord with the prediction for central galaxy quenching in terms of the bulge - disk parameters from the LGalaxies model (see Fig. 3, left panel).

Of course, the bulge - disk parameters are inter-correlated with each other (as is almost invariably the case with extragalactic data sets). Additionally, only two out of the four parameters are needed to extract the full set. Consequently, one might wonder as to whether the redundancy in the parameter set impacts the results in some manner. To explore this issue, we have rigorously tested the extraction of Random Forest feature importances from highly inter-correlated parameters. We find that the identification of the most important variable is highly robust up to a level of inter-correlation of $\rho \sim 0.99$ (see Appendix B.2, and \citealt{Piotrowska2021}). In our data, the bulge - disk parameters are correlated at a level $\rho < 0.85$, comfortably below this threshold. Moreover, we have also tested investigating just two bulge - disk parameters at a time, avoiding any redundancies in the sample. We find identical rankings to the full analysis shown here, confirming that the superiority of bulge mass to the other parameters is entirely stable to the way these data are presented to the Random Forest classifier. 

We have also tested whether differential measurement uncertainty could lead to the result of bulge mass being most important for quenching erroneously. We find that the result of Fig. 4 (left panel) is stable up to an order of magnitude of random Gaussian noise added to the bulge component alone. This is far higher than the total error on any parameter (typically $\sim$0.2-0.3 dex), and hence is much much higher than the maximum allowed differential error between bulge mass and any other variable. As such, our conclusion that bulge mass regulates quenching is completely robust to the accuracy of these measurements.

The overall performance of the Random Forest classifier is excellent, yielding an accurate prediction in $\sim$90\% of cases, with the vast majority of this predictive power originating from bulge mass. Taking into account the uncertainties inherent within all of the relevant measurements (including SFR, $M_*$, $B/T$ and so forth) this is a truly remarkable level of performance. Ultimately, this suggests that the parameters contained within the photometric bulge - disk set are very nearly an optimal grouping of parameters for predicting quenching.

\subsubsection{The expanded parameter set}

We take advantage of the high spectroscopic coverage in the SDSS to add three more parameters of potential interest, with measurements directly or indirectly achievable only via spectroscopy\footnote{The relevance of this statement will become more apposite when considering the SDSS in comparison to CANDELS in Section 5.}. Specifically, we add i) central stellar velocity dispersion ($\sigma_\star$); ii) local galaxy over-density evaluated at the 5th nearest neighbour ($\delta_5$); and iii) the dark matter halo mass ($M_H$), inferred from abundance matching to the total group or cluster stellar mass (see \citealt{Yang2007, Yang2009} for full details). We then retrace all of the steps of the Random Forest classification as in the preceding sub-section.

In the right panel of Fig. 4, we present the results from a Random Forest classification of galaxies into star forming and quenched categories based on the wider set of parameters. In blue we show the full raw sample with pure S\'{e}rsic cuts applied (exactly the same sample as shown in the same shade of blue in the left-hand panel of Fig. 4). Additionally, we investigate a sample where all but face-on (axial ratio $b/a > 0.9$) disks are removed, leaving all bulge-dominated systems intact. The logic of this test is that it restricts the measured aperture $\sigma_\star$ to an accurate probe of the intrinsic stellar velocity dispersion, rather than a contaminated measurement of intrinsic dispersion with stellar rotation into the plane of the sky (see \citealt{Bluck2016} for a detailed discussion on this point). It is important to note that this does not dramatically alter either the analysis goals or the performance, since an equal number of quenched and star forming galaxies are chosen for both training and testing in the Random Forest. Consequently, the sample size is significantly reduced but the relative ratio in bulge-to-disk dominance is approximately preserved.

In both samples under investigation in Fig. 4 (right panel), $\sigma_\star$ is now found to be clearly the most important parameter governing quenching, comfortably beating even bulge mass by a very wide margin. Indeed, once $\sigma_\star$ is available to the classifier, there is very little for bulge mass to offer in terms of predicting quenching. This strongly implies that the connection between bulge mass and quenching is not causal, but rather a result of inter-correlation with $\sigma_\star$. In terms of interpretation, it is very well established that central velocity dispersion is an excellent predictor of supermassive black hole mass (e.g. \citealt{Ferrarese2000, McConnell2011, McConnell2013, Saglia2016}). Furthermore, central velocity dispersion is known to correlate more strongly with dynamically measured black hole masses than bulge mass (see \citealt{Saglia2016, Piotrowska2021}). As such, we can interpret velocity dispersion as a plausible proxy for black hole mass.

Utilising the above reasoning, we can make another key test of the LGalaxies AGN feedback quenching model. The agreement is truly remarkable: both the model and the observations agree that black hole mass (or its best known proxy, $\sigma_\star$) is by far the most predictive parameter of central galaxy quenching. A very similar result is found in \cite{Piotrowska2021} by comparing three cosmological hydrodynamical simulations (all of which utilise different AGN feedback prescriptions for quenching), and investigating a wide range of alternative black hole mass calibrations. Although in that publication no morphological or density parameters were included, so in this sense we expand on that work here.

The overall performance of the classification increases in the full sample to AUC = 0.93 (almost to the theoretical limit for our data), but decreases slightly in the face-on sample very slightly to AUC = 0.88 (in testing). Hence, we conclude that stellar velocity dispersion adds information not contained within the original bulge - disk parameters as a whole, as well as outperforming each parameter individually in terms of predictive power. Note that the slight reduction in performance for the face-on test is primarily just a result of utilising a much smaller training sample (as we have tested by restricting the original sample to a similar size).

All of the other parameters considered are of very low importance to quenching, once $\sigma_\star$ is made available to the classifier. It is particularly interesting to highlight the near total lack of predictive power in stellar mass, halo mass, morphology, and local density. All of these parameters have been considered as potential causal drivers of quenching (e.g. \citealt{Baldry2006, Dekel2006, Driver2006, Cameron2009, Martig2009, Peng2010, Peng2012, Woo2013, Omand2014, Gensior2020}). Yet, our Random Forest analysis clearly demonstrates that these parameters cannot be causally related to central galaxy quenching, since they offer no novel information relevant to the classification once central velocity dispersion is known.

The classification rankings of parameters from the widened parameter set are in close accord with \cite{Teimoorinia2016} where we utilised an artificial neural network approach. However, the separation between parameters is much clearer in the present Random Forest approach, to the point where the current analysis is far more constraining of quenching in these data. More concretely, in \cite{Teimoorinia2016} we found a difference in performance between $\sigma_\star$ and $M_B$ of only a few per cent, whereas here we find a difference at the level of $\sim$a factor of ten (or greater in the face-on disk sample), i.e. 1000 per cent or more(!) The reason for this profound difference is due to the competitive nature of the Random Forest, which assesses each parameter in terms of how predictive it is of quenching relative to the other parameters in the set. In a sense, the difference between ANN and Random Forest is analogous to the difference between total and partial correlations, the latter being much more powerful for ascertaining the underlying dependence in correlated data. It is also worth highlighting that the ANN prescription in \cite{Teimoorinia2016} is akin to the Individual Parameter mode of the Random Forest, which we found in Appendix B.2 to be the least successful of the machine learning options for extracting causality. As such, the updated analysis presented here is a substantial improvement on the prior results in the literature.

The strong dependence of central galaxy quenching on central velocity dispersion is also consistent with \cite{Bluck2016} where we utilise area statistics to demonstrate the tightness of the $f_Q - \sigma_\star$ relation, and its invariance to variation in other parameters. However, yet again, the present analysis yields a much clearer (and more statistically robust) result than our prior study. Finally, we note that there are several other papers which highlight the importance of stellar velocity dispersion to quenching (e.g. \citealt{Wake2012, Terrazas2016}) broadly in agreement with the results of this sub-section. But these prior papers do not critically assess whether central velocity dispersion is superior to all of the parameters considered here, and importantly do not utilise a machine learning technique which is proven to be capable of extracting the causality hidden within complex data sets.


\begin{figure*}
\includegraphics[width=0.49\textwidth]{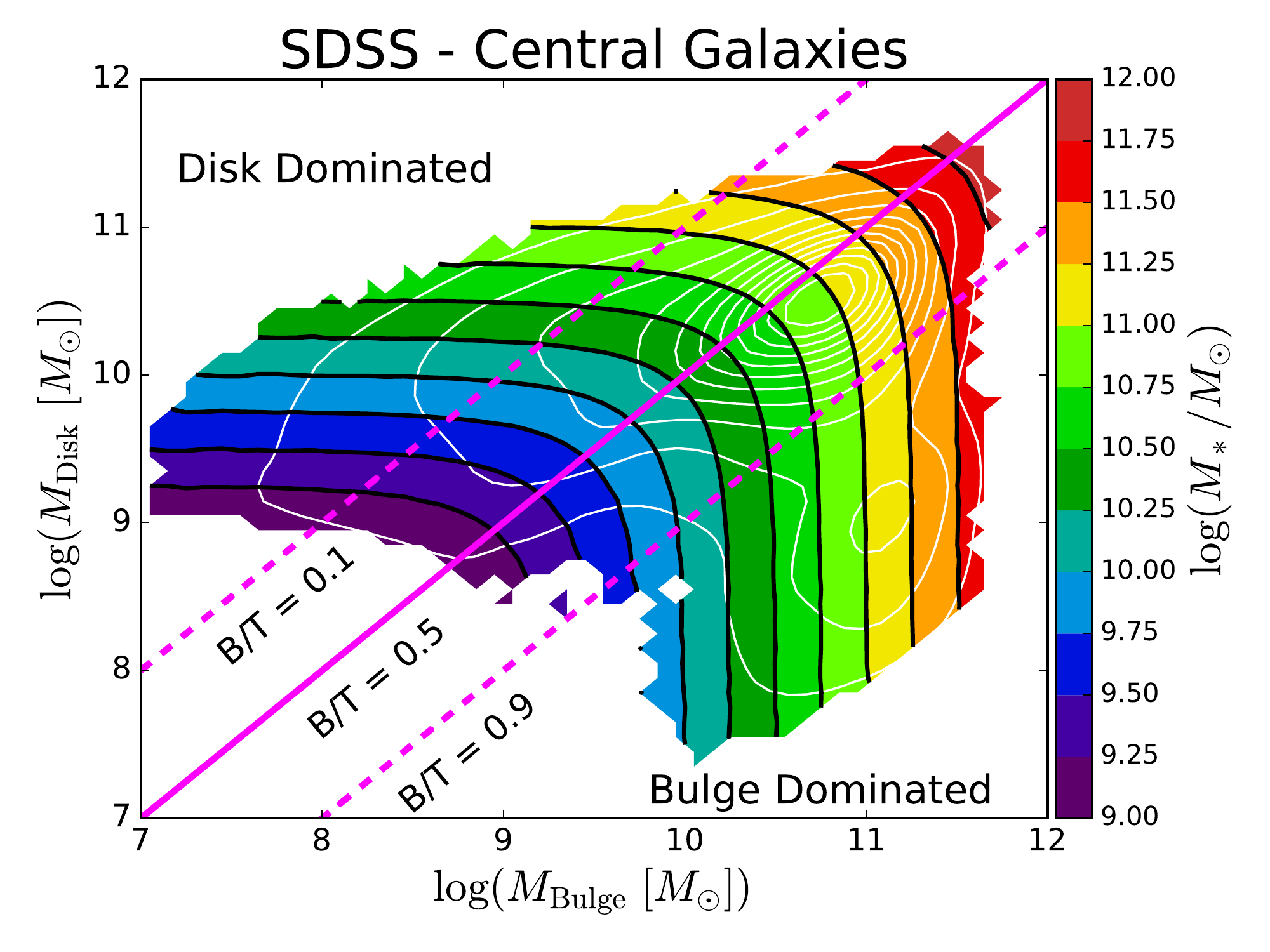}
\includegraphics[width=0.49\textwidth]{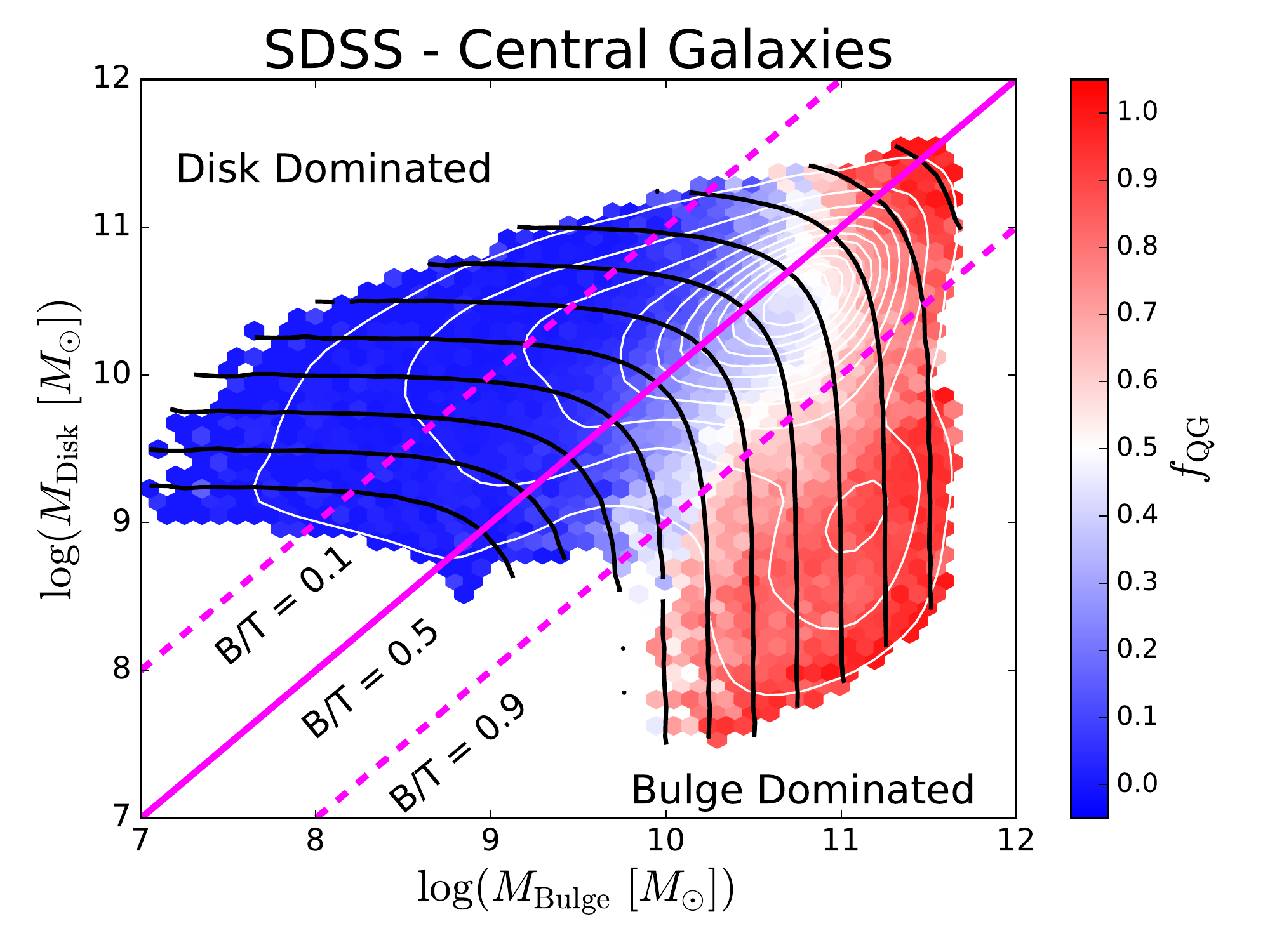}
\caption{Visual test on the primacy of bulge mass for predicting quenching. Disk mass vs. bulge mass, colour coded by total stellar mass (left panel) and the fraction of quenched galaxies (right panel). In both panels iso-mass and iso-morphology (i.e. $B/T$) lines are drawn to guide the eye, shown in black and magenta (respectively). Quenched central galaxies typically have high bulge masses. Note also that at a fixed stellar mass galaxies may be either star forming or quenched dependent upon $B/T$; and yet at fixed $B/T$ galaxies may be either star forming or quenched dependent upon $M_*$. Moreover, there is little-to-no visual dependence of quenching on disk mass explicitly. Hence, quenching proceeds most closely with increasing bulge mass. This provides a simple visual confirmation of the primary result from the RF analysis in the left panel of Fig. 4. }
\end{figure*}


\subsection{Visual tests on the random forest results}

Although the Random Forest results from the preceding section are very clear, we appreciate that not all of our intended readers will be familiar and comfortable with machine learning. As such, in this sub-section we present two quick visual assessments of the SDSS data which confirm the key results.

\subsubsection{The bulge - disk plane}

In Fig. 5 we show the location of galaxies in the bulge mass - disk mass plane, as indicated by white density contours. It is important to appreciate that the four bulge - disk parameters are not all independent. In fact, knowing the value of any two parameters enables the computation of the remaining two. Hence, by viewing the bulge - disk plane we can also ascertain how quenching proceeds with total stellar mass and $B/T$ structure, in addition to bulge and disk mass.

In Fig. 5 (left panel) we overlay iso-mass and iso-morphology lines, shown in black and magenta respectively. Additionally, we colour code the bulge - disk plane by the value of total stellar mass (as indicated by the colour bar). Thus, regions of a given colour all have the same total stellar mass, but different structures. Towards the top-left of the plane resides disk-dominated and pure disk galaxies; whereas towards the bottom-right of the plane resides bulge-dominated and pure spheroidal galaxies. Obviously, to increase bulge mass in this diagram one moves in the positive direction along the $x$-axis, and to increase disk mass one moves along the positive direction of the $y$-axis. 

In the right-hand panel of Fig. 5, we reproduce the $M_D - M_B$ 2-dimensional distribution of SDSS central galaxies. Here we colour code each small hexagonal region of the plane by the quenched fraction within that region of parameter space (as indicated by the colour bar). In a sense, this approach produces a map of quenching in the bulge - disk parameter space, enabling us to see which regions are suitable for star formation, and in which regions star formation is suppressed. 

Viewing the right-panel of Fig. 5, it is clear that quenching proceeds much more directly with bulge mass than with disk mass. In fact, varying disk mass at a fixed low bulge mass has no impact whatsoever on the fraction of quenched galaxies, and varying disk mass at a fixed high bulge mass has only a small impact on the quenched fraction. Conversely, varying bulge mass at any value of disk mass has a huge positive effect on the fraction of quenched galaxies. It is also clear from this diagram that pure disks (lying above the top dashed magenta line) are invariably star forming; and pure spheroids (lying below the bottom dashed magenta line) are invariably quenched, even at a fixed stellar mass. Yet, composite galaxies (residing between the two magenta lines) are star forming at low masses but quenched at high masses. Thus, we conclude that bulge mass is visibly the most effective of the bulge - disk parameters for governing quenching. 

This visual assessment clearly supports the main conclusion from Fig. 4 (left panel). However, note that the present analysis is entirely qualitative in nature, whereas the Random Forest classification yields a fully quantitative ranking of these parameters.

\subsubsection{The bulge mass - velocity dispersion plane}

In Fig. 6 we show the relationship between stellar velocity dispersion and bulge mass for the sample where disks are restricted to being face-on, i.e. the same sample as shown in the magenta bars on the right-hand panel of Fig. 4. The reason for this cut is to restrict $\sigma_\star$ to a sub-sample where it is an accurate measurement of intrinsic stellar dispersion, instead of an imperfect measurement of rotation into the plane of the sky convolved with intrinsic dispersion. The location of galaxies in the plane is indicated by density contours (shown in white) and the median relation (+/- 1 $\sigma$ dispersion) is shown as a solid (dashed) black line. It is clear that there is a very strong and reasonably tight relationship between the two parameters. In fact, the correlation strength between $\sigma_\star$ and $M_B$ is $\rho_{\rm Spearman}$ = 0.8 and the mean dispersion is just $\sigma_{\rm disp}$ = 0.13 dex. Note also that the dispersion tightens significantly towards higher masses, where there is in general a lower contribution from rotation in bulge structures.

It is expected that there should be a very strong correlation between velocity dispersion and the mass of the bulge, since the bulge structure is often thought to be pressure supported. Nonetheless, intrinsic dispersion in the $\sigma_\star - M_B$ relation is expected due to varying radii of galaxies. In practice, this implies that increasing $\sigma_\star$ at a fixed $M_B$ is a result of decreasing $R$ (assuming constant morphology). This is equivalent to increasing the core stellar mass density of the galaxy. Therefore, there is a natural expectation for a tight physical connection between these two parameters, but with meaningful physical information contained within the scatter.


\begin{figure}
\includegraphics[width=0.49\textwidth]{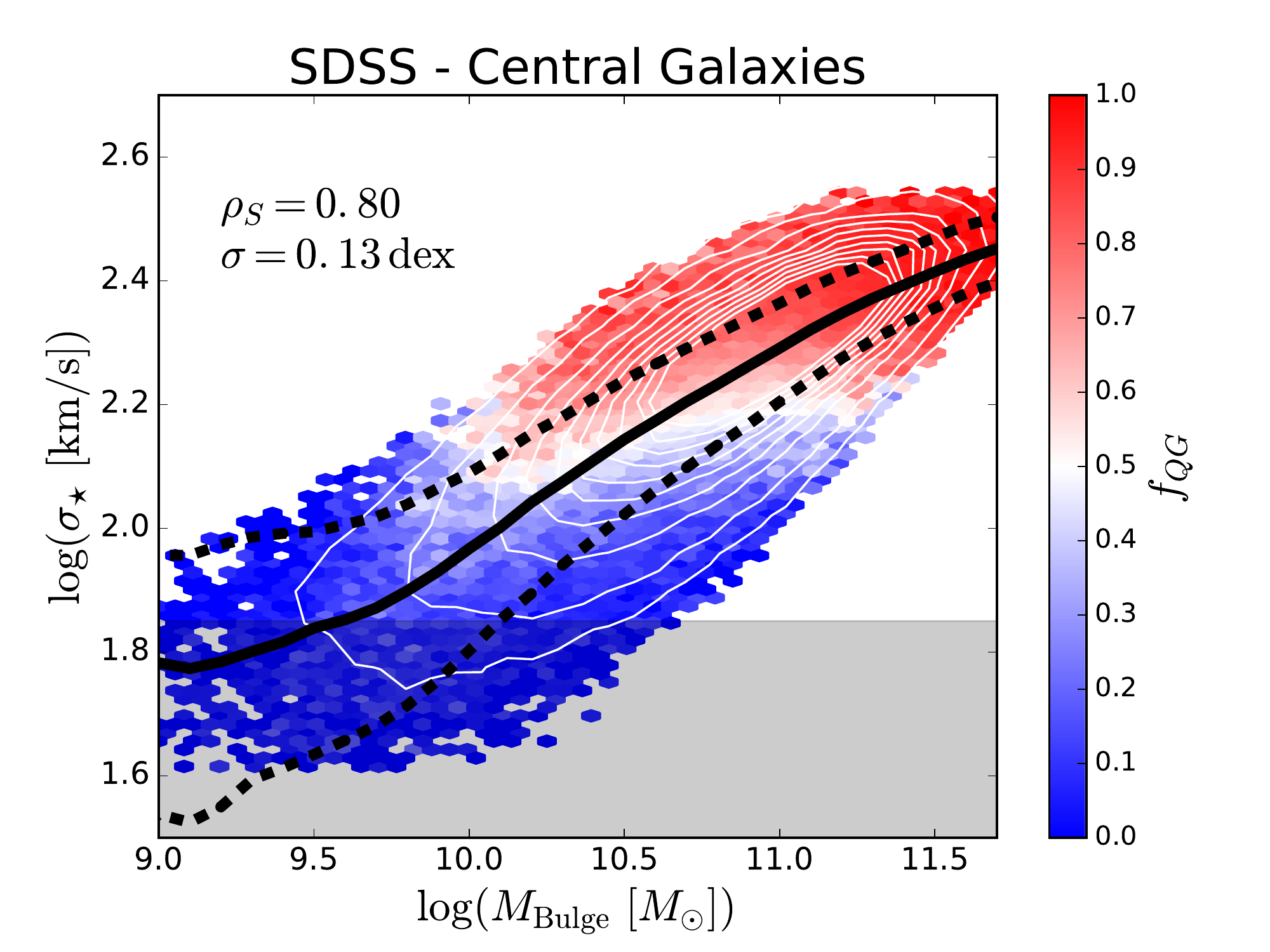}
\caption{Stellar velocity dispersion plot as a function of bulge mass for SDSS central galaxies. This plot is restricted to disk-dominated galaxies which present face-on ($b/a > 0.9$), in addition to all bulge-dominated galaxies. The plot is subdivided into small hexagons which display the quenched fraction within each region of the parameter space, as labelled by the colour bar. The median relation (+/-1$\sigma$) is shown as a solid (dashed) black line. The correlation strength between $\sigma_\star - M_B$ ($\rho_S$) and the rms dispersion ($\sigma$) are both displayed on the plot. The region in which velocity dispersion is poorly constrained by the SDSS data is shaded in grey. Clearly, the optimal route to maximise quenching through this 2D plane is through increasing $\sigma_\star$ (not $M_B$). This provides a simple visual confirmation of the primary result from the RF analysis in the right panel of Fig. 4.}
\end{figure}


\begin{figure*}
\includegraphics[width=0.33\textwidth]{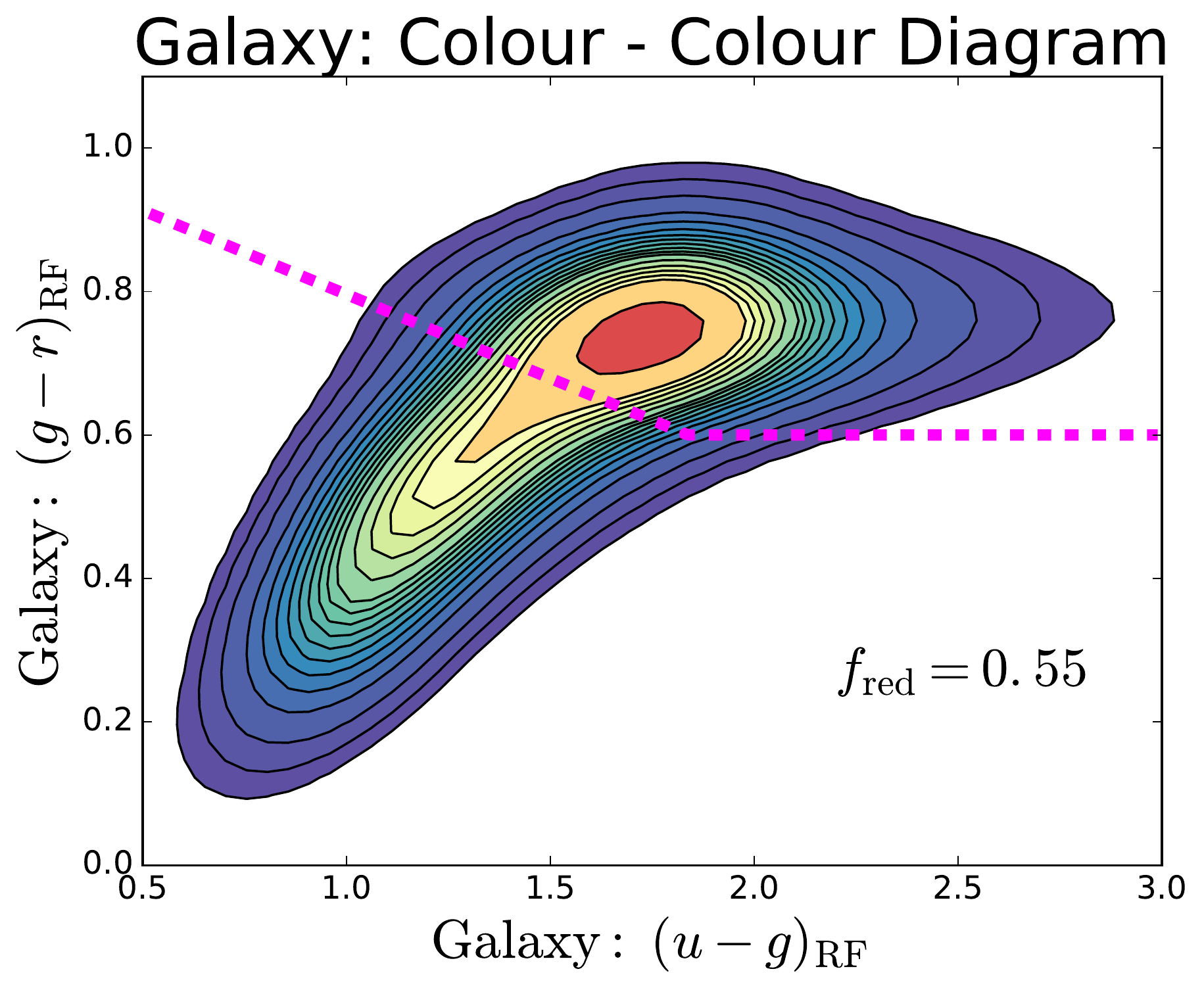}
\includegraphics[width=0.33\textwidth]{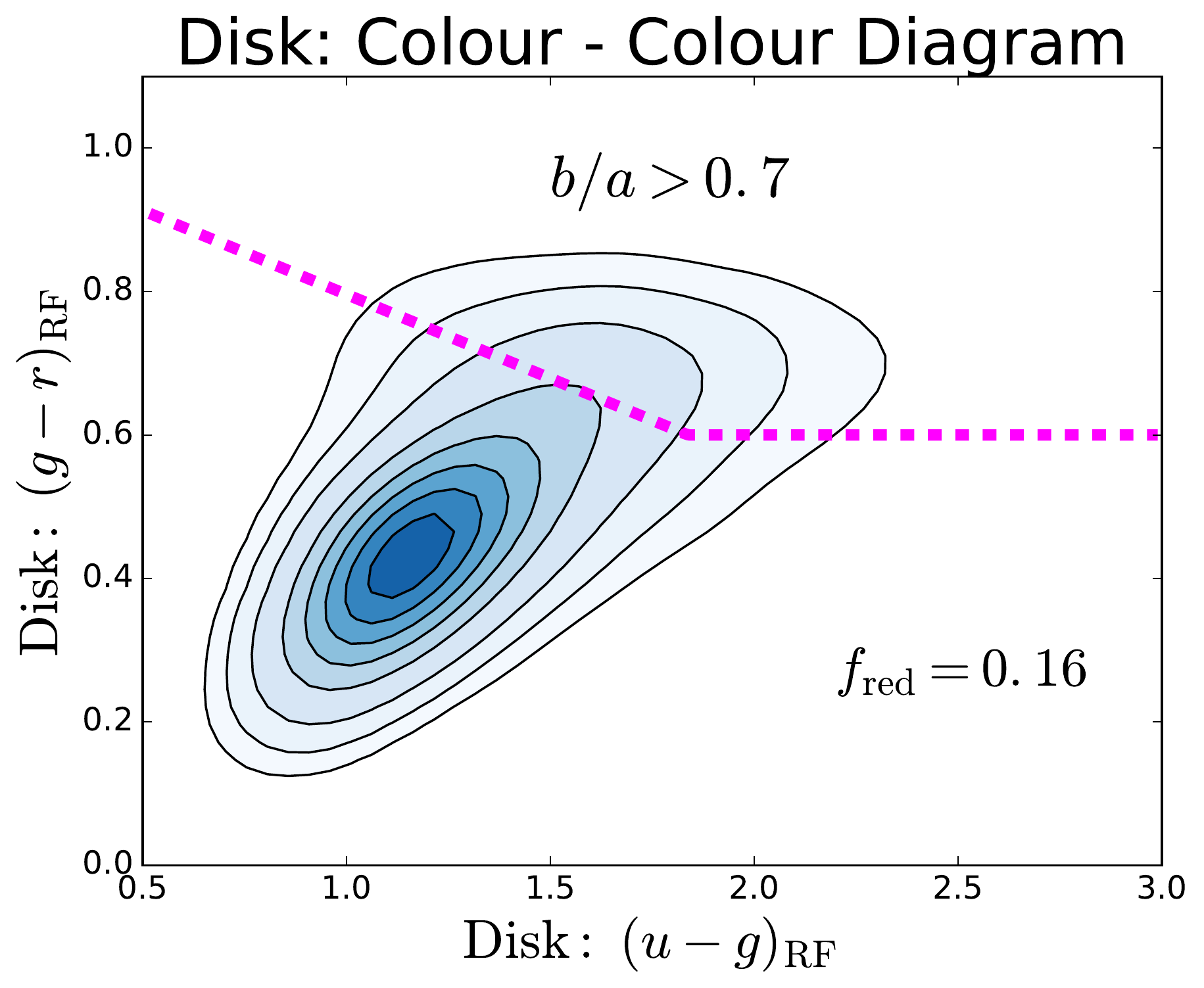}
\includegraphics[width=0.33\textwidth]{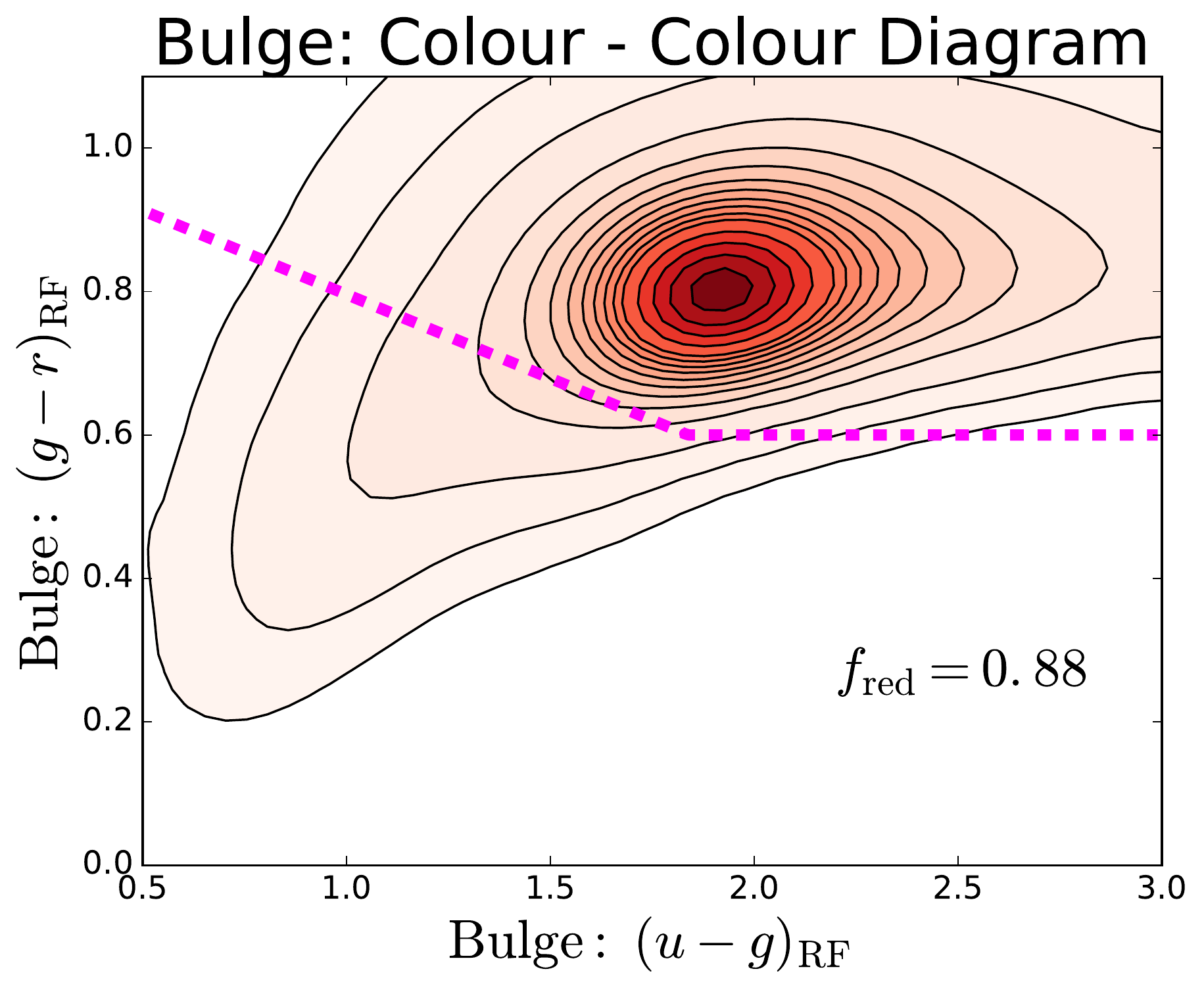}
\caption{The location of galaxies (left panel), disks (middle panel) and bulges (right panel) in rest-frame (g-r) -- (u-g) colour space. For the disk sample, we remove highly inclined galaxies (restricting to $b/a > 0.7$ systems) to mitigate the impact of dust extinction. In each panel, linearly spaced density contours are shown as light black lines with coloured shading. The adopted threshold for selecting quenched and star forming systems is shown as a dashed magenta line on each panel (which is introduced and motivated in Section 3.1.3). Additionally, the fraction of galaxies, bulges and disks which are identified by the cut to be quiescent are shown on the lower right of each panel.  }
\end{figure*}


\begin{figure*}
\includegraphics[width=0.99\textwidth]{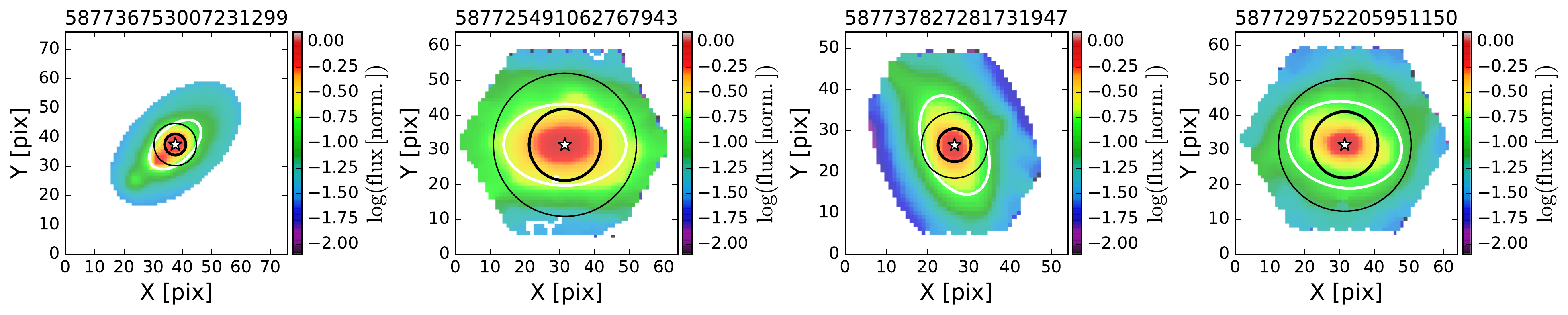}
\caption{Examples of four randomly chosen composite (bulge + disk) galaxies from our sample of MaNGA central galaxies. Each panel shows the pseudo-V-band flux (in normalised units), with a white star indicating the centre of the galaxy. Overlaid on each galaxy flux map is the bulge effective radius ($R_b$, inner black circle), 2 $\times$ the bulge effective radius (outer black circle), and the elliptical half light radius ($R_e$) of the galaxy as a whole (white ellipse). The SDSS object ID is provided as a title on each panel. In MaNGA, we define bulge spaxels to reside at $r < R_b$ and disk spaxels to reside at $r > 2R_b$, leaving a buffer zone where contamination from both components is likely high.}
\end{figure*}

The very strong observed correlation (and reasonably small scatter) between $\sigma_{\star}$ and $M_B$ in Fig. 7 acts as an important quality check on these data. It is important to stress that bulge mass is derived entirely from photometric data (see \citealt{Mendel2014}) and yet the velocity dispersion is derived entirely from spectroscopic data (see \citealt{Bernardi2007}). Hence, the recovery of the strong expected correlation between these parameters is a significant vindication of both methods. Clearly, the measurements of bulge masses used in the SDSS achieve a good consistency with the SDSS kinematics, enhancing the confidence with which we can use these measurements.

In Fig. 6, we sub-divide the $\sigma_\star - M_B$ plane into small hexagonal regions, each colour coded by the fraction of quenched galaxies (as indicated by the colour bar). This approach enables us to visually assess where star forming and quenched galaxies are located in this parameter space. It is obvious from visually inspecting Fig. 6 that quenching proceeds vertically in the plot, i.e. as a function of $\sigma_\star$ not $M_B$. Varying bulge mass at a fixed velocity dispersion has little impact on the fraction of quenched galaxies, yet varying velocity dispersion at a fixed bulge mass has a dramatic impact on the quenched fraction. 

The above result provides strong qualitative support to the result in Fig. 4 (right panel) that central velocity dispersion is far superior for predicting quenching than bulge mass, which is itself the most predictive of the bulge - disk parameters (see Fig. 4 left panel and Fig. 5). However, whilst Fig. 6 gives a clear visual impression of the dominant quenching parameter in this pairing, such a method clearly falls short when many parameters need to be explored together, or a quantitative comparison is required. Addressing these issues simultaneously was a key motivation for our Random Forest analysis (see Fig. 4 and associated text).


\subsection{The quenching of bulges \& disks in the SDSS \& MaNGA}

In the previous parts of this Section, we have established that our Random Forest classification technique is capable of recovering the known quenching results for galaxies in the local Universe (especially as in \citealt{Bluck2014, Bluck2016, Teimoorinia2016}). Moreover, we have placed these prior results on a much stronger statistical footing, completely ruling out the possibility of measurement uncertainty, sample selection, or analysis method choices to lead to the results erroneously. Briefly, we confirm that bulge mass is the most predictive parameter of quenching in photometric data, yet central velocity dispersion becomes the most important variable for predicting quenching in spectroscopic data.  

Furthermore, in Appendix B.2 we have thoroughly demonstrated the power of our machine learning approach, utilising a Random Forest classifier in the All Parameter mode, to extract causality from complex and highly inter-correlated data sets. Consequently, we are now in an excellent position to extend our novel method into unchartered observational territory in the local Universe. To this end, in this sub-section, we perform a thorough analysis of quenching within galactic structural sub-components, i.e. bulges and disks.

\subsubsection{Extraction methods for bulges \& disks}

By utilising the photometric bulge - disk decompositions of SDSS galaxies from \cite{Simard2011} and \cite{Mendel2014} in conjunction with our colour - colour classification (see Fig. 2), we can explore how bulge and disk structures quench, in addition to their host galaxies. Additionally, utilising the spatially resolved spectroscopy in the MaNGA survey (\citealt{Bundy2015}) along with star formation rate surface densities (computed in \citealt{Bluck2020a}) we can explore the quenching of bulge and disk regions within galaxies, offering a complementary analysis of bulge and disk quenching.

In Fig. 7 we show the distribution of SDSS galaxies (left panel), disks (middle panel), and bulges (right panel) in rest-frame (g-r) -- (u-g) colour space. Additionally, we show the optimal decision boundary for selecting quenched and star forming systems in this colour space as dashed magenta lines (which are derived in Section 3.1.3). For galaxies, there is a roughly even split between star forming and quiescent systems, as ascertained by their colours. For disks, we restrict systems to have an axial ratio of $b/a > 0.7$, i.e. removing galaxies which are significantly inclined relative to the plane of the sky. This significantly reduces the potential impact of dust extinction on the disk colours (see e.g. \citealt{Simard2011}). As expected, the majority of disks are found to be star forming. However, there is still a significant fraction of $\sim$15\% of disks which are found to be red enough to be classified as quenched via this method (even when presenting approximately face-on). Conversely, the vast majority of bulges are found to be red (as expected); but there are $\sim$10\% of bulges which have blue enough colours to be identified as actively star forming. Hence, we can meaningfully ask why some bulges are star forming and some disks are quenched. 

The power of the MaNGA survey comes from the use of spatially resolved spectroscopy to deduce physical properties across the face of nearby galaxies. Whilst the sample size is much smaller than the SDSS (by a factor of $\sim100$), the added spatial information in the sample yields a vast quantity of data which is ideal for probing the inner workings of galaxies. Furthermore, the star formation rates for spaxels in the MaNGA sample are fully dust corrected (see \citealt{Bluck2020a}) and hence this provides a valuable check on the SDSS colour-based approach. In terms of this paper, the pertinent question that we can answer with MaNGA (which is not possible with the SDSS) is, how are star formation and quenching dependent on spatial location within local galaxies? 

To answer this question, we adopt the star formation rate surface densities of spaxels derived in \cite{Bluck2020a} to separate each region within galaxies into star forming and quenched classes (see Section 3.1). We then separate spaxels into bulge and disk classes in order to investigate the quenching of star formation for these two structural components of galaxies. This enables a powerful test to the bulge - disk analysis of SDSS galaxies via rest-frame colours, incorporating spectroscopically derived SFRs at multiple locations within a representative sample of SDSS galaxies. 

To separate spaxels belonging to bulge and disk structures in MaNGA we adopt a similar approach to \cite{Lin2017}. Specifically, we collate all spaxels within one effective bulge radius ($r < 1R_b$) provided that the galaxy has $B/T > 0$ (after applying the pure S\'{e}rsic cuts discussed in Section 3), labelling this set as `bulge spaxels'. Pure spheroids (with $B/T = 1$) are considered to be made up exclusively of bulge spaxels, regardless of their location, and are added to the bulge spaxel group. Additionally, we collate all spaxels residing at distances larger than two effective bulge radii ($r > 2R_b$) provided that the host galaxy has $B/T < 1$ (after S\'{e}rsic cuts), labelling this set as `disk spaxels'. Pure disks (with $B/T = 0$) are naturally considered to be entirely made up of disk spaxels, regardless of their location within the galaxy, and are added to the disk spaxel group. Hence, in morphological composites (systems with $0 < B/T < 1$), we discard spaxels lying between 1 - 2 $R_b$, due to considering these regions as likely having a significant contribution from both the bulge and disk structure\footnote{We note that this approach effectively separates `bulge dominated' regions from `disk dominated' regions, but that some residual contamination is inevitable. The only way around this issue is to perform full spectrum kinematic bulge - disk decompositions, which is highly challenging and above the scope of the present work. However, note that in the photometric data a full colour decomposition is provided in both our SDSS and CANDELS analyses of bulge and disk quenching. The primary results form both of these alternative analyses agree closely with our MaNGA results, enhancing our confidence in the reliability of both methods.}. However, we also consider a grouping of all spaxels within galaxies, where the intermediate region in radius is included in our analysis.


\begin{figure*}
\includegraphics[width=0.49\textwidth]{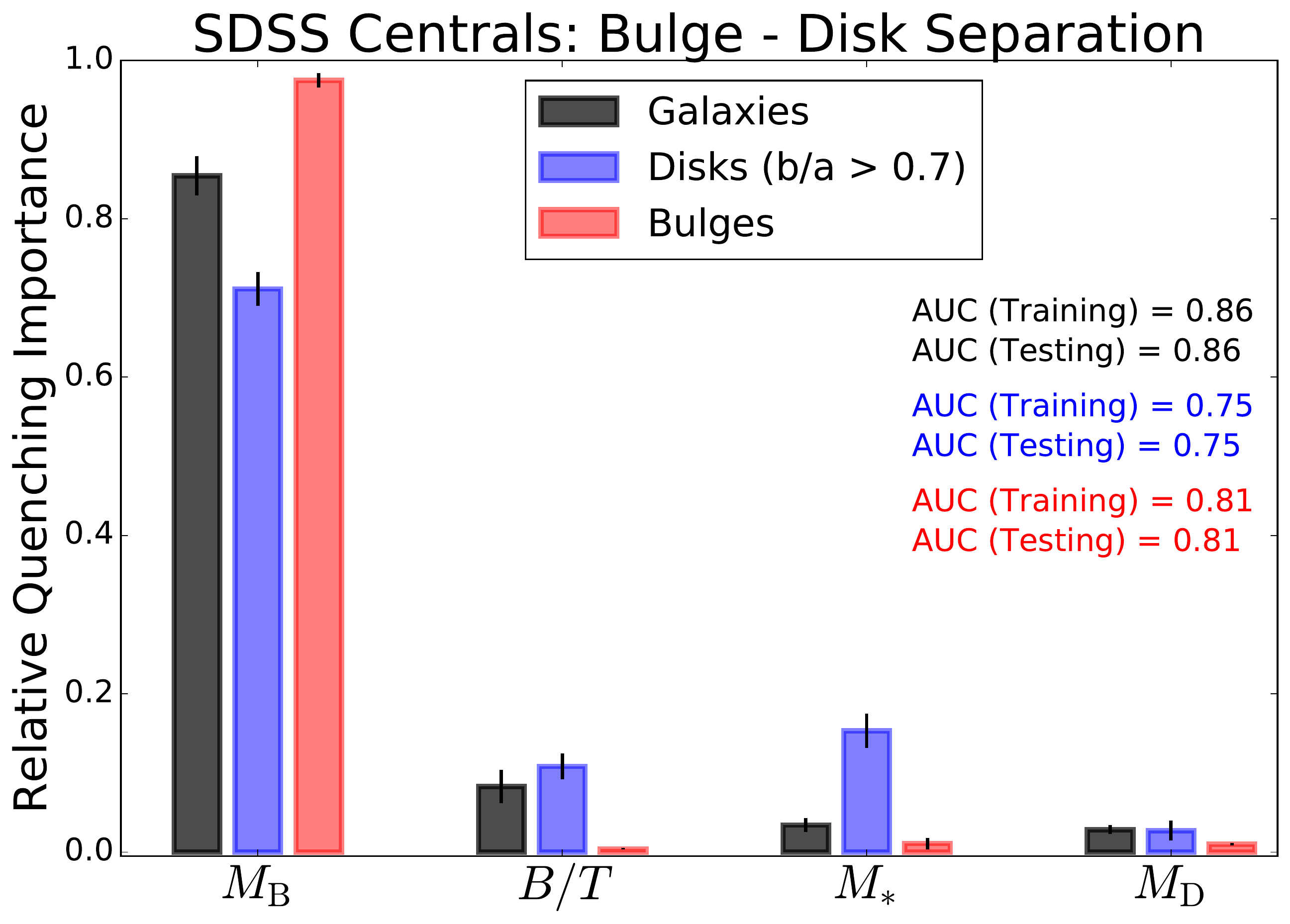}
\includegraphics[width=0.49\textwidth]{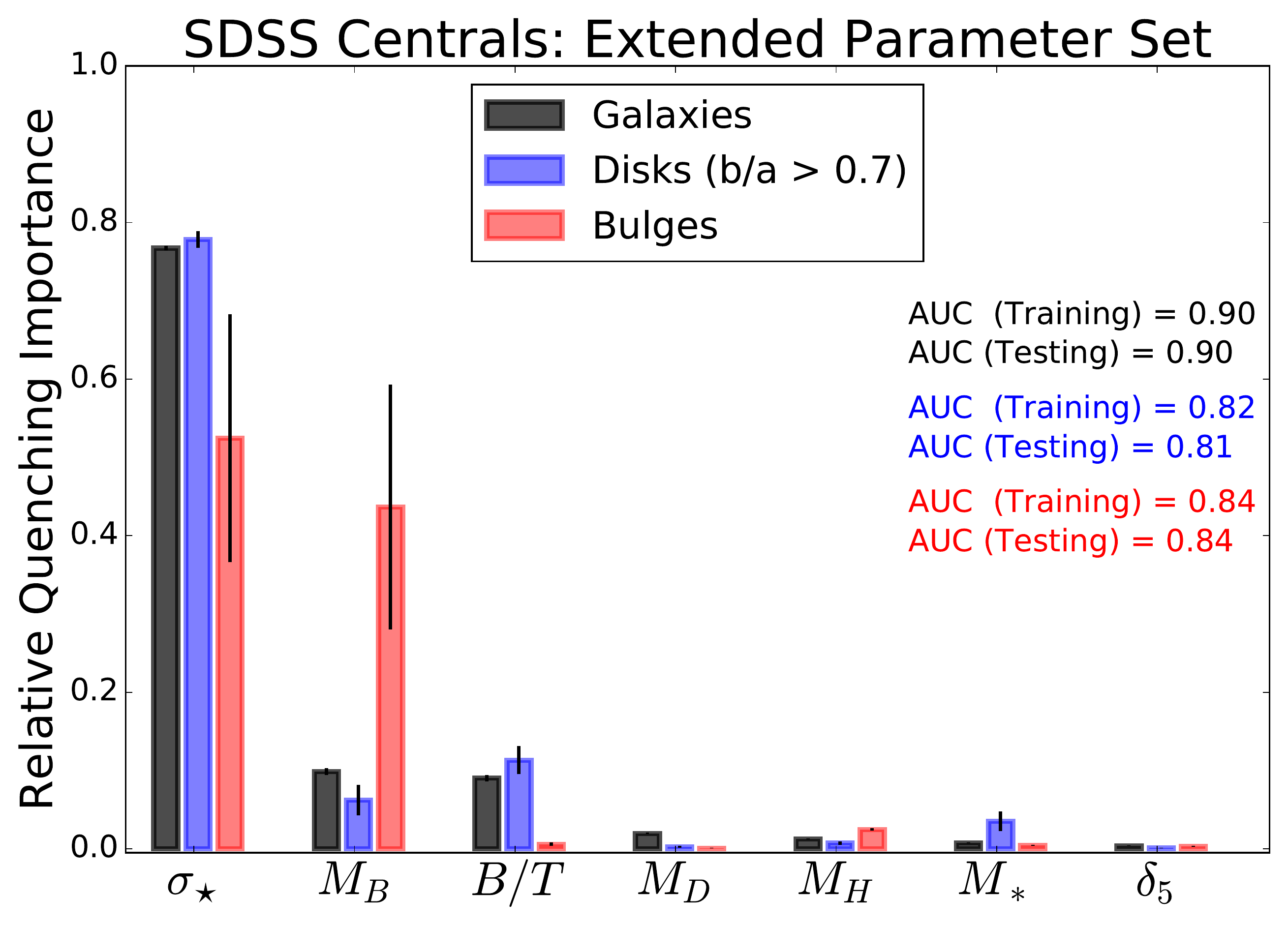}
\includegraphics[width=0.49\textwidth]{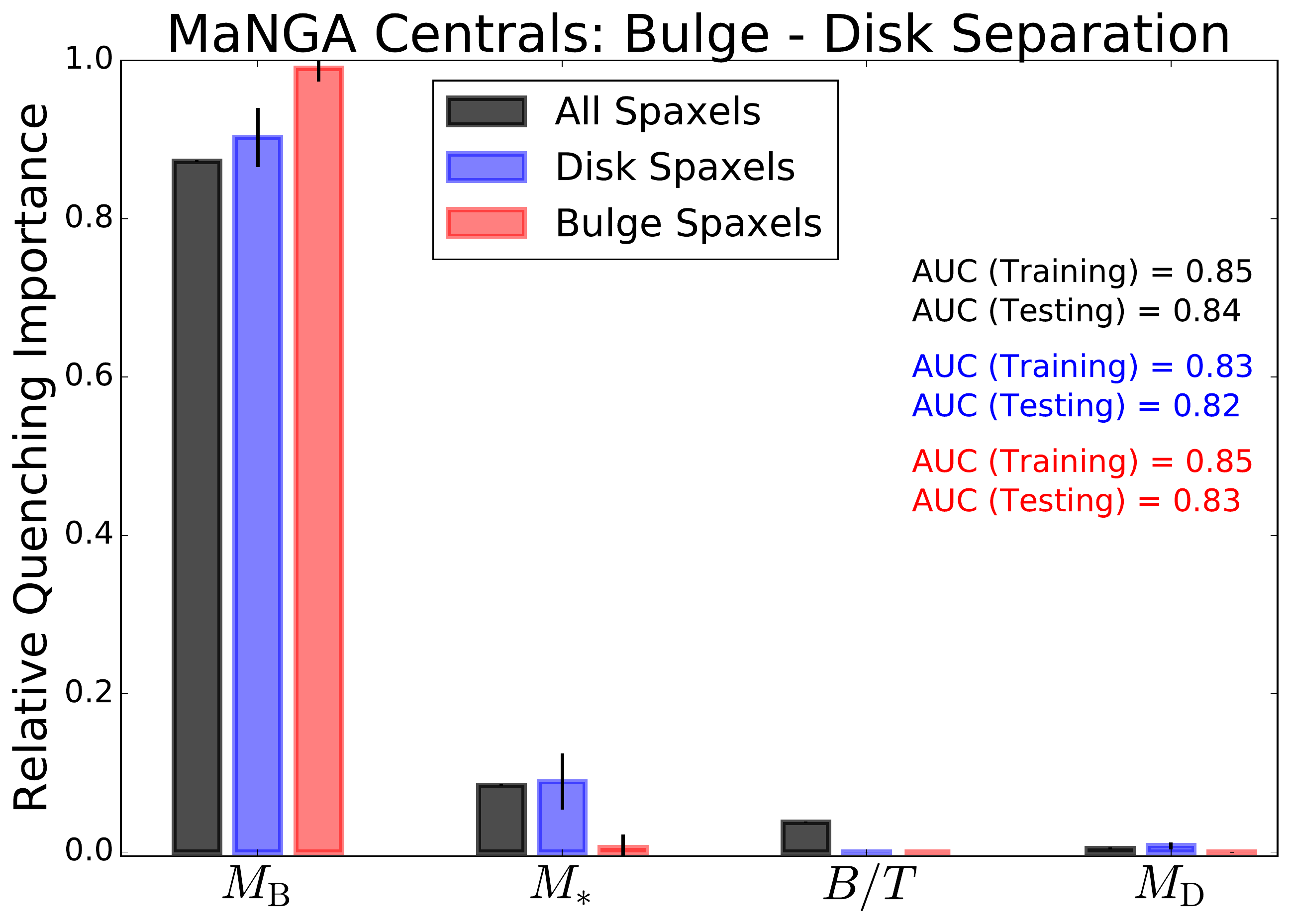}
\includegraphics[width=0.49\textwidth]{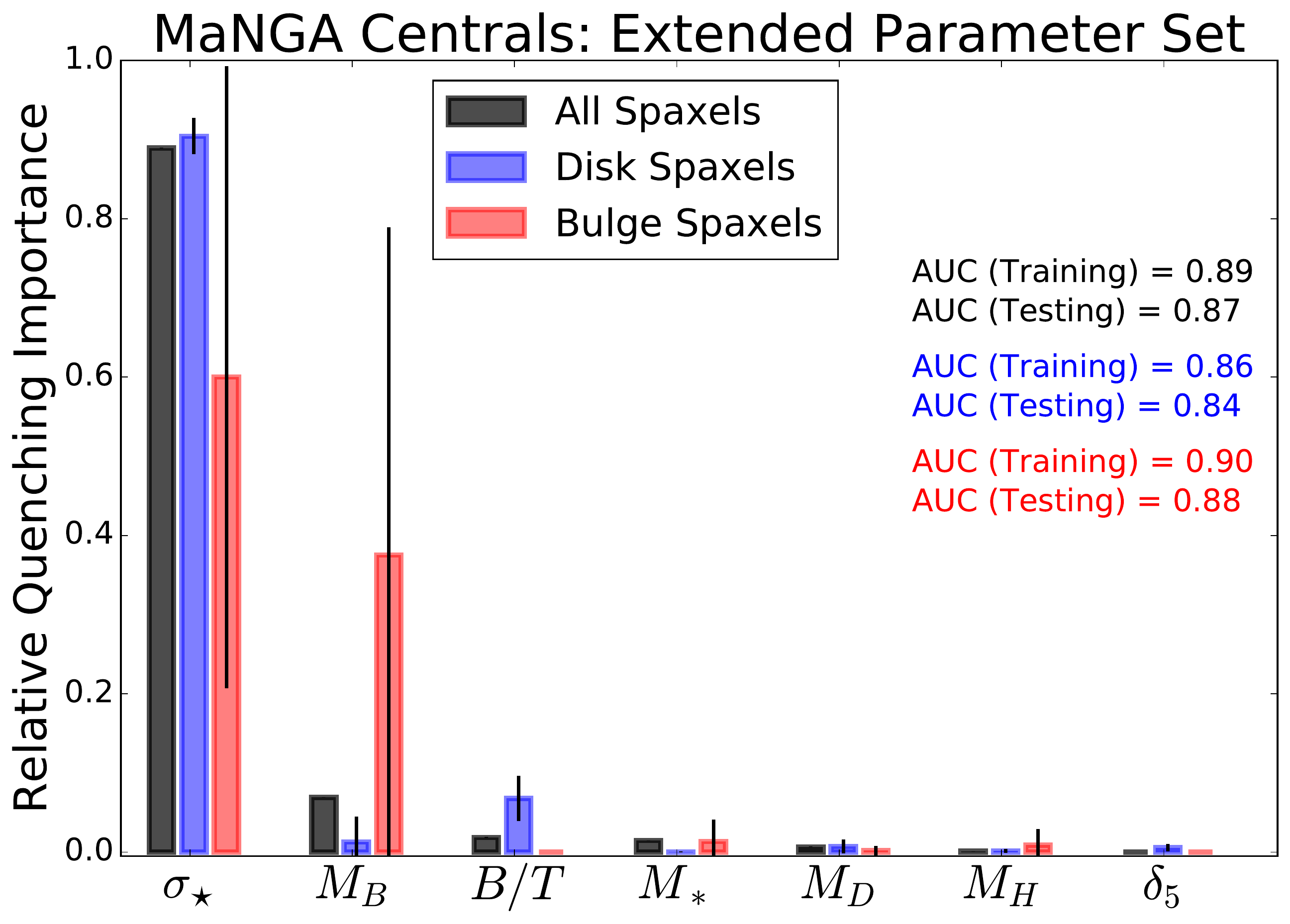}
\caption{Random Forest classification analysis to predict when local Universe central galaxies (shown in black), and their component disks (shown in blue) and bulges (shown in red), are star forming or quenched. On all panels the $y$-axis shows the relative importance of each parameter, and the $x$-axis labels each parameter in turn (ordered from most to least predictive of quenching). Error bars indicate the 1$\sigma$ dispersion on the relative importance from ten independent runs of the classifier. {\it Top panels: } Results based on colour selection in the full SDSS data set. {\it Bottom panels: } Results based on spatially resolved star formation rate surface densities in the MaNGA data set. The left-hand panels show results for the photometric bulge - disk parameters, and the right-hand panels show results for the extended parameter set. For galaxies as a whole, as well as for their bulge and disk components treated separately, bulge mass is clearly identified as the most predictive of the bulge - disk parameters. However, in the extended parameter set, central velocity dispersion replaces bulge mass as the most predictive parameter of quenching for galaxies as a whole, and for disks. Yet for bulges, bulge mass and central velocity dispersion tie for first place (within their errors). For galaxies as a whole, these results are essentially identical to Fig. 4, despite using significantly different methodologies (i.e. optical colours vs. SFRs).}
\end{figure*}

In Fig. 8 we illustrate our method to separate spaxels into bulge and disk structures. We present four randomly chosen galaxies (with an approximately equal bulge and disk mass for illustration purposes). We display normalised maps in pseudo V-band flux, with the centre of each galaxy indicated by a white star. We overlay the location of $1R_b$ as a bold black circle and the location of $2R_b$ as a lighter black circle. Bulge spaxels are defined to lie within the bold black circle; and disk spaxels are defined to lie beyond the light black circle. It is important to emphasize that we apply a mass surface density cut on the data (at $\Sigma_* > 10^7 M_{\odot} / {\rm kpc}^2$), as well as stringent $S/N$ requirements (of a continuum S/N $\sim$ 30 per voxel [spaxel grouping], see \citealt{Sanchez2016, Sanchez2016a}), which in practice restricts most galaxies to having spaxels within $\sim1.5R_e$, even if the field of view extends beyond this limit. Additionally, on each panel of Fig. 8 we overlay the ellipse at the semi-major axis of the half-light radius ($R_e$, shown in white) for comparison. 

One limitation of the MaNGA survey is that the majority of galaxies are only probed out to 1.5 $R_e$. Consequently, our conclusions from the MaNGA data set can only be trusted out to this limit. Nonetheless, across the whole sample the majority of spaxels are found in disk components ($\sim$ 4 : 1; disk : bulge spaxels). On the other hand, the SDSS photometry extends out to $\sim 3-5 R_e$ with sufficient surface brightness to impact the bulge - disk fits (see \citealt{Simard2011}). Thus, the value in comparing bulge - disk quenching in the SDSS and MaNGA is that these two galaxy surveys are highly complementary. In MaNGA we have a limited field of view and greatly reduced number of galaxies; yet in the SDSS photometry the field of view is essentially complete, and the number of galaxies is much larger. Conversely, in MaNGA we leverage spatially resolved spectroscopy to accurately constrain the true star formation rate within each region of each galaxy (fully accounting for dust extinction), whereas in the SDSS we use rest-frame colours as a proxy for star forming state. Hence, if the results from the analyses of both surveys agree, we can be highly confident that the issues inherent with each survey alone do not impact our conclusions.

\subsubsection{Classification results for bulges, disks, and galaxies}

In Fig. 9 we present the results from several Random Forest quenching classifications applied to galaxies (shown in black), bulges (shown in red), and disks (shown in blue). The top row of Fig. 9 shows results for the SDSS, using bulge - disk decompositions in light and a colour based definition of quenching (see Fig. 2). As in the previous sub-section, we restrict the axial ratios of disks to $b/a > 0.7$, to mitigate the impact of dust extinction on disk colours. Additionally, we have tested alternative thresholds at $b/a$ = 0.5 and 0.9, as well as not applying any inclination cuts at all. All of the results for the SDSS disks remain stable to these analysis choices. This clearly indicates that dust extinction is not likely to be seriously impacting our conclusions. Interestingly, as we make more stringent inclination cuts, the significance of the results for disks actually increases. Finally, we remind the reader again that dust extinction is not an issue for our MaNGA disk sample, which thus provides a final independent test on the results for disks.

The bottom row of Fig. 9 shows results for MaNGA, using a spatial segregation of bulge and disk spaxels, and a star formation rate surface density based measurement of quenching (see Fig. 1). In the left panels of Fig. 9 we consider the bulge - disk parameters (i.e. the same parameters as in the left panel of Fig. 5), and in the right panels of Fig. 9 we consider the extended parameter list (i.e. the same parameters as in the right panel of Fig. 5). As before, the height of each bar indicates the relative importance of each parameter for predicting quenching, and the error on this statistic is given as the variance across ten independent training and testing runs. Parameters under consideration are labelled on the $x$-axis, and are ordered from most to least predictive of quenching for galaxies as a whole.

Considering first the bulge - disk parameter samples in the SDSS and MaNGA (left panels of Fig. 9), we see that bulge mass is clearly identified as the most predictive parameter of central galaxy quenching not only for galaxies as a whole, but also for their component bulge and disk structures treated separately. This is true in both the SDSS colour decomposition approach and in the MaNGA $\Delta \Sigma_{\rm SFR}$ approach. It is particularly interesting to note that bulge mass is far more predictive of disk quenching than disk mass. This suggests that some process(es) associated with the central most regions within galaxies regulate quenching throughout the entire system. Bulges and disks are very different in terms of their kinematics, spatial location, stellar populations, ages and colours, yet in terms of their quenching dependence they are essentially identical. 

To account for this observational fact one is naturally led to a global quenching solution - i.e. one which has the capacity to impact all regions within galaxies. The AGN heating solution of LGalaxies is a good example of such a process, engendering quenching from starvation of gas supply. We consider other possibilities to explain this result in the Discussion.

Another highly important feature evident from the left panels of Fig. 9 is that the dependence of quenching in bulges and disks on the bulge - disk parameters can be extracted utilising component colours, yielding a very similar result in the SDSS to the full spatially resolved spectroscopic analysis in MaNGA. This is especially important because at high redshifts there are very few IFU studies of galaxies on a statistical basis. As such, we can use a colour based approach in CANDELS in the next section, and be reasonably certain that the results we find will be stable to better measurements with IFU's in the coming years.

Considering now the right panels of Fig. 9, we see that in the expanded parameter set central velocity dispersion is overwhelmingly the most predictive parameter of quenching in galaxies as a whole, and in disk structures. However, for bulges, the Random Forest cannot distinguish between $\sigma_\star$ and $M_{B}$ as the most effective parameter (within the 1$\sigma$ uncertainty). The reason for this is that blue bulges are rare, and so are bulge spaxels in general (purely for geometric reasons). As such, there is much less data available to train the Random Forest on in the case of bulges. Yet, fortunately, this ambiguity is not particularly concerning from an interpretational point of view. Clearly, either bulge mass or central velocity dispersion is the best available quenching parameter in bulges, and both of these parameters are well known to be closely correlated with black hole mass (e.g. \citealt{Ferrarese2000, Haring2004, McConnell2013, Saglia2016, Terrazas2016, Piotrowska2021}). As such, these results are certainly consistent with the quenching of bulge and disk structures via AGN feedback. Clearly, the central regions within central galaxies govern the quenching in both bulge and disk structures, and hence in galaxies as a whole. This is a very important new result, which strongly implies that quenching is a global process. We demonstrate this fact in an even clearer manner utilising the spatially resolved spectroscopy of MaNGA in the following sub-section.

All of the other parameters are found to be of very little importance to the quenching of bulges and disks, once a measurement of the central regions within galaxies is made available to the classifier. It is particularly important to highlight the utter lack of connection between quenching in bulge and disk components and the total stellar mass, group halo mass, local density or even morphology. Therefore, these parameters cannot be causally connected to quenching. This offers powerful new constraints on which mechanisms may be viable, which we consider further in the Discussion.


\begin{figure*}
\begin{centering}
\includegraphics[width=0.95\textwidth]{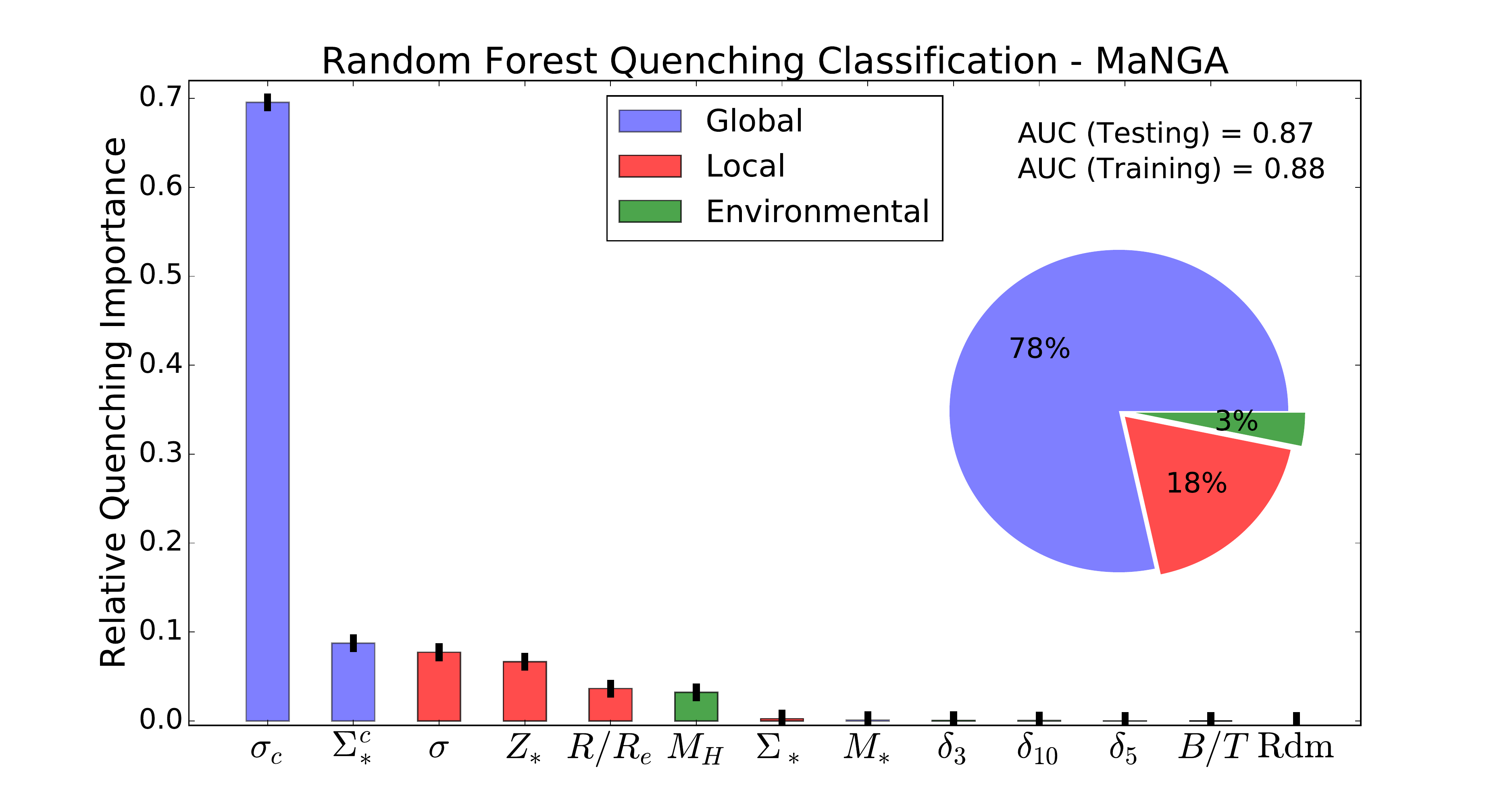}
\caption{Reproduction of fig. 9 (top panel) from \cite{Bluck2020a} utilising here the All Parameter RF classification mode (as opposed to the $\sqrt{N}$ Parameter [default]  model used previously). This figure shows the relative importance of global, local and environmental parameters for predicting quenching on spatially resolved scales in central galaxies observed in MaNGA. The $y$-axis displays the relative quenching importance of each variable in turn, and the $x$-axis labels each variable (shown in descending order of importance). From left-to-right, the parameters displayed are: central velocity dispersion ($\sigma_c$); central stellar mass surface density ($\Sigma_*^c$); velocity dispersion within each spaxel ($\sigma$), metallicity within each spaxel ($Z_*$); position of each spaxel within the galaxy ($R/R_e$); halo mass ($M_H$); stellar mass surface density within each spaxel ($\Sigma_*$); total stellar mass of the galaxy ($M_*$); local density evaluated at the 3rd, 10th and 5th nearest neighbours ($\delta_{3}$, $\delta_{10}$, $\delta_5$); bulge-to-total stellar mass ratio of the galaxy ($B/T$); and a random number (Rdm). Parameter bars are colour coded by whether the parameter is global (one parameter per galaxy, pertaining to the galaxy as a whole); local (one parameter per spaxel); or environmental (one parameter per galaxy, pertaining to the environment in which the galaxy resides). Uncertainties on the relative importances are given as the variance across 10 independent training and testing runs. It is clear that central velocity dispersion is by far the most predictive parameter of quenching on $\sim$kpc scales. Additionally, global parameters collectively are much more effective at predicting quenching on $\sim$kpc scales than local or environmental parameters (see the pie chart, inset).}
\end{centering}
\end{figure*}

\subsection{MaNGA: spatially resolved quenching \& star formation}

In this sub-section we take advantage of the kpc-scale spatial resolution of MaNGA to analyse quenching within local galaxies. From the previous sub-section, we know that the quenching dependence of bulges and disks are the same, both being governed by the bulge properties alone. However, this is a rather coarse division, and so it is certainly worth considering if a more refined spatial segregation would yield significant new insights. Additionally in this sub-section, we also compare spatially resolved quenching (in the full MaNGA sample) to spatially resolved star formation (in the star forming sub-sample). This is helpful because the spatial dependence on quenching and star formation turns out to be vastly different (see also \citealt{Bluck2020a}). 

We remind readers that the spatial extent of MaNGA galaxies is truncated at $\sim 1.5 R_e$. Consequently, the results and conclusions which follow in this sub-section explicitly relate to this spatial range. Nonetheless, the remarkable consistency in the quenching analyses of bulges and disks from the SDSS and MaNGA samples (in the preceding sub-section) encourage us that these results are also likely to be similar in a larger field of view.

\subsubsection{An updated random forest approach for classifying quenched \& star forming regions in MaNGA}

In \citealt{Bluck2020a} we investigated whether quenching (and star formation) is regulated by local (spatially resolved), global, or environmental phenomena. Our conclusion was that quenching is regulated by global physics, yet star formation is regulated by local physics operating within galaxies. One of our main analyses which arrived at this conclusion was conducted via RF classification of star forming vs. quenched spaxels, and comparing this to RF regression of $\Sigma_{\rm SFR}$ values in star forming systems (see fig. 9 in \citealt{Bluck2020a}). In \cite{Bluck2020a} we utilised the default mode of operation for RF classification and regression. For classification, this utilises $\sqrt{N}$ parameters, but for regression this utilises all parameters simultaneously. However, in Appendix B we found that switching to the All Parameter mode in classification improves the identification of causality in model data. Hence, it is extremely interesting to revisit our prior MaNGA classification analysis with the most effective method available.

In Fig. 10 we show a reproduction of fig. 9 (top panel) from \cite{Bluck2020a}. Here we utilise the All Parameter mode of the RF classifier, as opposed to the $\sqrt{N}$ parameter mode utilised in our prior work. Fig. 10 shows the relative quenching importance for a host of global, local and environmental parameters (as labelled by the $x$-axis). Central velocity dispersion is found to be overwhelmingly the most important parameter for predicting the quenching of regions within galaxies. All of the other parameters are of very low importance for regulating quenching, when $\sigma_c$ is made available to the classifier. Additionally, we group parameters by their type (i.e. as global, local or environmental). Collectively, global parameters vastly outperform both local and environmental parameters. Thus, quenching is a global phenomenon. This conclusion is consistent with our bulge - disk analyses in the SDSS and MaNGA (see Section 4.3).

In comparison to our prior result, both the most successful parameter and the most successful group remains invariant. Hence, our prior conclusions are stable to this new analysis. On the other hand, the present analysis places this result on a much firmer statistical footing. More specifically, we find that the importance of all parameters (except for $\sigma_c$) are significantly reduced in the All Parameter mode, displayed here in Fig. 10. The reason for this is that the RF is fully controlling for all nuisance variables here. Conversely, the importance of $\sigma_c$ rises dramatically in the present analysis. Leveraging our understanding from the RF tests (shown and discussed in Appendix B.2), we interpret this as importance being systematically shifted from the dominant parameter to spurious nuisance parameters when only a fraction of parameters are considered at each node.


\begin{figure*}
\includegraphics[width=0.49\textwidth]{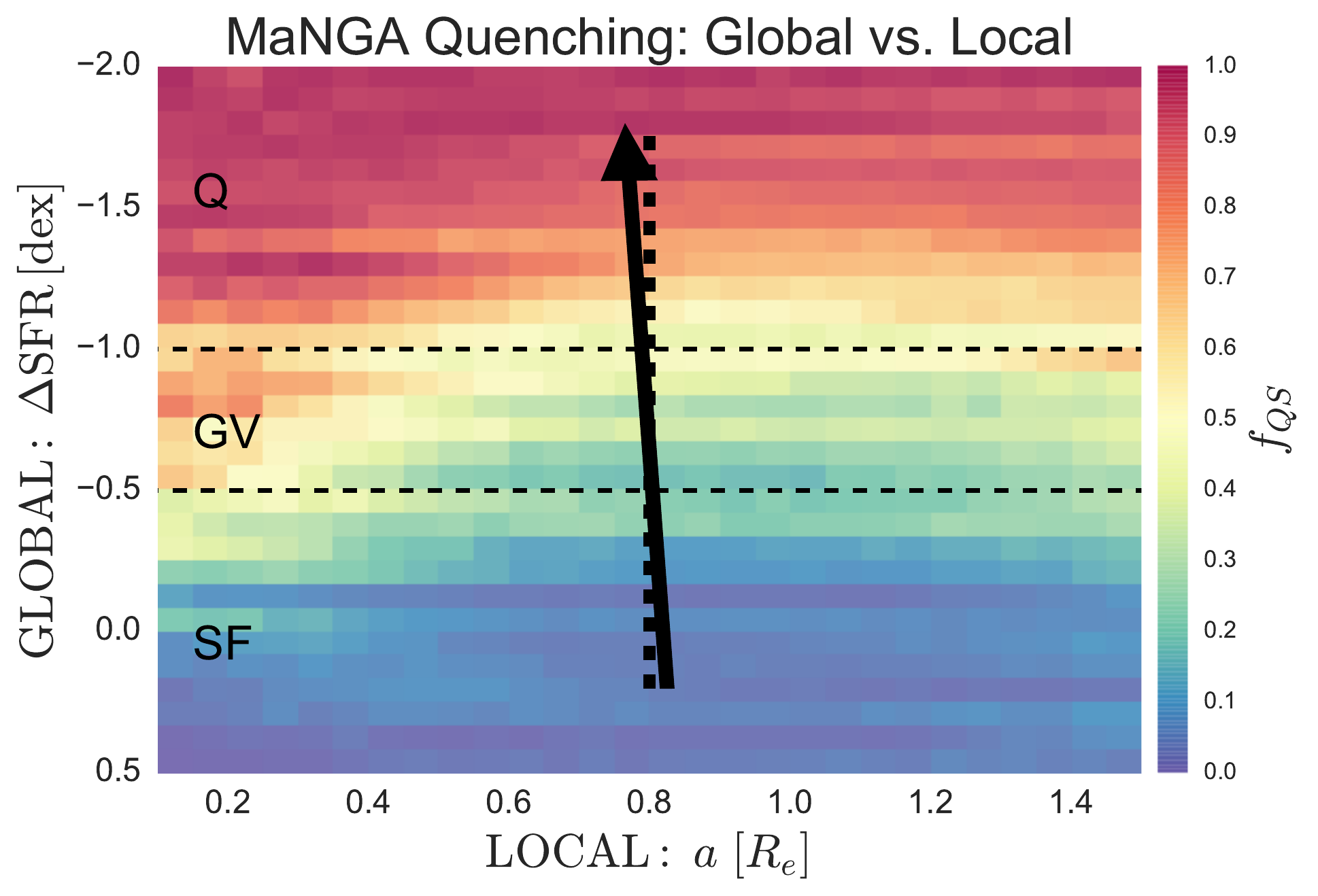}
\includegraphics[width=0.49\textwidth]{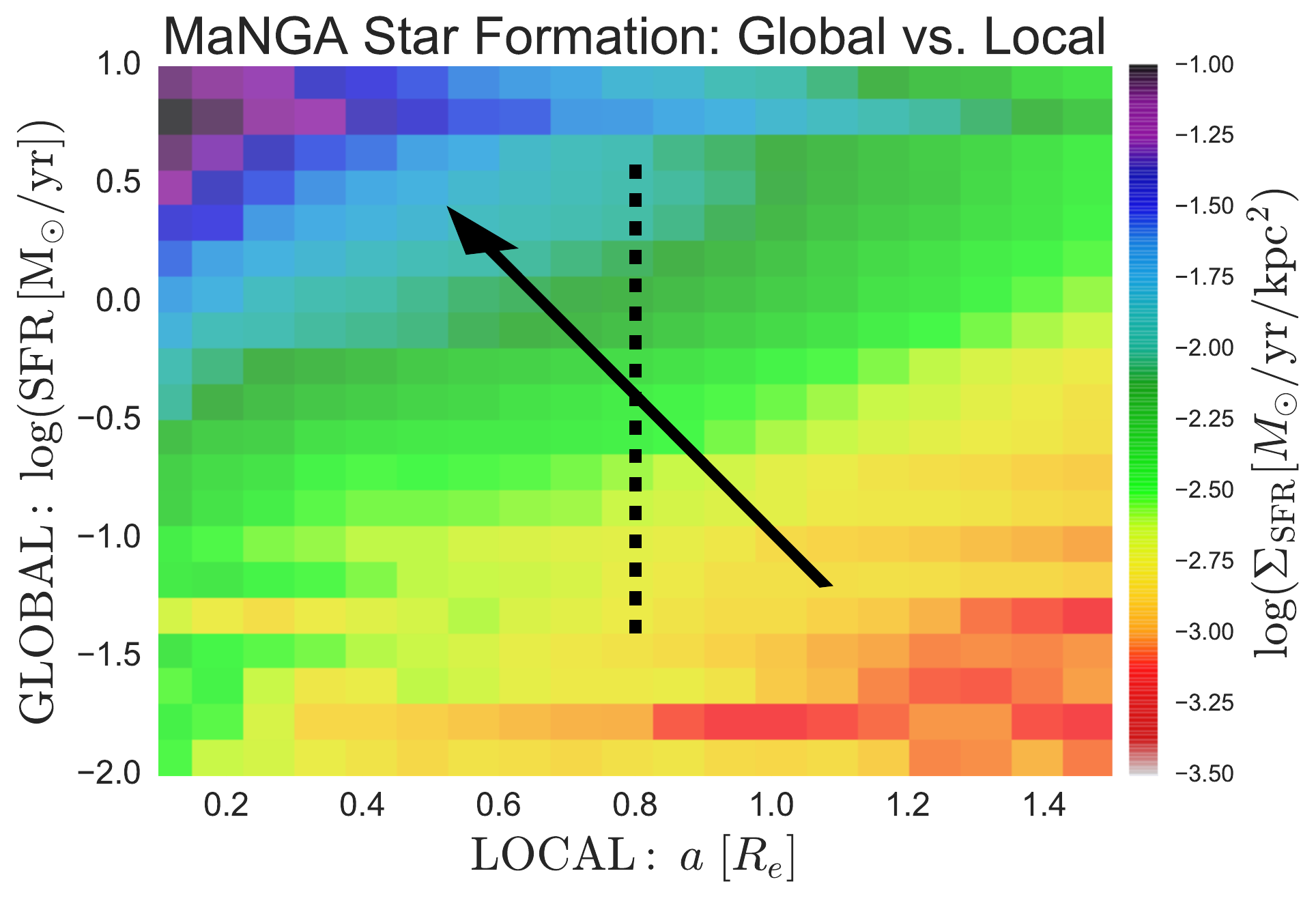}
\caption{Visual test of whether quenching and star formation are regulated by global or local physics. {\it Left panel: } Global vs. local quenching colour map. The fraction of quenched spaxels in each rectangular bin is displayed by colour (labelled by the colour bar), shown as a function of the global galaxy star forming state (quantified by $\Delta$SFR) and the position of spaxels within the galaxy (quantified by the deprojected elliptical radius, $a$). The location of the global `green valley' is indicated by dashed black lines. {\it Right panel: } Global vs. local star formation rate colour map. The mean star formation rate surface density for star forming systems in each rectangular bin is displayed by colour (labelled by the colour bar), shown as a function of global galaxy SFR and deprojected elliptical radius ($a$). There is evidently little radial dependence on quenching, except for a weak trend around the green valley. Alternatively, for star formation, there is a highly pronounced radial dependence evident at all values of global SFR. To quantify the relative dependence on global and local processes, on each panel we show an arrow which points in the direction of maximising quenched fraction ($\Theta_Q = -3\pm1^\circ$, left panel) and star formation rate ($\Theta_{SF} = -37\pm5^\circ$, right panel). Note that quenching proceeds almost vertically in the plane, indicating global dependence; whereas star formation proceeds diagonally, indicating significant local and global dependence.}
\end{figure*}

For example, in fig. 9 (top panel) in \cite{Bluck2020a}, the velocity dispersion within each spaxel ($\sigma$) is found to be the second most important variable (after $\sigma_c$), with a comparable absolute level of relative importance. Yet, here, we find that $\sigma$ has very little value for predicting quenching, once $\sigma_c$ is available to the classifier. Thus, the importance given to $\sigma$ in our prior analysis was spurious (as indeed we argued for in our previous paper). The value of the new approach is that it removes the need for post-hoc interpretation almost entirely from the RF classification results, directly pointing to the most probable causal relation (from the available parameters).

In terms of our present goals, the results from Fig. 10 confirm that quenching is a global process (i.e. that regions within galaxies are not frequently in differing star forming states to the galaxy as a whole). Moreover, the results from Fig. 10 are in beautiful agreement with Figs. 4 \& 9 (right panels) as to the most effective quenching parameter. Importantly, central velocity dispersion remains the best parameter for predicting quenching on kpc-scales as well as on galaxy-wide scales (and in bulge and disk structures). This can only be the case if the quenching of all regions within galaxies occur largely in concert. 

Since the default mode of RF regression is already the (most effective) All Parameter mode, we do not repeat the star forming regression analysis from the lower panel of fig. 9 in \cite{Bluck2020a} here. Nevertheless, it is worth recalling that for star formation (in star forming systems), local spatially resolved parameters are most effective for predicting $\Sigma_{\rm SFR}$, especially the stellar mass surface density within each spaxel ($\Sigma_*$). Hence, quenching and star formation are fundamentally distinct physical processes within galaxies - the latter is regulated by local physics within each region of each galaxy; whereas the former is regulated by galaxy-wide physics, governed primarily by the central-most regions within galaxies. This result is of enormous value in constraining the possible physical mechanisms at work in galaxy evolution across cosmic time. We defer to the Discussion a thorough consideration of which quenching mechanisms remain viable following these results.

\subsubsection{A visual comparison of quenching vs. star formation in MaNGA}

In Fig. 11 (left panel) we present a 2-dimensional colour map showing the fraction of quenched spaxels distributed according to the global star forming state of the galaxy ($\Delta$SFR) and the location of each spaxel within each galaxy ($a/R_e$, i.e. the elliptical radius in units of the half light radius). In star forming systems, the fraction of quenched spaxels is low everywhere in radial extent (out to 1.5 $R_e$, the typical limit to the MaNGA field of view). Conversely, in quiescent systems, the fraction of quenched spaxels is very high everywhere in radial extent (out to 1.5$R_e$). Hence, star forming galaxies are typically star forming, and quenched galaxies are typically quenched, everywhere within 1.5 $R_e$\footnote{Note that this is not as trivial as it may seem, since the level of sub-galactic conformity is $>$85\%, far higher than the trivial level of 50\% needed for a simple definition of star forming or quenched (on average).}. However, for intermediate levels of global star formation, i.e. in the global `green valley', there is a weak but noticeable trend whereby there is a higher fraction of quenched spaxels in the centre of galaxies relative to their outer regions. 

Viewing Fig. 11 (left panel) in its entirety, it is clear that quenching progresses mostly as a function of the global star forming state of galaxies, rather than the position within each galaxy from which the spaxel is drawn. Thus, to leading order, quenching is a global, not local, process. This result agrees with our conclusion from Fig. 10 (and \citealt{Bluck2020a}), where we demonstrate that spatially resolved (local) parameters are less informative of whether spaxels within galaxies are star forming or quenched than global (galaxy-wide) parameters. Additionally, this result is consistent with \cite{Bluck2020b}, where we show that green valley galaxies exhibit the signature of inside-out quenching in centrals, but the vast majority of galaxies are either star forming or quenched throughout the entire radial extent probed by MaNGA.

In Fig. 11 (right panel) we present instead a 2-dimensional colour map showing the mean star formation rate surface density ($\Sigma_{\rm SFR}$) in star forming galaxies as a function of the global SFR of galaxies and the location of spaxels within each galaxy ($a/R_e$). The change from viewing the fraction of quenched spaxels to the mean star formation rate surface density is striking. Whilst there is also a strong global dependence in the sense that galaxies with higher SFRs host spaxels with higher $\Sigma_{\rm SFR}$ values (unsurprisingly), there is also a very pronounced radial dependence such that the highest star formation rates are found in the centre of star forming galaxies. These results are consistent with \cite{Bluck2020a, Bluck2020b}, but are presented in a novel manner here. Viewing the right panel of Fig. 11 in its entirety, it is clear that star formation has both a local and a global component.

To be more quantitative about our conclusions from Fig 11, we utilise the vector decomposition of partial correlations technique (introduced in \citealt{Bluck2020a}) to quantify the optimal direction to move through the global - local plane in order to: i) maximise the fraction of quenched spaxels (left panel); and ii) maximise $\Sigma_{\rm SFR}$ (right panel). We refer to these angles as the quenching angle ($\Theta_Q$) and the star formation angle ($\Theta_{SF}$), respectively. Explicitly, we construct the quenching angle as the arctangent of the ratio of the partial correlation of quenched fraction with $\Delta$SFR (at fixed $a/R_e$) and the partial correlation of quenched fraction with $a/R_e$ (at fixed $\Delta$SFR). The star formation angle is similarly defined by simply replacing quenched fraction with $\Sigma_{\rm SFR}$. That is, we compute:

\begin{equation}
\Theta_{Q/SF} = {\rm tan}^{-1} \bigg( \frac{\rho_{\, \mathrm{GLOBAL}; \, \mathrm{LOCAL}}}{\rho_{\, \mathrm{LOCAL}; \, \mathrm{GLOBAL}}} \bigg)
\end{equation}

\noindent where the numerator indicates the partial correlation between quenching (or star formation) and the global parameter (whilst holding the local parameter fixed), and the denominator inverts this. Uncertainties on angles are inferred through bootstrapped random sampling of the data. See \cite{Bluck2020a} for full details on this method. 

The quenching angle is found to be $\Theta_Q = -3\pm1^{\circ}$, and the star formation angle is found to be $\Theta_{SF} = -37\pm5^{\circ}$. Both angles are measured clockwise from vertical, in the $\Delta {\rm SFR} - a/R_e$ and ${\rm SFR} - a/R_e$ planes, respectively. Hence, quenching is almost purely a global process (with a slight trend to enhanced quenching at the centre of galaxies in the green valley region). Conversely, star formation has a roughly equal global and local dependence, and hence the location within the galaxy is a highly important parameter for determining the local level of star formation in (globally) star forming systems.

\subsection{Section summary}

In conclusion to the entirety of this section, we find that the quenching of galaxies, bulges, and disks are all governed primarily by the conditions of the central-most regions within local galaxies (i.e. bulge mass and central velocity dispersion). These results are identical in the SDSS and MaNGA samples, even though the methodology for bulge - disk separation and identifying quenched regions are very different between the two surveys. This is of great practical value for analysing the quenching of galaxies, bulges, and disks at higher redshifts because it demonstrates that a colour based approach yields results that closely mirror a more sophisticated spatially resolved spectroscopic approach. We apply this colour based approach to the largest galaxy survey at intermediate-to-high redshifts in the next section.

Additionally, we have explored the possibility for significant variation in quenching on sub-galactic scales (within $1.5 R_e$), taking advantage of the unprecedented spatial coverage of MaNGA. Yet, to leading order, quenching is found to be a global process, impacting all parts of a galaxy in concert. The one important caveat to this statement is that quenching galaxies (i.e. those with intermediate global levels of star formation) are found to have a significant (though subtle) secondary trend with radius, such that their centres are more frequently quiescent than their outskirts (in agreement with, e.g. \citealt{Tacchella2015, GonzalezDelgado2016, Ellison2018, Medling2018, Bluck2020b}). On the other hand, star formation in star forming galaxies does show a pronounced dependence on radius (as also commented upon in \citealt{Bluck2020a, Bluck2020b}). Taken together, these results imply that quenching must be a global process, which requires a galaxy-wide (or even halo-wide) physical mechanism; whereas star formation is a local process, highly dependent upon the physical conditions within each region of each galaxy.



\begin{figure}
\includegraphics[width=0.49\textwidth]{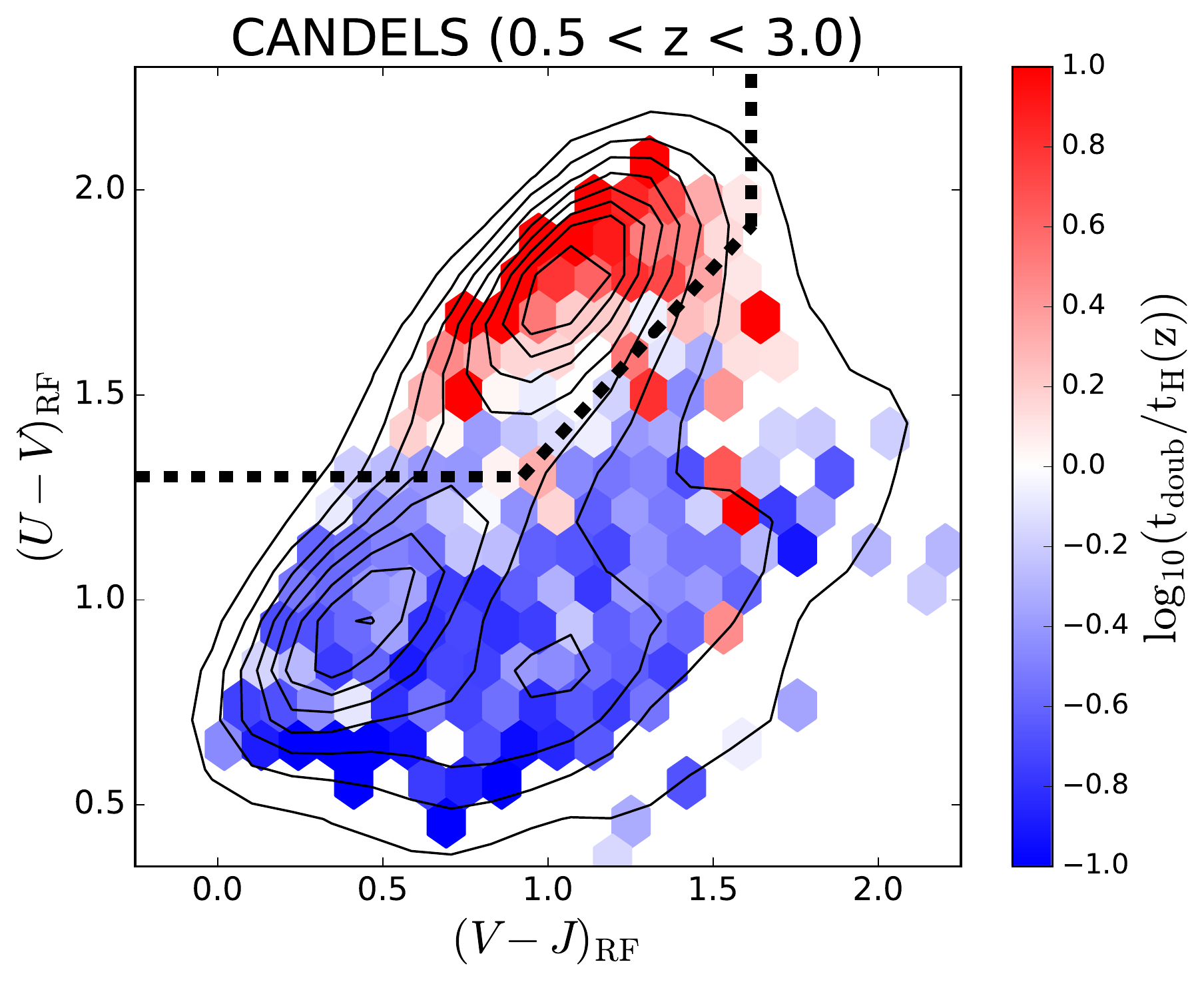}
\caption{UVJ colour - colour diagram for CANDELS galaxies. The location of galaxies in colour space is shown by black density contours. Each hexagonal region of the plane is colour coded by the mean doubling time (1/sSFR) expressed in units of the Hubble time ($t_H(z)$). The threshold for quenching from \cite{Williams2009} at z = 1 (the median redshift of CANDELS galaxies in our sample) is shown as a dashed magenta line. Note that there is good qualitative agreement between colour and SED sSFR methods for identifying quenched galaxies. }
\end{figure}

\section{Quenching at high redshifts}

In this section we expand our analysis of quenching to high redshifts, utilising data from CANDELS and especially the bulge - disk decomposition catalogue of \cite{Dimauro2018}. Our goal is to assess which parameters are most effective at predicting (and hence regulating) quenching at high redshifts, and compare these results to their low redshift analogues from the SDSS and MaNGA (discussed in Section 4). In so doing, we establish that there is a remarkable consistency in the parameters which regulate quenching at high and low redshifts, indicating a high likelihood for a stable quenching mechanism operating over cosmic time. 

Moreover, we also analyse the quenching of bulge and disk structures (treated separately) in the distant Universe for the first time. Following our results in the previous section, we expect our colour-based analyses to be consistent with upcoming wide-field IFU spectroscopic surveys in the coming years (e.g., from JWST and the ELTs). Consequently, we achieve a preview of spatially resolved quenching in observational data at moderate-to-high redshifts in a statistically representative sample of galaxies now.

\subsection{Identifying quenched galaxies throughout cosmic time}

In order to investigate quenching in the CANDELS data set we must first construct a method to classify galaxies into star forming and quenched categories, which is effective in the photometric data available at high redshifts. As discussed in Section 3, the two broad choices for classifying galaxies into star forming and quenched categories come from rest-frame optical/NIR colours and from star formation rates. In this section we utilise UVJ rest-frame colours to identify red galaxies, bulges and disks, which are furthermore not likely to be a result of extensive dust extinction (see \citealt{Williams2009}). Additionally, we briefly consider SFRs derived from SED fitting (since spectroscopy is not available for the vast majority of the CANDELS sample) as a test to the UVJ colour method. Both the UVJ colours and the SED SFRs are corrected for dust extinction by assuming a \cite{Calzetti2000} extinction law, following the standard chi-square minimisation technique comparing measured magnitudes against a grid of model galaxy spectra with varying star formation histories, stellar masses, extinction levels, metallicities and redshifts (see \citealt{Dimauro2018} for full details).

In Fig. 12 we show the location of CANDELS galaxies in rest-frame (U-V) - (V-J) colour space with linearly spaced density contours (shown as black lines). There is marked bimodality in the location of CANDELS galaxies in this colour space, such that there is a red peak and a blue peak, with an extended region lying off to the lower-right. Numerous prior studies have classified high redshift galaxies leveraging the appearance of bimodality in the UVJ colour - colour diagram (e.g. \citealt{Williams2009, Patel2012, Muzzin2013, Belli2015, Schreiber2018}). The essential idea is that, in the absence of dust extinction, more quiescent galaxies appear redder due to hosting older stellar populations. Additionally, the leading order impact of dust obscuration may be taken into account by excising the region in UVJ colour space where extinction dominates. In this work we utilise the redshift evolving colour cuts of \cite{Williams2009}. We display the cut at the median redshift of the CANDELS sample (z $\sim$ 1) as dashed magenta lines in Fig. 12. Visually, these cuts do a very reasonable job of separating the blue and red peaks in the density distribution.

We cross-validate the UVJ colour - colour selection method with a method based on sSFR. To this end, in Fig. 12 we colour each small hexagonal region within the UVJ plane by the mean doubling time expressed in units of the Hubble time ($t_{\rm doub}/t_H(z)$). The doubling time is given simply by the inverse of sSFR (evaluated from SED fitting in \citealt{Dimauro2018}), and is expressed in units of the Hubble time to establish a redshift invariance in the measurement (e.g. \citealt{Tacchella2019}). Colour selected quenched objects have noticeably higher $t_{\rm doub}/t_H(z)$ values and, conversely, colour selected star forming galaxies have much lower $t_{\rm doub}/t_H(z)$ values (as expected). This comparison establishes a broad qualitative consistency between these two methods for identifying quenched and star forming galaxies in CANDELS.

It would be possible to take the sSFR method as being primary, and then to construct the optimal linear decision boundary for quenching in UVJ colour space, in exact analogy to our approach in $ugr$ colour space for SDSS galaxies (see Section 3.1.3). However, the CANDELS SFRs are based on SED fitting to typically just 5 wavebands, and hence are much less accurate and reliable than the SDSS (and MaNGA) spectroscopic based SFRs. As such, we believe it is more robust (and straightforward) to use the colour based method in this section. To achieve a fair comparison with the SDSS data we also utilise a colour based method for the low-z data in this section as well. Nonetheless, we also test reproducing all of our high-z results with photometric SFRs. All of the results and conclusions are identical to what we present here with UVJ colours.


\subsection{CANDELS: random forest classification}

\subsubsection{Ensuring a fair comparison}


\begin{figure*}
\includegraphics[width=0.33\textwidth]{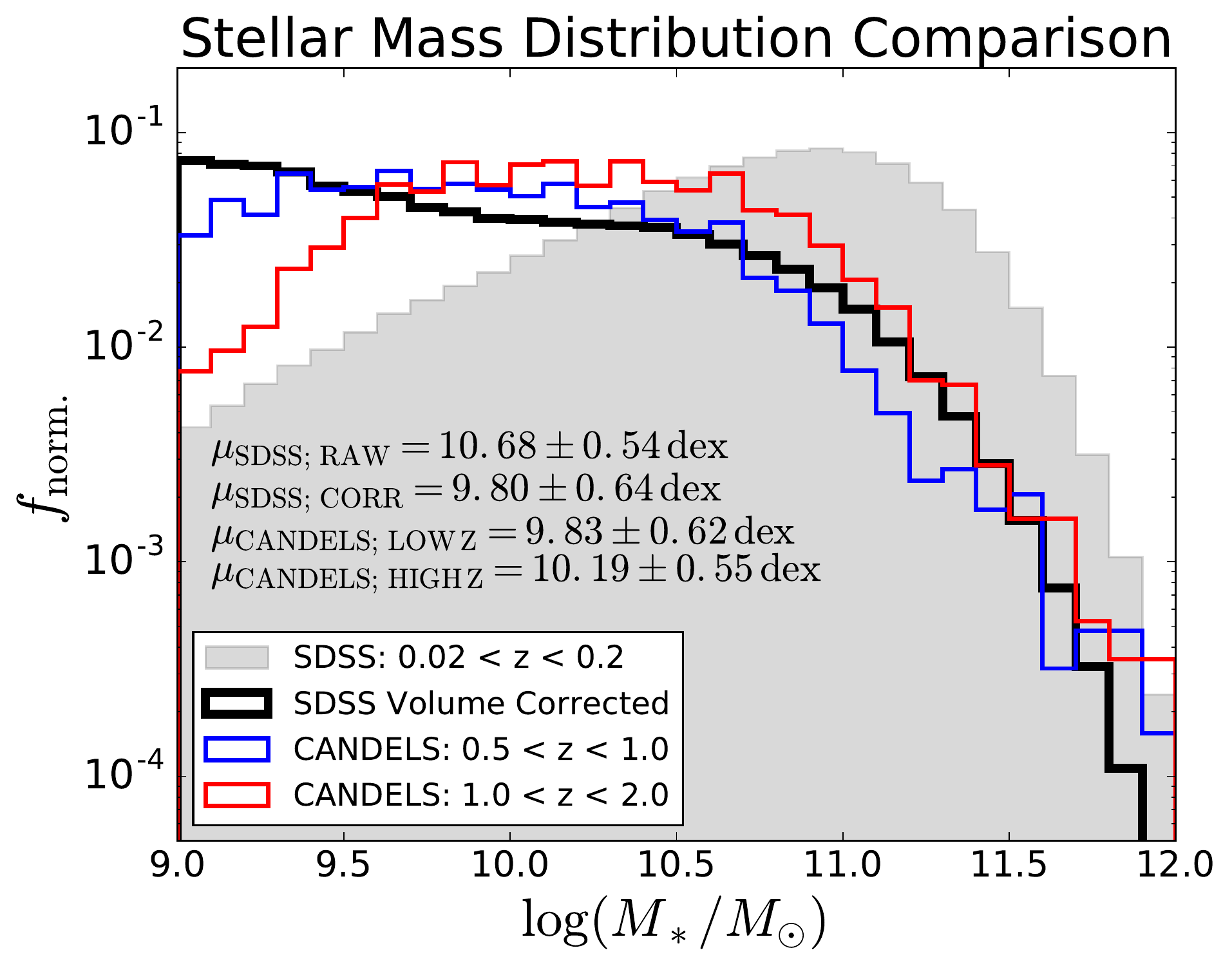}
\includegraphics[width=0.33\textwidth]{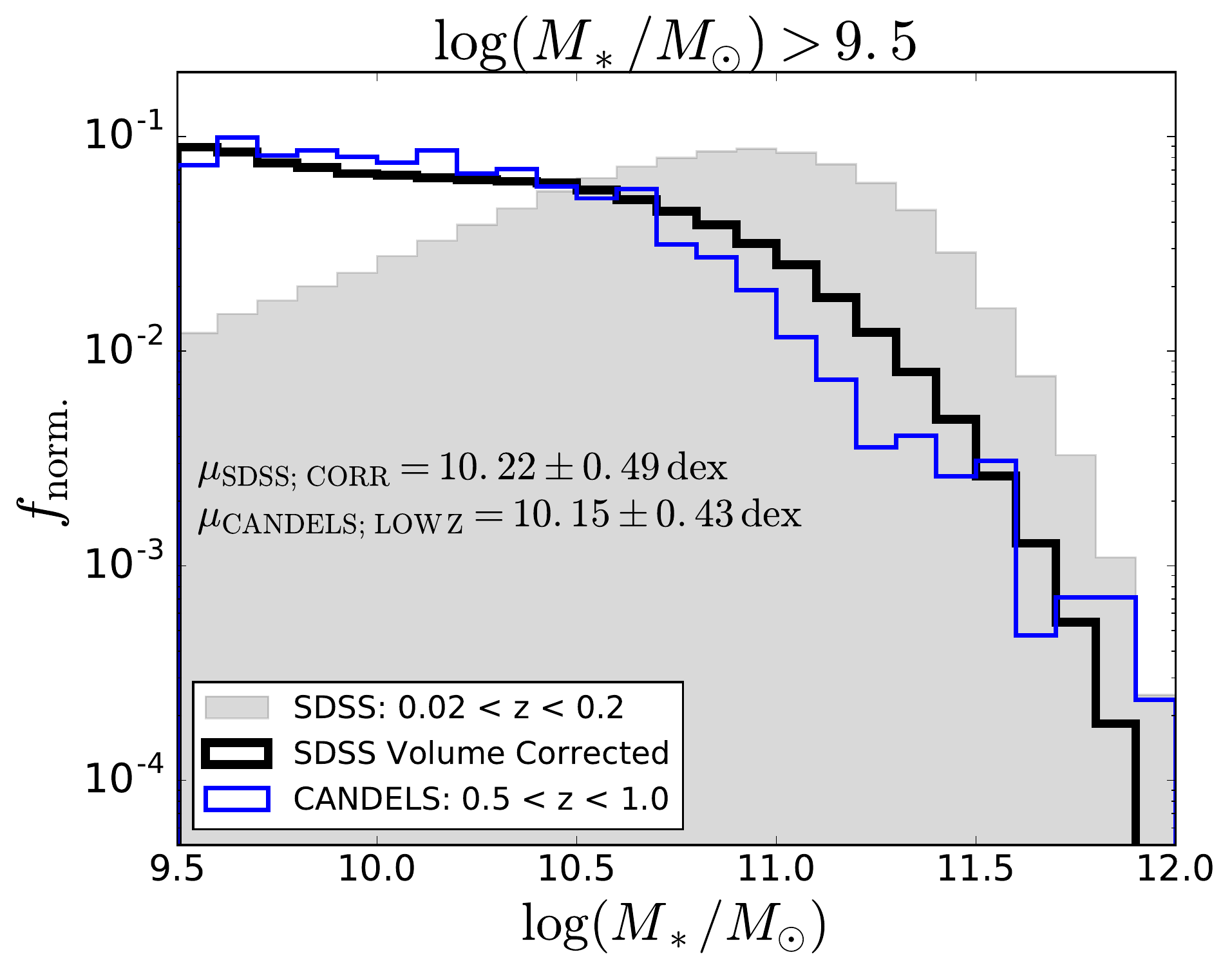}
\includegraphics[width=0.33\textwidth]{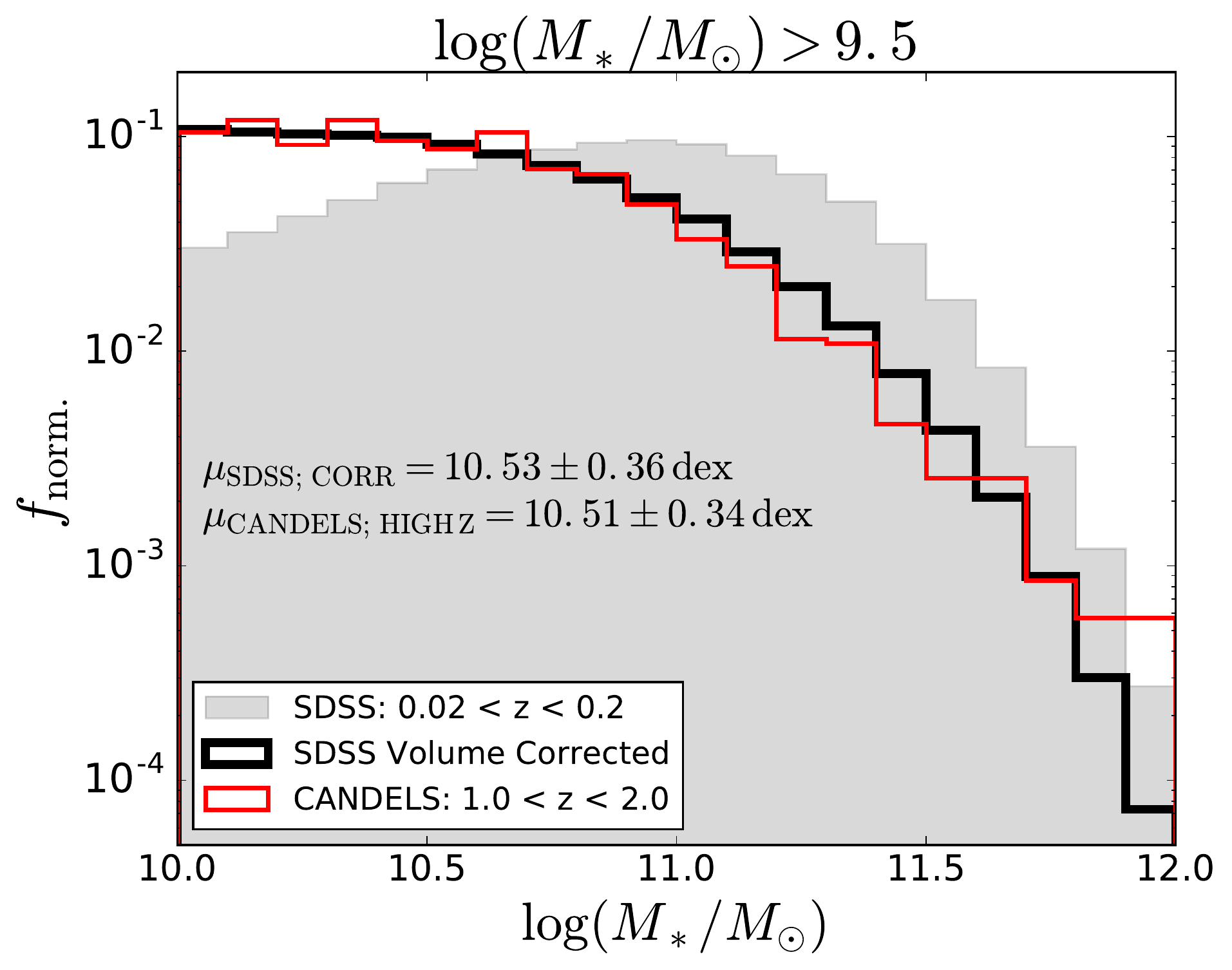}
\caption{Distribution of stellar mass in the SDSS and CANDELS datasets. {\it Left panel: } The normalised stellar mass distribution for the full mass range in the SDSS compared to the intermediate ($0.5 < z < 1.0$) and high ($1.0 < z < 2.0$) redshift ranges in CANDELS. The SDSS mass distribution is shown in a raw (shaded histogram) and volume corrected (solid black line) form. Note that there is much better agreement between the volume corrected SDSS mass distribution and the CANDELS data, due to the much greater depth of CANDELS. Nonetheless, there is systematic incompletion at low masses in CANDELS. {\it Middle panel: } Comparison of intermediate redshift CANDELS data to the SDSS at a stellar mass limit of $\log(M_*/M_{\odot}) > 9.5$. {\it Right panel: } Comparison of high redshift CANDELS data to the SDSS at a stellar mass limit of $\log(M_*/M_{\odot}) > 10$. In the middle and left panels, there is reasonable qualitative agreement between the mass distributions, and the mean and standard deviation of each sample agree comfortably within 0.1dex in both cases. We use the mass restricted samples for our quantitative comparison between quenching in the two surveys, to avoid issues from differing stellar mass distributions.  }
\end{figure*}

In this sub-section we perform a Random Forest classification of CANDELS galaxies into star forming and quenched categories based on UVJ colour - colour cuts (as illustrated in Fig. 12). Given that CANDELS is primarily a photometric survey, we restrict our analysis here to the bulge - disk parameters, taken from the photometric SED fitting catalogs of \cite{Dimauro2018}. Furthermore, the lack of accurate spectroscopic redshifts for the vast majority of galaxies in CANDELS prevents us from categorising galaxies as centrals or satellites. Consequently, we re-analyse the SDSS data for the full galaxy sample to enable a fair comparison between redshifts. However, central galaxies dominate at all redshifts (e.g. \citealt{Henriques2015}) and hence the following analyses are in any case most probably representative of the central galaxy population (but not the satellite galaxy population). Indeed, in the SDSS, centrals outnumber satellites by $\sim$4:1 (see \citealt{Yang2007, Yang2009}). Additionally, we perform the SDSS RF classification runs based on our colour selection (see Fig. 2) to enable a high level of consistency in the methodology.

Another issue which is potentially very important is that we select a similar sample of galaxies at both high and low redshifts in order to make a meaningful comparison. To this end, in Fig. 13 (left panel) we show the distribution in stellar mass of SDSS galaxies for the raw counts (filled grey histogram) and for the $1/V_{\rm max}$ weighted sample (open black histogram). For comparison, we overlay the raw stellar mass distributions for CANDELS in two redshift ranges: $0.5 < z < 1.0$ (shown in blue) and $1.0 < z < 2.0$ (shown in red). It is immediately clear that the CANDELS raw distributions agree much more closely with the volume corrected SDSS data than the raw SDSS data. This is easy to understand since the CANDELS data is much deeper than the SDSS, by several orders of magnitude. Indeed, at high masses, the stellar mass distribution of CANDELS and the (volume corrected) SDSS are in very good accord. However, at low masses there is significant incompletion in the CANDELS data, which gets progressively more severe at higher redshifts.

As a result of observing the systemic incompleteness relative to the volume corrected SDSS sample in CANDELS, we apply the following stellar mass cuts to both the CANDELS data and the SDSS data for comparison: $\log(M_*/M_{\odot}) > 9.5$ for $(0.5 < z < 1.0)$; and $\log(M_*/M_{\odot}) > 10.0$ for $(1.0 < z < 2.0)$. In the centre and right panels of Fig. 13 we show the comparison between volume corrected SDSS data and the raw CANDELS data for the two mass cuts, respectively. After applying these cuts the mean stellar mass and standard deviation of the stellar mass distributions are very similar between the SDSS and CANDELS (both agreeing within 0.1 dex, i.e. well within their respective uncertainties). Consequently, we adopt the above mass cuts for our comparison, and restrict our SDSS analyses here to the volume corrected sample. This ensures that the distribution of galaxies in stellar mass is very similar, despite the very different depths and areas of the two surveys.

\subsubsection{Classification results}


\begin{figure}
\begin{centering}
\includegraphics[width=0.49\textwidth]{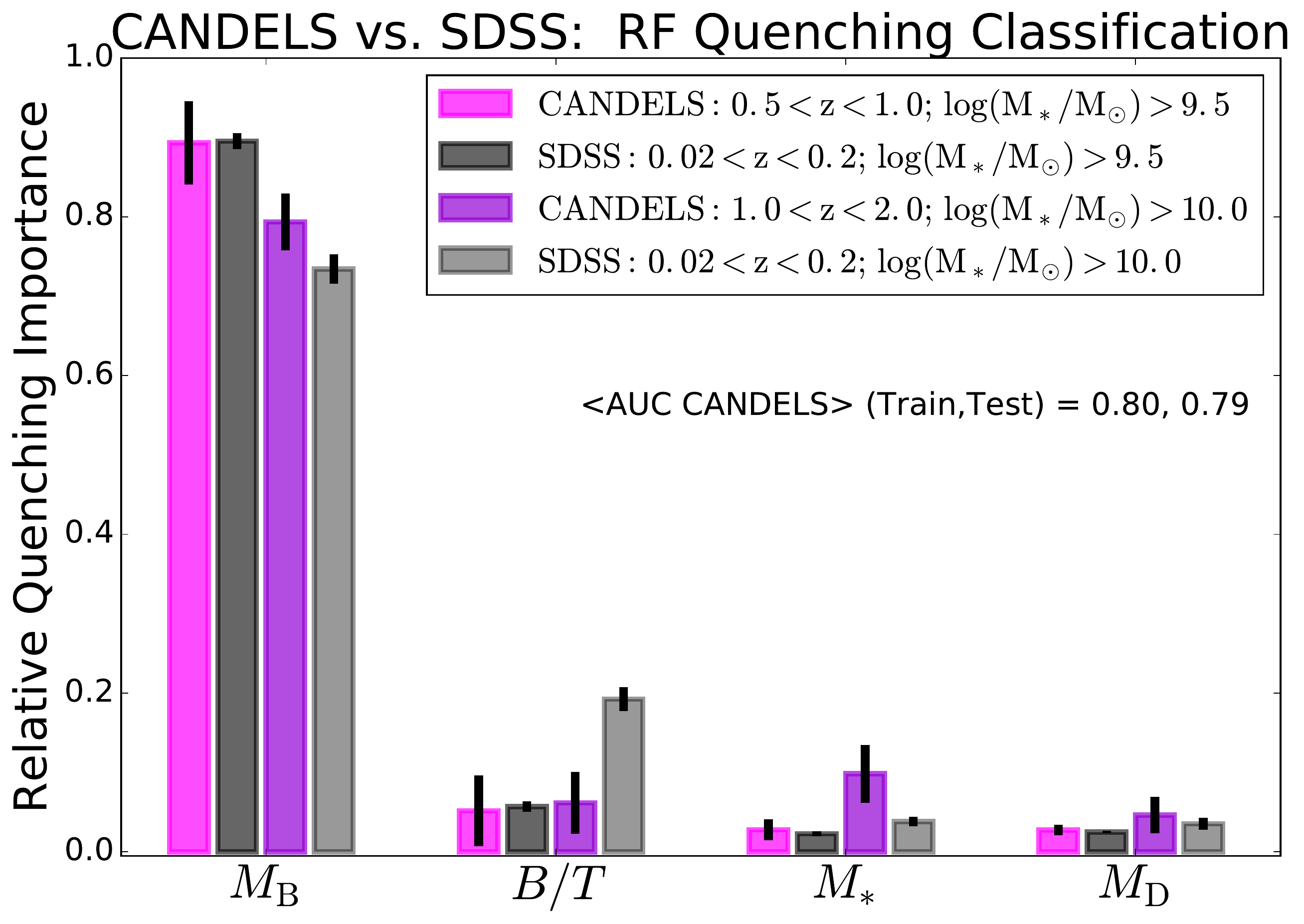}
\caption{Random Forest quenching classification analysis showing results for the bulge - disk parameters in CANDELS (shown in shades of purple) compared to the SDSS (shown in shades of grey). The structure of this plot is similar to Fig. 5 (left panel). The two surveys are compared at various mass cuts (as noted in the legend), ensuring that differences in their stellar mass distributions are minimised. In all datasets, the full galaxy sample is used here (i.e. no restriction to centrals is made) and a colour based definition of quenching is utilised, ensuring a high level of consistency between the samples. Bulge mass is found to be the clear best parameter for predicting quenching in both the high and intermediate redshift CANDELS data, and in the low redshift SDSS data.}
\end{centering}
\end{figure}

In Fig. 14 we present the results from a Random Forest quenching classification of CANDELS galaxies, based on training and validating with the UVJ colour - colour classification method. The photometric bulge - disk parameters are assessed in terms of their relative ability to predict quenching. The CANDELS data set is split into an intermediate and high redshift range (shown in light and dark purple, respectively). For the intermediate redshift range we restrict the sample to $\log(M_*/M_{\odot}) > 9.5$ and compare to the full SDSS galaxy sample (with volume correction applied) utilising the same stellar mass cut. For the high redshift range we restrict the sample to $\log(M_*/M_{\odot}) > 10.0$ and compare the full SDSS galaxy sample (with volume correction applied) utilising the same stellar mass cut. As shown in the previous sub-section, these restrictions ensure a very similar stellar mass distribution for both comparisons.

For both CANDELS redshift ranges, bulge mass is found to be clearly the most effective parameter for predicting quenching in galaxies. The $B/T$ morphology, total stellar mass and disk mass are all found to be of very little importance to quenching, once bulge mass is available to the classifier. In comparison to the SDSS, the results are essentially identical, also clearly establishing bulge mass as by the far the most predictive parameter for quenching in galaxies. Thus, there is no significant evolution in the dependence of quenching on the bulge - disk parameters from cosmic noon to the present epoch. This strongly suggests that quenching is due to a stable mechanism operating throughout cosmic time.

Remarkably, the LGalaxies semi-analytic model predicts that quenching should be regulated primarily by bulge mass at both high and low redshifts (see Fig. 3), with little-to-no evolution in this parameter set across cosmic time. This prediction is precisely recovered in the multi-epoch observational data. Consequently, the preventative AGN feedback model in LGalaxies is a viable explanation to the observational results presented here in Fig. 14. We review the possibility for other physical mechanisms to also give rise to these observational results in the Discussion.

The results from Fig. 14 are in good qualitative agreement with the results from \cite{Lang2014}, who find that the bulge mass - quenched fraction relationship is tighter than the equivalent relationships with stellar mass, $B/T$ morphology and disk mass. Essentially this is the high-z equivalent of the result found at low-z in \cite{Bluck2014}. However, here we utilise a much larger sample of high redshift galaxies, incorporating almost a factor of three times more bulge - disk decompositions in our analysis. Methodologically, we utilise a sophisticated machine learning algorithm capable of extracting causal insights from complex inter-correlated data, which is completely novel to this particular problem. As such, we significantly expand on the early work of both \cite{Bluck2014} and \cite{Lang2014} here. Nonetheless, the fact that there is very good qualitative agreement between these papers is highly encouraging. Moreover, in the next sub-section we explore the quenching of bulges and disks separately in CANDELS for the first time.

As with the SDSS analysis in Section 4, the CANDELS data are extremely stable to differential measurement uncertainty and so it is not feasible that the superiority of bulge mass over the other bulge - disk parameters could be an artefact of the accuracy with which these data are measured. Additionally, as mentioned above, we have tested an alternative definition of quenching based on sSFR values (from spectroscopy in the SDSS and SED fitting in CANDELS). The results are essentially identical to those shown here for the colour based selection. Hence, our conclusion, that bulge mass is the most predictive parameter of quenching at high redshifts, is extremely stable to the method used for identifying quenched systems.

\subsubsection{A visual test on the classification results}

As a final test, in Fig. 15 we repeat the visual assessment of the location of quenched galaxies in the bulge - disk plane from Section 4.2.1, applied here to the CANDELS data. We assess the full redshift range in CANDELS together because there is no significant variation in the dependence of quenching on the bulge - disk parameters from cosmic noon to the present epoch (see Fig. 14), and the much smaller data size in CANDELS relative to the SDSS results in the full range being a much clearer visual representation. The structure of Fig. 15 is identical to Fig. 5 (right panel), so we do not review it again here.

The CANDELS data in Fig. 15 appears more separated into distinct regions than the SDSS distribution (shown in Fig. 5). There are two reasons for this. The first is that the CANDELS data set is much smaller than the SDSS, by a factor of over twenty. Hence, rare objects are much more frequently observed in the SDSS than in CANDELS, filling in the gaps between the density peaks. Secondly, the CANDELS bulge - disk decompositions have a pre-classification via machine learning, such that some systems are deemed to be pure disks or spheroids in advance of photometric fitting. Galaxies without a bulge or disk component are assigned a nominal low value of the absent component (1/100th the total stellar mass) for display purposes. Alternatively, in the SDSS, the pure Sersic cuts are applied post fitting, and so we do not have to worry about displaying pure spheroids or disks (although we have tested both applying and not applying Sersic cuts in Section 4).


\begin{figure}
\includegraphics[width=0.49\textwidth]{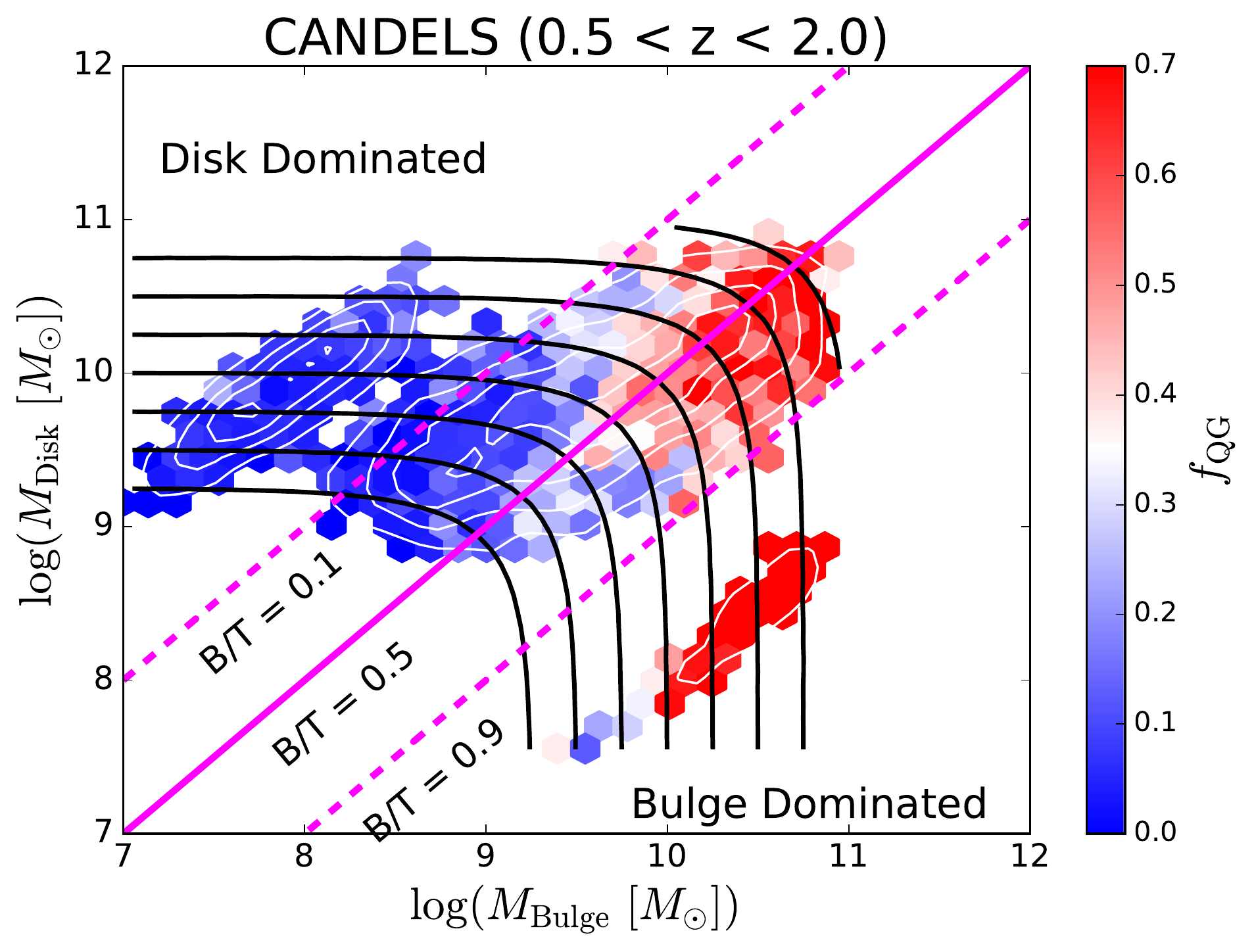}
\caption{The distribution of CANDELS galaxies in the $M_D$ - $M_B$ plane for the full redshift range ($0.5 < z < 2.0$). The location of galaxies is indicated by density contours (shown in white). Note that the tri-modality of the contour distribution is a result of us relocating pure spheroids and pure disks to a nominal position of $B/T$ = 0.99 and 0.01 respectively (for display purposes). Iso-mass and iso-morphology contours are shown on both panels as black and magenta lines, respectively. The $M_D$ - $M_B$ plane is colour coded by quenched fraction (as indicated by the colour bar labels). Clearly, quenching progresses primarily with bulge mass in these data, which confirms the primary RF result shown in Fig. 14. It is also instructive to compare this figure with Fig. 5 for the SDSS.}
\end{figure}

Viewing the entirety of Fig. 15, it is clear that quenching progresses primarily with bulge mass (not disk mass). Furthermore, it is also clear that at a fixed stellar mass (i.e. along an iso-mass line), galaxies may be either star forming or quenched dependent upon their morphology; whereas at a fixed morphology (i.e. along an iso-morphology line), galaxies may be either star forming or quenched dependent upon their mass. Thus, neither total stellar mass nor $B/T$ structure are capable of accurately constraining quenching alone at high redshifts. This result is identical to our low-$z$ findings in Fig. 5 for the SDSS (see Section 4). Moreover, this provides a simple visual confirmation of the Random Forest result in CANDELS from Fig. 14.

\begin{figure*}
\includegraphics[width=0.33\textwidth]{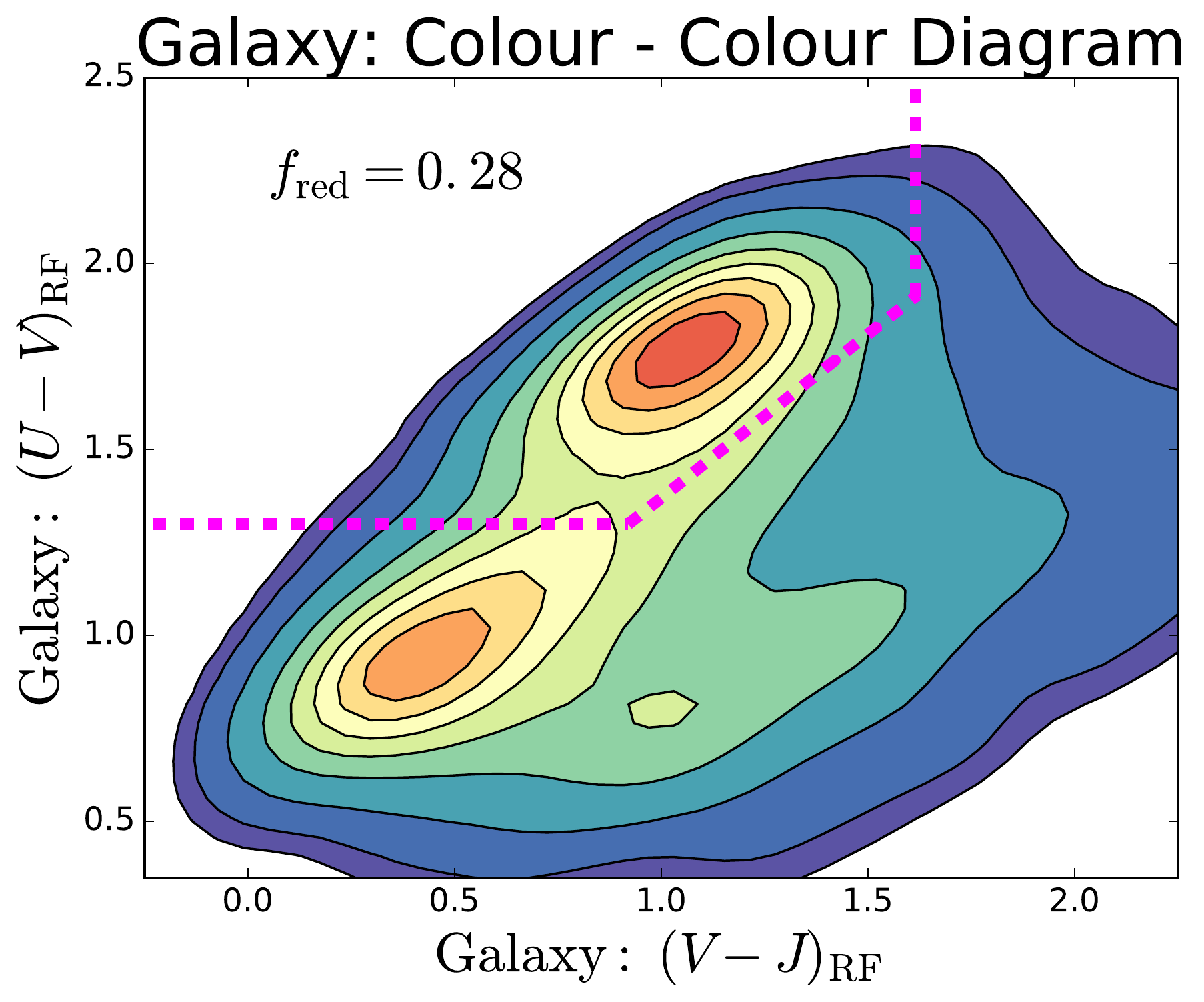}
\includegraphics[width=0.33\textwidth]{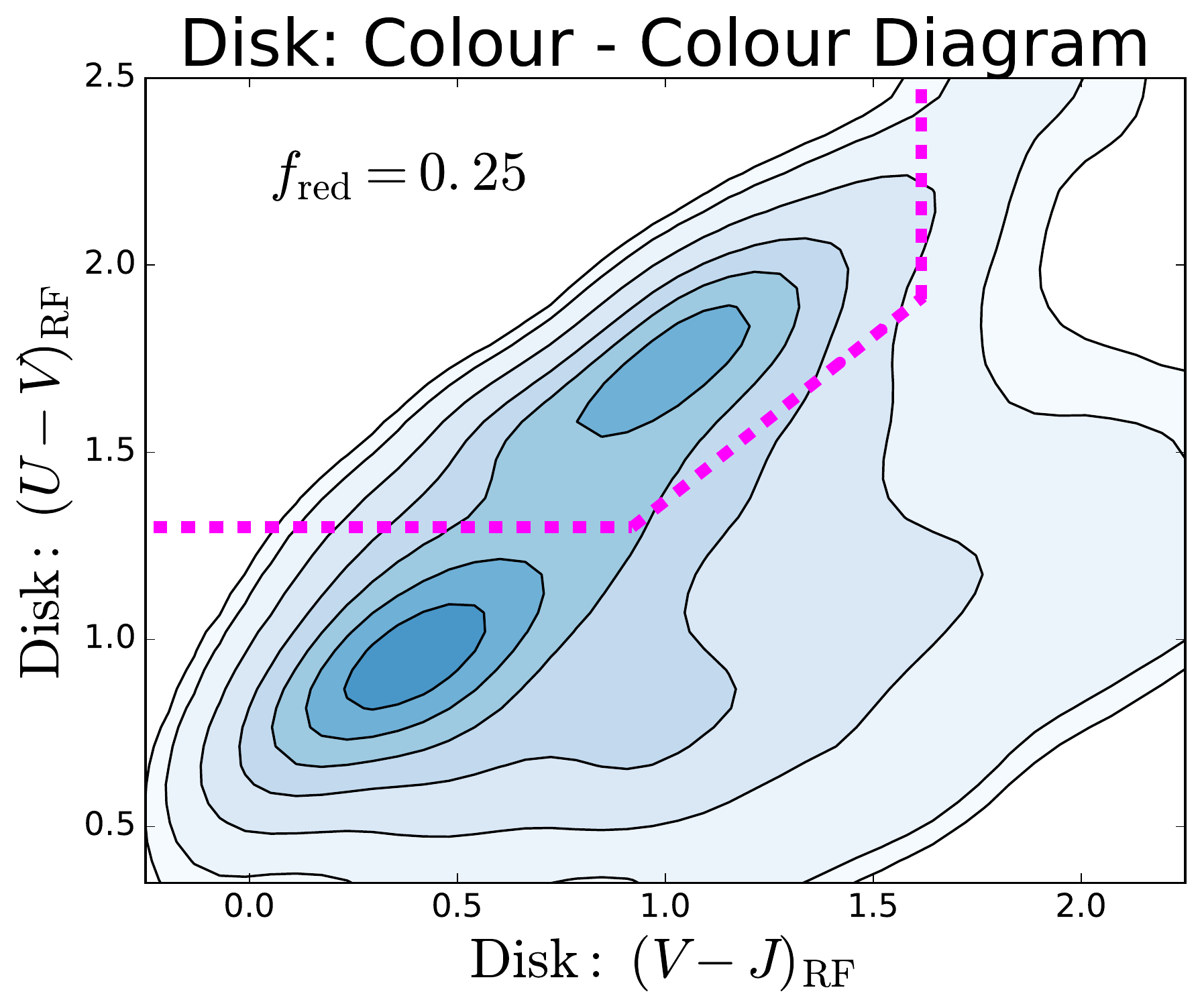}
\includegraphics[width=0.33\textwidth]{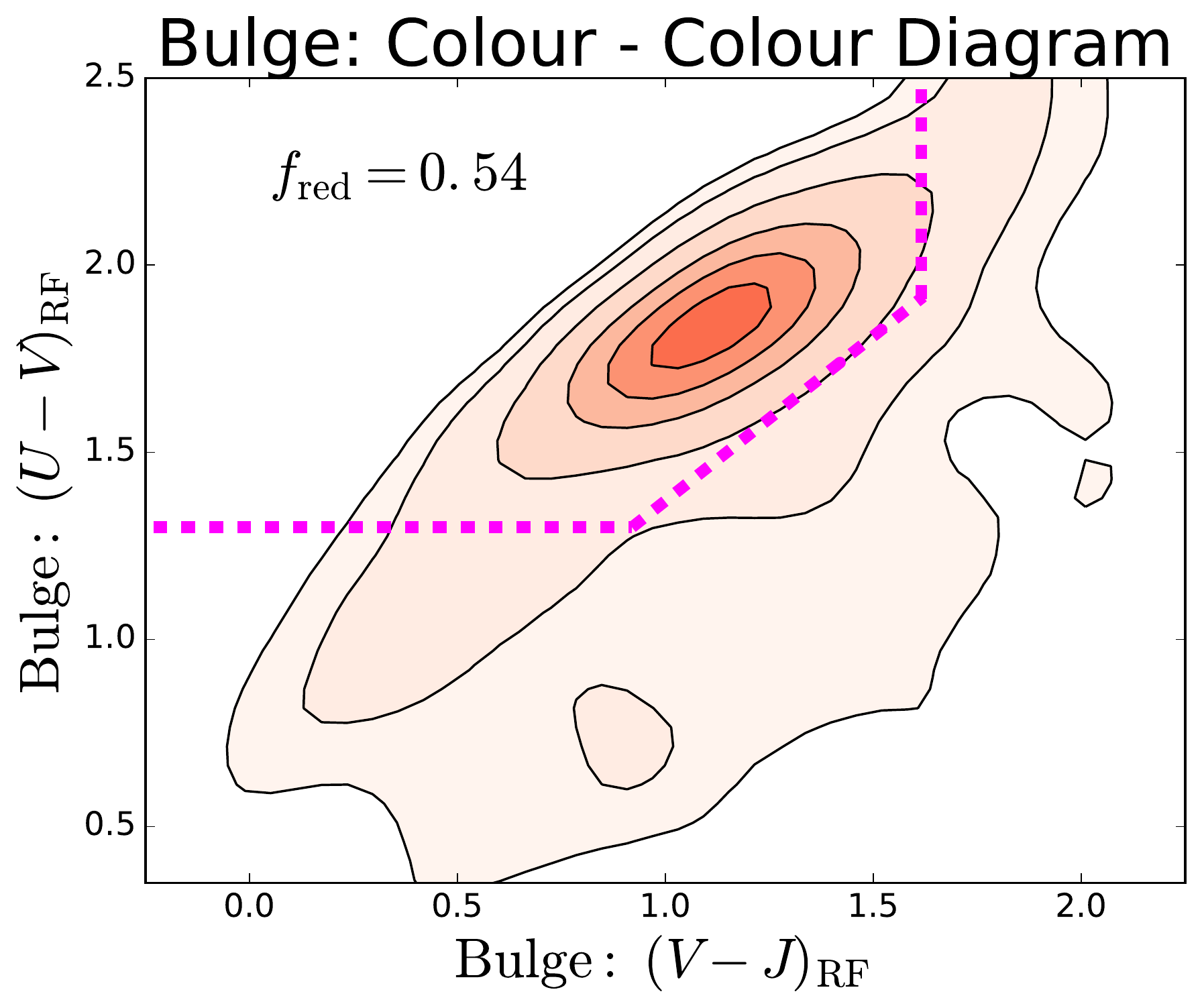}
\caption{Location of CANDELS galaxies (left panel), disks (middle panel), and bulges (right panel) in rest-frame (U-V) -- (V-J) colour space. The threshold for quenching from Williams et a. (2009) at z = 1.0 is shown as a dashed magenta line on each panel. Additionally, the fraction of red galaxies, disks and bulges is overlaid on the upper left of each panel.}
\end{figure*}

\begin{figure}
\includegraphics[width=0.49\textwidth]{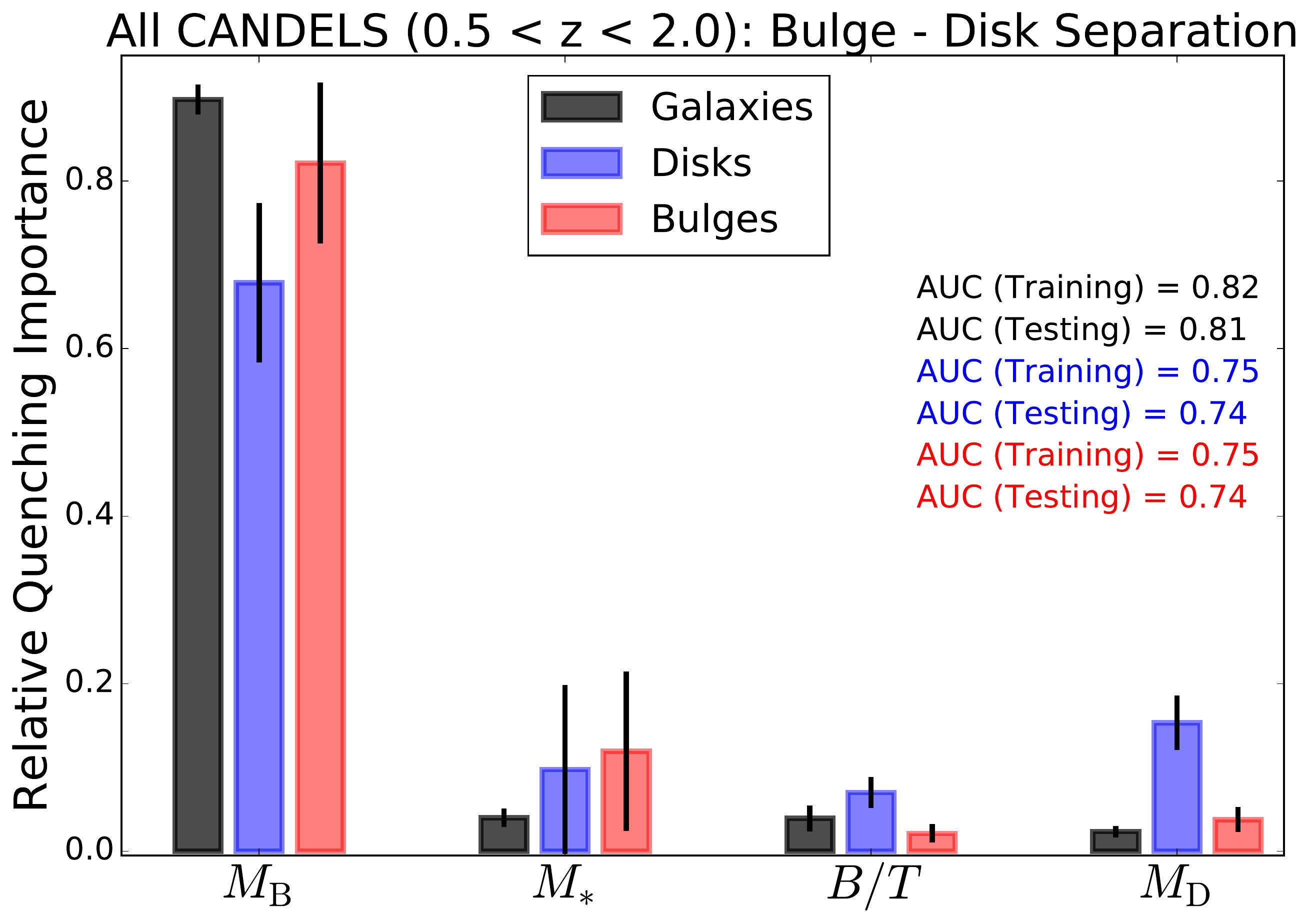}
\caption{Random Forest quenching classification for bulge and disk structures in CANDELS for the full redshift range ($0.5 < z < 2.0$), based on UVJ colours. Results for galaxies treated as a whole are shown in black; results for bulge structures and disk structures treated separately are shown in red and blue, respectively. For both bulge and disk structures, bulge mass is clearly found to be the most important parameter governing quenching at high redshifts. Note also that all other parameters are consistently found to be of very little predictive power, even disk mass in disk structures. Compare to Fig. 9 for the low redshift results, which show a very similar trend. }
\end{figure}


\subsection{Bulge \& disk quenching in CANDELS}

Utilising the bulge - disk decomposition catalogs of \cite{Dimauro2018} we have access to UVJ rest-frame colours not only for galaxies as a whole, but also for their bulge and disk structural sub-components. Hence, we can additionally explore the quenching of bulges and disks separately within the CANDELS data (exactly as with the SDSS bulge - disk decompositions, see Section 4.3). This is a completely novel analysis in these data, which gives us a unique opportunity to investigate quenching on sub-galactic scales at this epoch for the first time.

In Fig. 16 we present the location of galaxies (left panel), disks (middle panel) and classical bulges (right panel) in rest-frame UVJ colour - colour space. The quenching threshold is shown as dashed magenta lines on each panel, taken from Williams et al. (2009). Note that there is a weak redshift dependence on this threshold, which we incorporate into all analyses, but for display purposes we show only the cut at the median redshift of the CANDELS data.

On each panel of Fig. 16 we display the fraction of red objects, i.e. the fraction of objects which are identified to be quiescent by our adopted UVJ decision boundary (see Fig. 12). In the full CANDELS sample, 28\% of galaxies are identified to be red, with 72\% of galaxies identified to be blue. We find that the vast majority of disks are blue (75\%) but with a significant minority of quiescent systems (25\%). For bulges, there is a roughly even split between blue and red systems, unlike in the SDSS where the vast majority of bulges are red. Since there are significant numbers of both red disks and blue bulges in the CANDELS data set, we can meaningfully explore which parameters are most informative about the quenching of bulges and disks at high redshifts.

To this end, in Fig. 17 we present a Random Forest classification analysis of quenching in the full CANDELS redshift range for bulge and disk structures treated separately. This is the first time the quenching of bulges and disks has been considered separately in observational data at this epoch. We analyse the full redshift range because the number of red disks is so low that in conjunction with the much lower number of galaxies in CANDELS, relative to the SDSS, this particular analysis is not yet possible to perform in narrower redshift bins. Nevertheless, the remarkably high consistency between the redshift bins in CANDELS for the galaxy quenching analysis (see Fig. 14) encourages us that there is unlikely to be significant deviations within this redshift range for bulge and disk quenching as well. Additionally, we add to Fig. 17 the galaxy quenching analysis for the full CANDELS redshift range. No mass cuts are applied here since we are not seeking an explicit like-for-like comparison with the SDSS. As such, this also acts as a test on the stability of our results from CANDELS to sample selection choices (i.e. whether or not to impose stellar mass cuts).

For both bulge and disk structures (as well as for galaxies treated as a whole), bulge mass is consistently found to be by far the most predictive parameter of quenching at a very high confidence level. Thus, the quenching of all parts of high redshift galaxies are regulated by the central regions alone. All other bulge - disk parameters are of very little importance for quenching, once the bulge mass is known. This completely rules out stellar mass or morphology as causal drivers to the quenching of galaxies, bulges, or disks. Remarkably, this finding at high redshifts in CANDELS is essentially identical to the low redshift results in the SDSS and MaNGA (see Fig. 9, left panels). 

We conclude that the close dependence of bulge and disk quenching on the bulge component alone has been stable throughout the bulk of cosmic history. Furthermore, we also conclude that quenching acts globally within galaxies, quenching all parts of the galaxy in in concert, in a timescale which is short relative to the Hubble time. 

To our knowledge this is the first evidence for a globally acting quenching mechanism at high redshifts in the literature. In comparison to low redshift studies, it agrees closely with our results in Section 4 and with \cite{Bluck2020a, Bluck2020b}. We look forward to confirming these photometric findings in wide-field spectroscopic surveys targeting cosmic noon in the coming years (especially with VLT-MOONS, JWST IFU surveys, and observations with the proposed ELTs).


\section{Discussion - the quenching of galaxies, bulges, and disks since cosmic noon}

\subsection{Theoretical routes to quenching: AGN feedback vs. morphological stabilization}

As a result of the observational results from Sections 4 \& 5, we seek a theoretical mechanism (or set of mechanisms) which yield a very tight dependence of quenching on the central regions within galaxies. Furthermore, this mechanism must have been in place since at least cosmic noon, and have evolved little to the present epoch. Additionally, the quenching mechanism must be effective at ceasing star formation throughout entire galaxies, quenching first the bulge and then the disk. Given the use of centrals at low redshifts (with the explicit lack of environmental dependence found in Fig. 4), and the expectation for centrals to dominate the galaxy sample at high-$z$ as well, we are seeking a quenching mechanism independent of environment, i.e. an intrinsic quenching mechanism. Hence, we may ignore the host of environmental quenching mechanisms from ram pressure stripping to galaxy - galaxy harassment (e.g. \citealt{Bosch2008, Wetzel2013, Woo2013, Bluck2016, Bluck2019}) in the following discussion.

Before we consider in detail two highly probable solutions, we first rule out a number of less likely alternatives. First, it has been suggested that identifying a close dependence of quenching on the central regions of galaxies could be explained a causally (see \citealt{Lilly2016}). The argument appeals to the size evolution witnessed in the galaxy population from $z \sim 2$ to the local Universe (e.g. \citealt{Trujillo2007, Buitrago2008, Bluck2012, Newman2012}), whereupon galaxies are typically smaller for their mass and morphological type at earlier epochs. This relates to quenching because once a galaxy ceases to form stars the stellar structure is essentially stable, modulo late-time mergers. Hence, galaxies which quenched early in the history of the Universe will have smaller sizes for their mass and consequently denser cores, potentially consistent with higher bulge masses and central velocity dispersions. 

However, at low-$z$ in \cite{Bluck2016} we established that galaxies in the green valley also appear to have higher bulge masses (and central velocity dispersions) than star forming systems. Yet, in the green valley, systems are quenching contemporaneously, and hence progenitor bias arguments of this sort do not apply. Moreover, in this work we have found that a close dependence of quenching on bulge mass persists up to at least $z \sim 2$. Consequently, the progenitor bias argument is much less likely to be valid considering the little time allowed for further evolution in the size - mass relation before the Big Bang. Furthermore, no constraints on size evolution (or lack thereof) beyond $z\sim2$ are currently available. Finally, we note again (as in \citealt{Bluck2020a}) that this scheme offers no explanation for quenching, and thus one is still compelled to speculate as to the actual physical mechanism(s) responsible. As such, we do not consider this possibility any further here.

Supernova feedback can have a profound influence on star formation in low mass systems (e.g. \citealt{Cole2000, Guo2011, Henriques2019}), blowing out gas from the galaxy and inputting heat into the surrounding gaseous halo. However, at high masses, supernovae are not energetic enough to expel gas from the system, and generate substantially less energy than needed to stabilise the hot gas halo from cooling and collapse (e.g. \cite{Bower2006, Bower2008, Henriques2015, Henriques2019}). In this work, we have established that bulge mass is considerably more predictive of quenching than total stellar mass. Given that the total integrated energy released by supernovae must be proportional to the total stellar mass of the system (see \citealt{Bluck2020a} for an analytical derivation), this observational result clearly disfavours supernova feedback as a significant intrinsic quenching mechanism. 

Alternatively, gas may shock heat to the virial temperature upon infall into a massive dark matter halo (e.g. \citealt{Dekel2006, Dekel2009, Dekel2019}). Clearly, quenching via this mechanism will led to a tight dependence on halo mass (as explored in \citealt{Woo2013, Woo2015}). However, in this work (and in \citealt{Bluck2016, Bluck2020a, Bluck2020b}) we establish that halo mass is much less predictive of quenching than bulge mass or central velocity dispersion at low-$z$ (see Fig. 4). Thus, virial shock heating cannot be the dominant quenching mechanism at low redshifts. At high-$z$ we do not have access to reliable estimates of halo masses and hence cannot make a concrete statement with respect to the role of virial shock heating at these epochs. Nonetheless, the stable dependence of quenching on bulge mass across cosmic time suggests that the mechanism for quenching must also have been relatively stable, and so we predict that halo mass (once available) will not significantly impact the parameterization of quenching in the early Universe.

Now we come to one of two highly viable solutions to the quenching problem. There exists a tight relationship between bulge mass and black hole mass (e.g. \citealt{Magorrian1998, Haring2004}), and an even tighter relation between central stellar velocity dispersion, $\sigma_\star$, and black hole mass (e.g. \citealt{Ferrarese2000, McConnell2011, McConnell2013, Saglia2016, Piotrowska2021}). As such, our observational finding of a very close dependence of quenching on bulge mass throughout cosmic time (and an even tighter dependence on $\sigma_\star$, when available) implies a close connection between quenching and the central supermassive black hole. Indeed, we may re-conceptualise our results as implying that galaxies quench when they host high mass black holes. Certainly, this statement is completely consistent with the findings of this paper. Using this interpretation, the results from Sections 4 \& 5 are in complete accord with the direct predictions from the LGalaxies quenching model (see Section 3.3, especially Fig. 3). It is also important to highlight that these results are highly consistent with predictions from cosmological hydrodynamical simulations as well (in Eagle, Illustris and Illustris-TNG; see \citealt{Piotrowska2021}).

An alternative to AGN feedback, which has garnered increased attention recently, is kinematic stabilisation (e.g. \citealt{Martig2009, Gensior2020}). This process may also be reasonably successful at explaining the tight dependence of quenching on bulge mass at all epochs, and $\sigma_\star$ at low-$z$. There are two versions of morphological stabilisation commonly considered in the literature: morphological `$Q$'-quenching as a result of increased gas dispersion and turbulence in bulge structures (e.g. \citealt{Gensior2020}); and morphological stabilisation of the gas disk from tidal torques induced by a central mass concentration, i.e. the bulge (e.g. \citealt{Martig2009}). Both of these approaches to kinematic stabilisation predict that galaxies with higher mass bulges (and higher $\sigma_\star$) will be more frequently quenched. Moreover, they also offer a natural explanation for the centres of galaxies reaching quiescence before their outskirts, due to kinematic effects scaling primarily with the local potential (see e.g. \citealt{Ellison2018, Bluck2020b}). This much is in complete accord with the observational findings of this paper.

The primary observational difference between preventative AGN feedback and kinematic stabilisation (in any mode) lies in the gas content of galaxies. Preventative `radio-mode' AGN feedback operates by stabilising the hot gas halo against cooling and collapse. The stabilised hot gas halo then additionally acts as a shield against cosmic inflows (e.g. \citealt{Vogelsberger2014a, Schaye2015}). Thus, galaxies quench as a result of a lack of gas, required as fuel for star formation. Consequently, a key prediction of radio-mode feedback is that the gas content of quiescent galaxies will be substantially lower than the gas content of actively star forming systems. On the other hand, kinematic stabilisation has no direct impact on the gas content of galaxies (neither preventing inflow nor expelling gas from the system). As such, the fundamental prediction from kinematic stabilisation is that the gas content of quiescent systems will be higher than star forming systems. This follows because gas may continue to be accreted into the system (from hot halo cooling and cosmic cold gas streams) and yet star formation is prevented so gas is not depleted. Conservation of baryon number thus requires the gas fraction to rise (by up to an order of magnitude or more). 

Therefore, to distinguish between kinematic and AGN quenching, one must look at the gas content of galaxies. It is now very well established that both the molecular and HI gas content of quiescent systems is lower than star forming systems by at least one order of magnitude (e.g. \citealt{Saintonge2016, Saintonge2017, Piotrowska2020, Piotrowska2021, Ellison2020, Ellison2021}). Hence, kinematic stabilisation alone cannot explain the existence, and stability, of quiescent systems in the Universe. Indeed, there is at least two orders of magnitude of discrepancy in the natural prediction for the gas fraction in quiescent systems between kinematic-only quenching and observations. Furthermore, this discrepancy is likely to be an underestimate since most studies of gas content as a function of quenching focus on the `green valley' rather than fully quenched systems. Looking specifically at red ellipticals, the molecular and HI gas fractions are typically found to be lower still, with many non-detections (e.g. \citealt{Saintonge2017, Dou2021}) and frequently no evidence of emission lines or dust extinction at all (e.g. \citealt{Piotrowska2020, Bluck2020a}). 

As a result of the clear evidence for substantial reduction in gas content during quenching, we conclude that kinematic stabilisation cannot be the sole (or dominant) intrinsic quenching mechanism. Some other process must be responsible for reducing the gas content, which is only exacerbated by the natural expectation for the gas content to rise in the absence of star formation. Put simply, some process must prevent accretion of gas into the system in order to be consistent with observations. Indeed, this is a very well known theoretical problem (see \citealt{Somerville2015} for a review). Preventative AGN feedback offers a natural explanation for this problem. Additionally, this scenario fits closely with one of our main observational results, i.e. that the central regions within galaxies regulate quenching throughout the entire system. 

However, the above arguments do not completely rule out secondary effects from kinematic stabilisation of gas collapse in quenching galaxies. Indeed, recent studies agree that both star formation efficiency (${\rm SFE} \equiv {\rm SFR}/M_g = 1/\tau_{\rm dep}$) and gas fraction ($f_g \equiv M_g/M_*$) decrease during quenching (see \citealt{Saintonge2016, Saintonge2017, Ellison2020, Brownson2020, Piotrowska2020, Piotrowska2021}). Kinematic stabilisation cannot explain the reduction of $f_g$; whereas preventative AGN feedback cannot explain the reduction in SFE (although some other variants of AGN feedback may be able to explain this, e.g. `kinetic mode' in Illustris-TNG, see \citealt{Weinberger2018, Zinger2020, Piotrowska2021}). 

As a result of the above discussion, there exists a possibility for a synergy between preventative AGN feedback and kinematic stabilisation, in order to yield fully consistent results with observations. This notwithstanding, kinematic stabilisation and AGN feedback are not on an entirely equal footing. The vast majority of the work done to quench systems is achieved by the mechanism which shuts down gas accretion from the circum-galactic medium (CGM); whereas the response function of the internal gas dynamics in the inter-stellar medium (ISM) is a much lower energetic process (and impacts a much lower mass of gas). Furthermore, in the absence of any morphological stabilisation, galaxies will still quench once their gas reservoirs are fully depleted; yet in the absence of AGN feedback, galaxies will never quench since their gas fractions will continue to rise to orders of magnitude higher levels, whereupon no viable stabilisation process can prevent collapse into new stars. 

As a result of the above argument, we conclude that AGN feedback in the preventative mode is the most probable intrinsic quenching mechanism, given our observational results and considering a wide variety of other results from the literature. This is precisely the mode which the LGalaxies model uses, and the predictions from it are met exceptionally well by the multi-epoch observations analysed in this paper.

\subsection{SDSS comparison with LGalaxies:\\ black hole mass vs. the bulge - disk parameters}


\begin{figure*}
\begin{centering}
\includegraphics[width=0.85\textwidth]{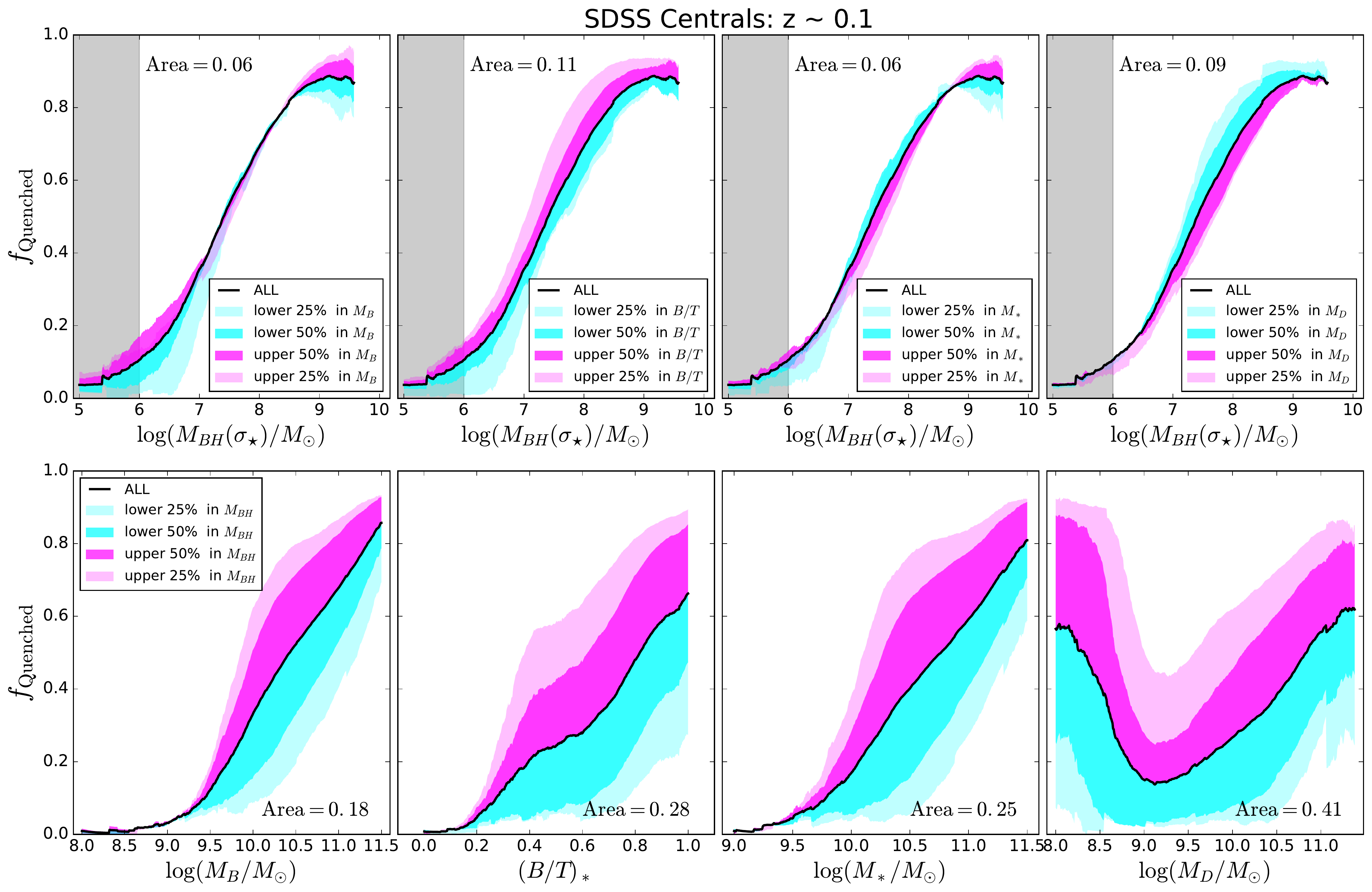}
\includegraphics[width=0.85\textwidth]{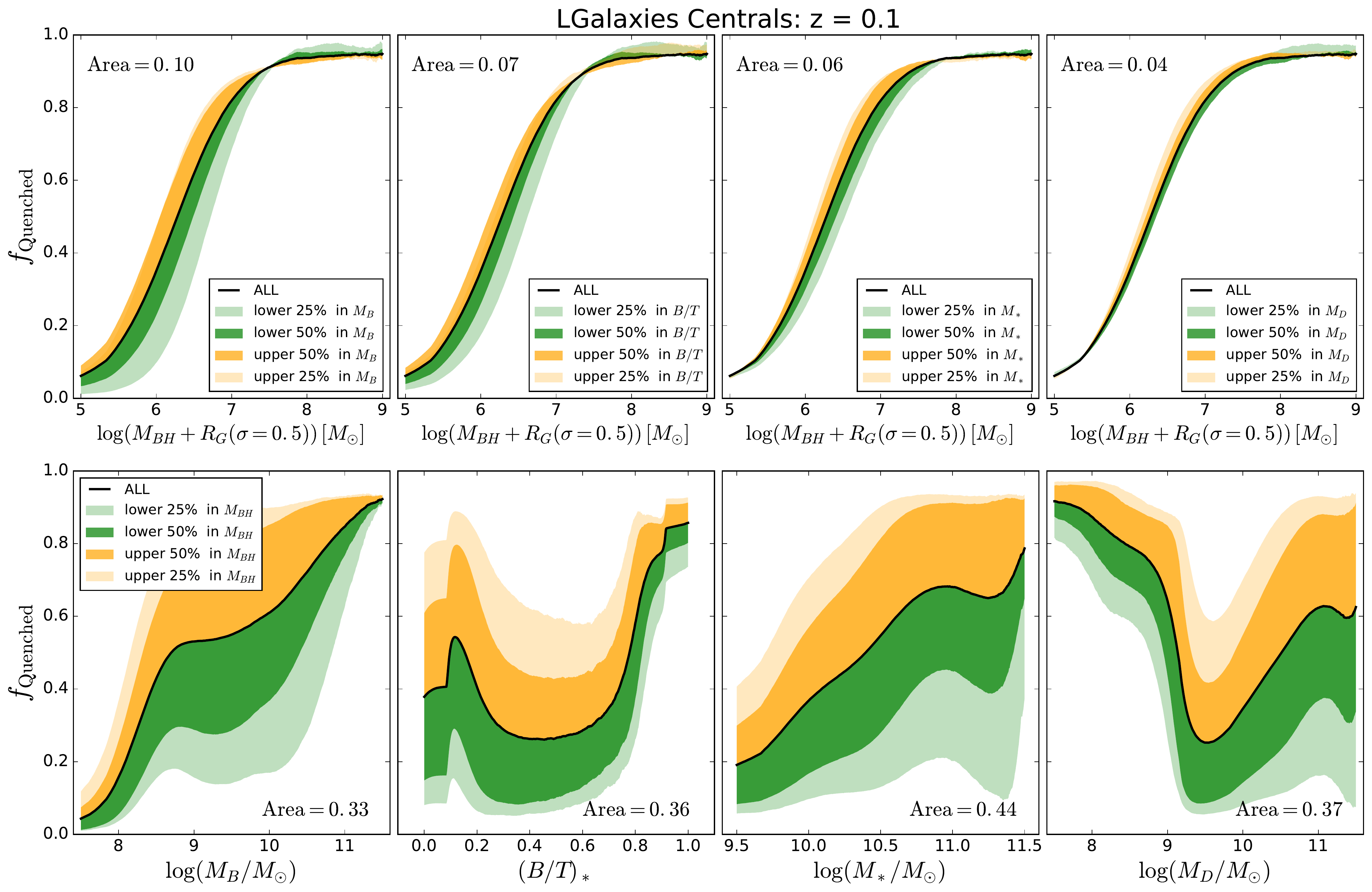}
\caption{Quenched fraction relationship with black hole mass, and each of the bulge - disk parameters. Results are shown for observations (top panels; SDSS at z $\sim$ 0.1) and for a semi-analytic model (bottom panels; LGalaxies at z = 0.1). The top row of each panel indicates the quenched fraction relationship with black hole mass, with the second row of each panel indicating the quenched fraction relationship with each of the bulge - disk parameters in turn. The quenched fraction relationships are subdivided into percentile ranges of a third variable (as indicated by the legends). We quantify the tightness of the relations by the area subtended by the upper and lower 50th percentiles in each third parameter (which is displayed on each panel). The typical uncertainties on the area statistic are $\leq$0.01 for the SDSS and $\leq$0.005 for LGalaxies. For the observational data, black hole mass is estimated through the $M_{BH} - \sigma_\star$ relation, and a volume correction is applied. For the model, black hole masses have noise added to approximate the calibration uncertainty in the observational estimate. For both observations and the model, quenching is defined consistently as $\Delta {\rm sSFR}(z)$ $<$ 1dex. Clearly, in both the SDSS and LGalaxies, the $f_Q - M_{BH}$ relationship is considerably tighter than for any other parameterisation.}
\end{centering}
\end{figure*}

The striking reduction in importance of bulge mass in Fig. 4 (right panel), once central velocity dispersion is made available to the classifier, is uncannily reminiscent of the result shown for LGalaxies in Fig. 3 (left panel) when black hole mass is made available to the classifier. Indeed, $\sigma_\star$ is often utilised as a proxy for black hole mass in large surveys (e.g. \citealt{Bluck2016, Bluck2020a, Bluck2020b}), due to the very tight empirical relationship found between it and dynamically measured black hole mass in the local Universe (see e.g  \citealt{Ferrarese2000, McConnell2011, McConnell2013, Saglia2016, Piotrowska2021}). 

To explore the plausible link between quenching and supermassive black holes further, we estimate black hole masses in the SDSS by (\citealt{Saglia2016}):

\begin{equation}
\log(M_{BH} \, [M_{\odot}]) = 5.25 \times \log(\sigma_\star \, [{\rm km/s}]) - 3.77,
\end{equation}

\noindent This calibration yields a scatter of just 0.46 dex against 96 dynamically measured black hole masses (\citealt{Saglia2016}), and is shown to have only weak dependence on morphology or structural class of host galaxy (see also \citealt{Bluck2016, Piotrowska2021}). We have tested utilising separate scaling relations for early and late morphological types, and pseudo and classical bulges, and the results are almost identical to what we present below for the unified calibration in eq. 5.

In Fig. 18 we present the quenched fraction relationship with black hole mass for the SDSS (top panels) and the LGalaxies z = 0.1 snapshot (bottom panels). In the SDSS, black hole mass is inferred from eq. 5. Additionally, we weight the observed quenched fraction statistic by $1/V_{\rm max}$ to approximate a volume complete sample, appropriate for comparison to LGalaxies. Following the rationale of Section 4.1, we also restrict disk-dominated galaxies to presenting face-on in order to reduce contamination from rotation into the plane of the sky. We statistically correct for this cut by further weighting by the inverse of the completeness (as in \citealt{Bluck2016}). We have checked that this approach yields consistent mass functions to \citealt{Thanjavur2016} for the full data set. Additionally in Fig. 18 top-panel, we present the quenched fraction relationship for each of the photometric bulge - disk parameters, for the new weighted sample. Each of the quenched fraction relationships is split into percentile ranges of a third variable, as indicated by the legends; and the area subtended between upper and lower 50th percentiles is shown on each panel.

For LGalaxies in Fig. 18, we incorporate Gaussian random noise into the black hole mass values in the following manner:

\begin{equation}
M_{BH} \hspace{0.2cm} \rightarrow \hspace{0.2cm} M_{BH} + R_{G}(\mu=0; \sigma=0.5)
\end{equation}

\noindent i.e. we take a random draw from a Gaussian distribution centred on zero with a standard deviation of 0.5\,dex, chosen to mimic the total uncertainty on the indirect estimates in the SDSS (0.46 dex scatter and 0.1\,dex intrinsic measurement uncertainty, added in quadrature). As a result, the observations and model are placed on precisely the same footing: both are volume complete, and both have similar average uncertainties on the black hole measurements. We also use an identical definition of quenched of $\Delta {\rm sSFR}$ $<$ 1\,dex for both; and set the hyper-parameters in the area fraction plots of Fig. 18 (i.e. bin size, smoothing factor and integration limits) to identical values. This ensures a very high level of consistency between the model and the observational data, enabling robust comparison.


\begin{figure*}
\includegraphics[width=0.49\textwidth]{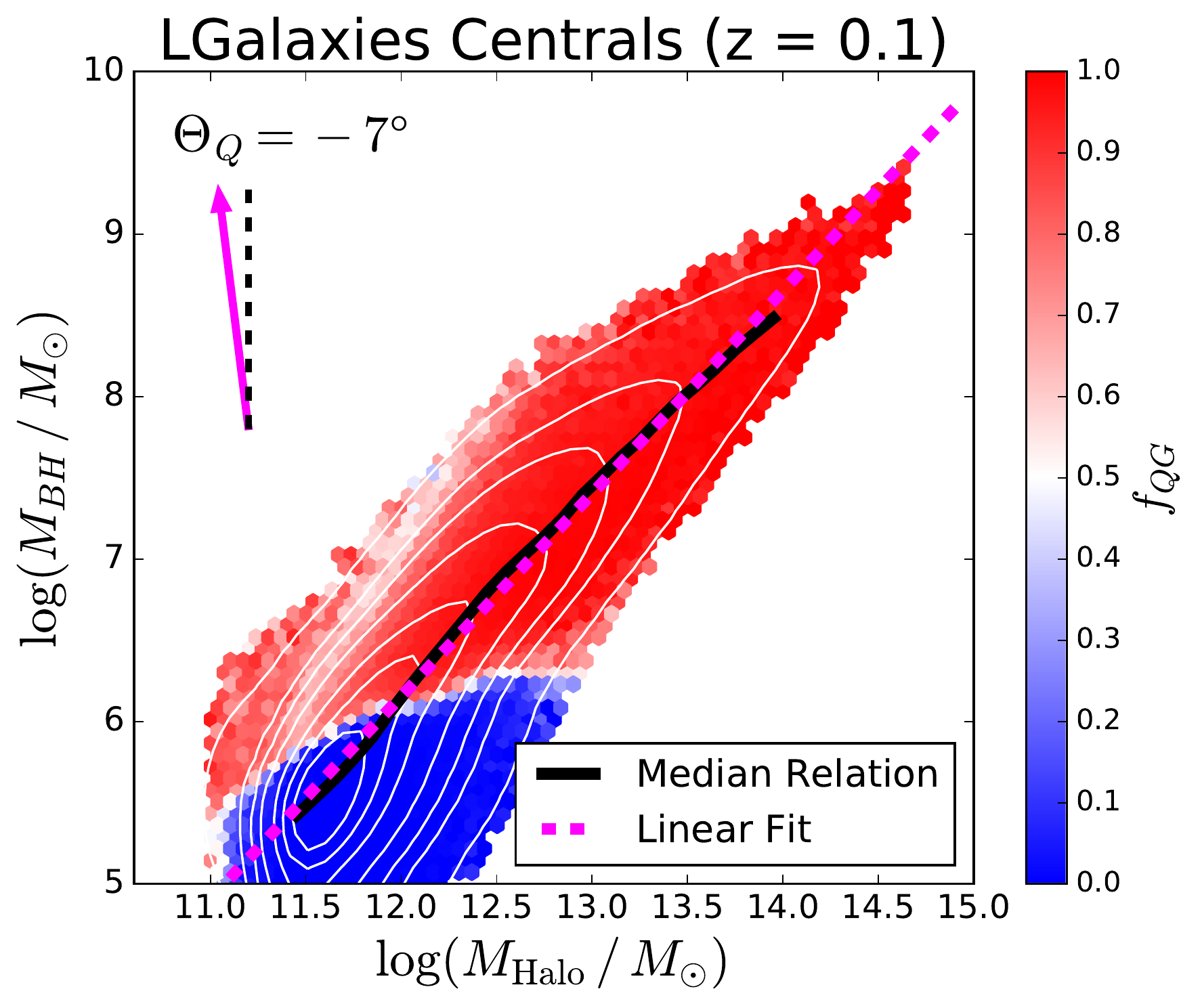}
\includegraphics[width=0.49\textwidth]{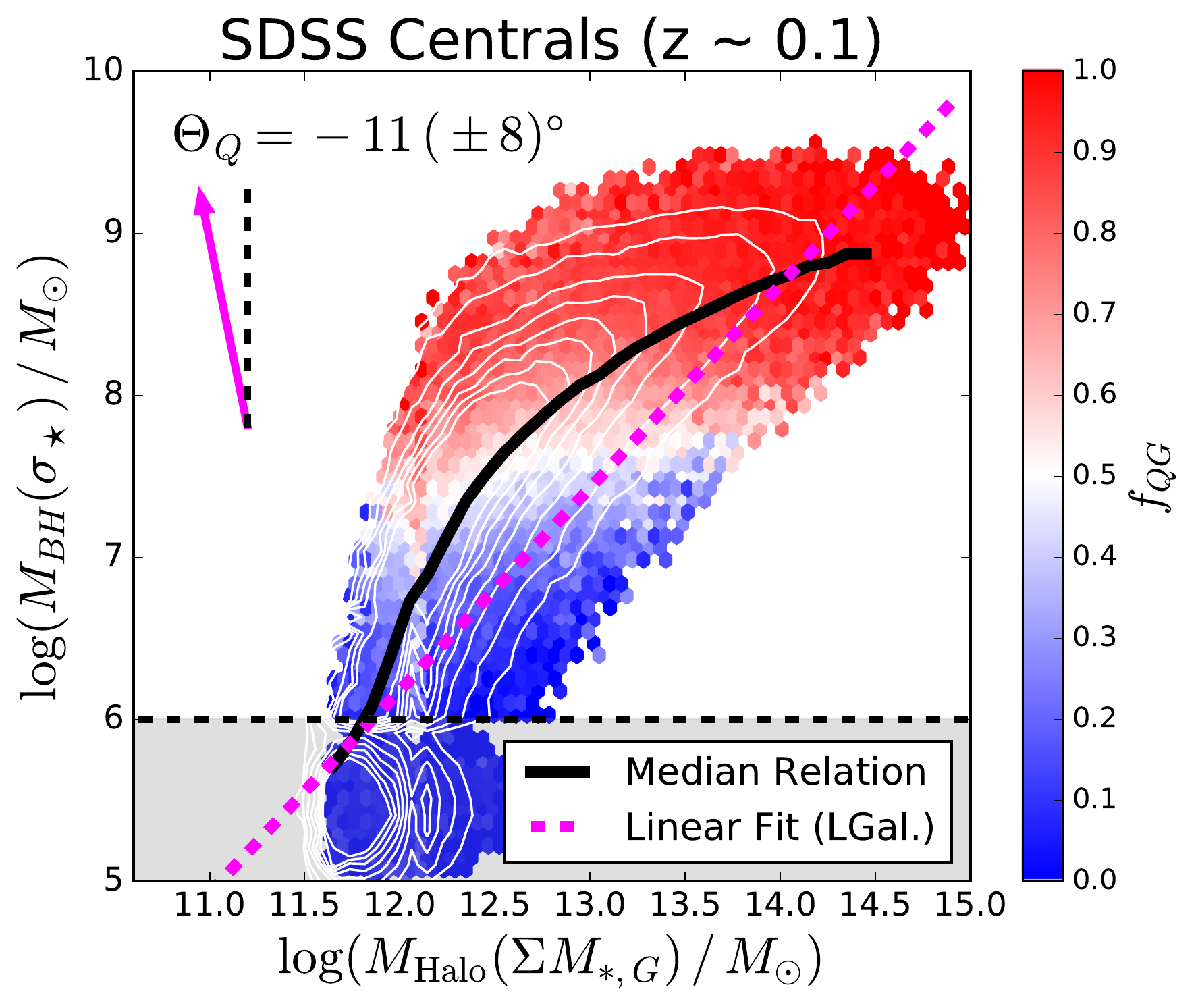}
\caption{Black hole mass -- halo mass relationship in LGalaxies (left panel) and in the SDSS (right panel) for central galaxies at z $\sim$ 0.1. On both panels, the $M_{BH} - M_{\rm Halo}$  plane is subdivided into small hexagonal regions, each colour coded by the quenched fraction of galaxies within that region of parameter space (as labelled by the colour bar). The optimal direction to travel through the plane in order to maximise quenching is indicated by a magenta arrow, with an orientation quantified by the quenching angle ($\Theta_Q$). For the observational data, uncertainties on the quenching angle are estimated from bootstrapped random sampling within the measurement uncertainties. On both panels, density contours are shown in white (which are $1/V_{\rm max}$ volume weighted for the SDSS), and the median relations are shown by solid black lines. On the right panel, the region in which SDSS velocity dispersions become unreliable (used here to estimate $M_{BH}$) is shaded out. On both panels we display a linear least squares fit to the LGalaxies $M_{BH} - M_{\rm Halo}$ relation, utilised in the right panel to aid in comparison. Clearly, quenching proceeds primarily with black hole mass in both the LGalaxies model and in the SDSS observations, with a subtle secondary anti-correlation with halo mass.}
\end{figure*}

Viewing the entirety of Fig. 18, there is an astonishing level of qualitative agreement between the SDSS observational results and the LGalaxies model predictions at z = 0.1. The $f_Q - M_{BH}$ relation is considerably tighter than for any other parameterization, for both observations and simulations. This general result was seen before just for the comparison between $M_{BH}$ and $M_{*}$ in Bluck et al. (2016). Additionally the result for this parameter pairing is completely consistent with an analysis of a much smaller sample of dynamical black hole masses in \cite{Terrazas2016, Terrazas2017}. Quantitatively, the area statistics demonstrate that parameterizing quenching with black hole mass in the SDSS leads to tighter relations by a factor of 2.5 - 5 (depending on the bulge - disk variable). In LGalaxies, black hole mass (reduced in accuracy by 0.5 dex) leads to tighter relations by a factor of 3 - 7 (depending on the bulge - disk variable). Hence, even at a quantitative level, the improvement in parameterizing quenching with black hole mass over any of the bulge - disk parameters is reasonably similar.

Most of the relationships probed in Fig. 18 are qualitatively similar between the LGalaxies model and the SDSS observations, with a general caveat that the impact of varying $M_{BH}$ at fixed bulge - disk parameters tends to be larger in the model than in the data (even after appropriately scrambling the black hole masses). Nonetheless, the general trends are clearly very similar at a qualitative level. The one exception to this is with $B/T$ structure, where the $f_Q - B/T$ relation in LGalaxies is non-monotonic as opposed to the monotonic relation in the SDSS. The morphologies of galaxies in the model are known to be too disk-dominated (compared to observations), and hence we suspect that this unusual feature originates in this problem (see \citealt{Bluck2019} for a detailed discussion). Additionally, at a quantitative level, we note that the black hole mass threshold of quenching ($M_{BH,Q}$) is significantly lower in LGalaxies than in the SDSS, with $M_{BH,Q}$(LGal, z=0.1) = $10^{6.3\pm0.2} M_{\odot}$ and $M_{BH,Q}$(SDSS, z=0.1) = $10^{7.2\pm0.2} M_{\odot}$. A similar offset was noted before in \cite{Bluck2016} as well. Therefore, there is very good qualitative agreement between the model and observations, but there are nonetheless significant quantitative differences. 

In conclusion to this sub-section, we find that the LGalaxies semi-analytic model successfully predicts many of the key observational results in this paper. The model achieves this through the use of radio-mode AGN feedback, which leads to global quenching of galaxies via starvation of gas supply. More specifically, in lieu of black hole mass, bulge mass is found to be the key parameter driving quenching in the model across cosmic time, exactly as in observations. Moreover, using $\sigma_\star$ as a proxy for black hole mass in the SDSS, we find a remarkable agreement between the tightness of the quenched fraction relationship with black hole mass in both observations and simulations. As such, we conclude that our observational results are consistent with AGN feedback in the radio mode, at least as implemented in LGalaxies. Furthermore, given the arguments presented in Section 6.1, we conclude that AGN feedback is the most probable explanation for galaxy quenching, across the vast majority of the age of the Universe.

\subsection{Quenching in the causal parameter space:\\ $M_{BH}$ driven heating vs. $M_{\rm Halo}$ driven cooling}

According to the LGalaxies semi-analytic model, the intrinsic quenching of central galaxies is determined solely by the mass of the supermassive black hole, the mass of the group or cluster dark matter halo in which the galaxy resides, and the epoch at which the galaxy is situated (see Appendix A for an analytic derivation). The dependence on redshift may be easily removed by selecting galaxies at a single snapshot in cosmic history. At a fixed epoch, black hole mass and halo mass remain as the only two causal parameters in the model. Given the great success of the model in the a causal bulge - disk parameter space in comparison to multi-epoch observations, it is highly instructive to attempt to explore quenching in the causal black hole mass - halo mass plane.

To achieve this comparison, we restrict to the SDSS sample where there is wide spectroscopic coverage. We utilise dark matter halo masses inferred through abundance matching from \cite{Yang2007, Yang2009}, as in Section 4. Additionally, in order to explore the plausible link between quenching and supermassive black holes, we estimate black hole masses in the SDSS utilising eq. 5 (see the previous sub-section).

In the left panel of Fig. 19 we show the $M_{BH} - M_{\rm Halo}$ relation in LGalaxies for the z = 0.1 snapshot, restricting to central galaxies. Additionally, we subdivide the plane into small hexagonal regions each colour coded by the quenched fraction within that region of parameter space (as indicated by the colour bar). Clearly, quenching progresses predominantly as a function of black hole mass, rather than halo mass, in the model (as found in Fig. 3). However, the threshold black hole mass for quenching systematically rises with halo mass (exactly as explained analytically in Appendix A). We quantify the optimal route through the $M_{BH} - M_{\rm Halo}$ plane in order to maximise quenching utilising the ratio of partial correlations (see \citealt{Bluck2020a} for a detailed explanation). We find the quenching angle in LGalaxies to be $\Theta_Q = -7^{\circ}$, i.e. very close to vertical (indicating pure $M_{BH}$ dependence), with a slight negative correlation with halo mass at a fixed black hole mass. Conceptually, this is explained by higher mass haloes experiencing higher cooling rates and hence requiring more energy input from AGN feedback to quench.

In the right panel of Fig. 19 we show the same plot for central galaxies observed in the SDSS at $z \sim 0.1$. Clearly, quenching proceeds primarily as a function of black hole mass, rather than halo mass, exactly as predicted in LGalaxies (see also \citealt{Bluck2016, Bluck2020a, Bluck2020b} for similar results). Again, we quantify the optimal route through the $M_{BH} - M_{\rm Halo}$ plane in order to maximise quenching. For the SDSS, $\Theta_Q = -11\pm8^{\circ}$, where the errors are inferred via bootstrapped random sampling within the measurement uncertainties. Note that this angle is very slightly different to \cite{Bluck2020a}, but is qualitatively very similar. The reason for the small deviation is a change in sample selection in the current work (to ensure a fair comparison with LGalaxies). 

There is excellent agreement between the quenching angles for LGalaxies and the SDSS (compare the arrows in both panels on Fig. 19). In both the model and the observational data, quenching is clearly predominantly a function of black hole mass, with a subtle secondary dependence on halo mass. The secondary dependence is such that increasing halo mass actually decreases the probability of galaxies being quenched slightly, at a fixed black hole mass. This effect is naturally explained in the radio-mode heating paradigm (see Appendix A). However, it is worth comparing to another paradigm of quenching here: virial shock heating (e.g. \citealt{Dekel2006, Dekel2014, Woo2013, Woo2015, Dekel2019}). 

In the virial shock heating paradigm, quenching is predicted to scale fundamentally with halo mass (see e.g. \citealt{Woo2013, Bluck2020a}), and hence have essentially no dependence on black hole mass at fixed halo mass. Therefore, as a result of our observational results, we can rule out halo mass quenching as the primary driver of quenching in the local Universe (see also \citealt{Bluck2016, Bluck2020a} for similar conclusions). This notwithstanding, it is important to stress that in order for AGN heating to be effective, a hot statistic atmosphere must have developed, and it is assumed in the LGalaxies model that this is achieved through virial shocks. Thus, the point is not that virial shocks do not matter for quenching, but rather that they cannot by themselves account for the lack of cooling in massive haloes (which is the fundamental prerequisite for quenching).

Despite the excellent agreement between LGalaxies and the SDSS in terms of the importance of black hole mass and halo mass for quenching in local galaxies (as illustrated by the arrows on each panel), there are nonetheless very clear differences between the two panels of Fig. 19. Most strikingly, the threshold black hole mass for quenching is approximately an order of magnitude higher in the observations compared to the model (as also seen in Fig. 18). This suggests that the coupling strength of accretion to black hole mass (i.e. $k_{\rm AGN}$ in eq. A3) must be reduced in the model. 

However, since the model is already tuned to get the $f_Q - M_*$ relation approximately correct (as in \citealt{Peng2010, Peng2012}), this implies that the $M_{BH} - M_{*}$ relation must also change. One final constraint comes from the fact that the model is also tuned to get the $M_{BH} - M_{B}$ relation approximately correct (as in \citealt{Haring2004}). Consequently, the only remaining scaling law to tweak is that of morphology - mass ($B/T - M_*$). Happily, the required change in the morphology - stellar mass relation is consistent with the observation in \citealt{Bluck2019} that galaxies are overly disk dominated in LGalaxies compared to the SDSS at low redshifts. As such, increasing the black hole mass of quenching is likely to be achievable in a self-consistent manner within the model. That is, the mass of black holes within galaxies will increase, but the efficiency of energy output per unit black hole mass will decrease to compensate. 

The need for a change of this sort is also clearly evident in Fig. 19 by considering the $M_{BH} - M_{\rm Halo}$ relation directly. Black hole mass increases much more steeply with halo mass in the SDSS than in LGalaxies, followed by a significant levelling off at high halo masses. This can be seen most clearly by comparing the median relation in the SDSS to the linear fit in LGalaxies (on the right panel of Fig. 19). Improving the model by more accurately reproducing this important scaling relation, under the additional constraint of preserving the $f_Q - M_*$ relation, would naturally increase the threshold black hole mass of quenching in the model (precisely as needed).


\section{Summary}

We present an analysis of the quenching of galaxies, bulges and disks across cosmic time in three observational galaxy surveys, and one simulated data set. More specifically, we analyse quenching in the local Universe utilising the SDSS and MaNGA IFU survey. At low redshifts, we utilise the public stellar mass bulge - disk decomposition catalogue of \cite{Mendel2014} and photometric bulge - disk decompositions from \cite{Simard2011}. To compare to high redshifts, we utilise data from the CANDELS survey, especially the public stellar mass bulge - disk decomposition catalogue of \cite{Dimauro2018}. We also make detailed comparisons to the LGalaxies semi-analytic model (\citealt{Henriques2015}) to test our machine learning based analysis method, and to aid in the interpretation of our observational results.

We develop a sophisticated machine learning approach utilising a Random Forest to classify galaxies, bulges, disks, and spaxels into star forming and quiescent categories. In Section 3 (and in Appendix B), we carefully test our novel statistical method, thoroughly demonstrating its ability to correctly extract the causal structure in simple and more complex classification problems, utilising mock data and a semi-analytic model (see Fig. 3 and Fig. B1). This method is of great value for analysing complex astronomical data, and has the potential for wide applications in the field.

We apply our Random Forest classification technique to multi-epoch galaxy surveys (SDSS and MaNGA in Section 4; CANDELS in Section 5) to study galaxy quenching. Our primary observational results are as follows:

\begin{enumerate}

\item In the local Universe, bulge mass is the most predictive parameter of galaxy quenching, out of the photometric bulge disk parameters (including bulge mass, disk mass, total stellar mass and $B/T$ morphology). This confirms the result of \cite{Bluck2014} and places it on a much firmer statistical footing. See Fig. 4 (left panel) \& Fig. 5.  \\

\item Bulge mass is also the most predictive parameter of galaxy quenching at intermediate and high redshifts as well. Hence, the dependence of quenching on these parameters has been stable since at least cosmic noon. This confirms and extends the result of \cite{Lang2014}, placing it on a much firmer statistical footing, in part by exploring a three-fold increase in the number of high-z galaxies. See Fig. 14 \& Fig. 15. \\

\item The quenching of bulge and disk structures (treated separately) are both primarily dependent on bulge mass alone. This suggests a common cause of quenching throughout galaxies, arising out of the inner-most regions. This is true at both high and low redshifts, which further highlights the stability of the quenching mechanism across cosmic time. Additionally, we demonstrate that a colour-based approach yields consistent results to a full spatially resolved spectroscopic approach in the local Universe (by comparing bulge and disk quenching in MaNGA to the SDSS). All of these results are entirely novel. See Fig. 9 \& Fig. 17. \\

\item Utilising spatially resolved spectroscopy from MaNGA, we demonstrate that quenching is primarily a global process, impacting all regions within galaxies in concert. Conversely, star formation has a strong spatially resolved dependence, indicating that the level of star formation in star forming systems is governed by local physics, which varies substantially throughout each galaxy. We provide a novel analysis here, which confirms and extends these important conclusions from \cite{Bluck2020a, Bluck2020b}. See Fig. 10 \& Fig. 11.\\

\item When available in spectroscopic data sets, central stellar velocity dispersion is even more predictive of quenching than bulge mass. Indeed, when central velocity dispersion is available there is essentially no importance given to any of the following parameters: total stellar mass, group halo mass, local galaxy density, $B/T$ morphology, bulge mass or disk mass. This confirms and extends the results in \cite{Bluck2016} and \cite{Teimoorinia2016}, placing our prior conclusions on a much more robust statistical footing. Moreover, these results completely rule out virial shock heating, supernova feedback, morphological stabilisation, or environmental effects as dominant quenching routes for central galaxies in the local Universe. The strength of this statement is only possible by using the unique capacity of our RF classifier to remove spurious correlations in the determination of quenching importance. See Fig. 4 (right panel), Fig. 6 \& Fig. 9.

\end{enumerate}

\noindent A natural explanation for the above observational results may be formulated by noting that black hole mass is highly correlated with bulge mass, and hence higher mass bulges host more massive supermassive black holes. The quenching of galaxies as a function of bulge mass may thus be explained as a consequence of more massive black holes yielding greater levels of historic feedback into galaxy haloes, preventing gas accretion into the system and hence suppressing star formation. The above hypothesis is further supported by the even greater predictive power of central velocity dispersion over quenching, when available in spectroscopic data sets. This follows because central velocity dispersion is now well established to be even more tightly correlated with black hole mass than bulge mass.

By comparing the observations to the LGalaxies semi-analytic model, we find that radio-mode AGN feedback offers a plausible mechanism to account for the observational results of this paper. Furthermore, the LGalaxies model makes several quantitative predictions which are closely mirrored in the observational data (see Figs. 3, 18 \& 19). More specifically, LGalaxies predicts that bulge mass should be the most predictive photometric parameter of galaxy quenching at both low and high redshifts. Moreover, LGalaxies also predicts that quenching should be globally regulated within galaxies, governed by the buoyancy of the hot gas halo, which itself is regulated by heating from radio-mode AGN feedback (see Appendix A for a detailed discussion). We argue that the high degree of success of the LGalaxies model (in comparison to multi-epoch observational data), indicates that the general paradigm of radio-mode (or preventative) AGN feedback is a plausible avenue for explaining the quenching of galaxies, bulges and disks across cosmic time.  

Finally, for a comparison of quenching in observations to cosmological hydrodynamical simulations we encourage readers to view \cite{Piotrowska2021}; and for a critical assessment of the dependence of quenching on kinematic parameters we encourage readers to view \cite{Brownson2022}. Briefly, the former establishes that the predictions from LGalaxies shown here are also remarkably similar to predictions from Eagle (\citealt{Schaye2015}), Illustris (\citealt{Vogelsberger2014, Vogelsberger2014a}) and Illustris-TNG (\citealt{Nelson2018}). The latter establishes that velocity dispersion is more predictive of quenching than any other kinematic parameter, including circular velocity, $V/\sigma$, dynamical mass, specific kinetic energy, or angular momentum. Hence, the primacy of velocity dispersion for predicting quenching, and its plausible link to AGN feedback through simulations, are both further enhanced by these parallel contemporaneous works from our group.


\section*{Acknowledgments}

We thank the anonymous referee for a highly constructive referee report, which has helped to improve the presentation of our work. We are very grateful to Jarle Brinchmann, Kevin Bundy, Paola Dimauro, Bruno Henriques, Marc Huertas-Company, J. Trevor Mendel, Sebastian S\'anchez and Luc Simard for making their value added galaxy catalogs publicly available, and additionally for helpful discussions on these data in many cases. AFLB acknowledges helpful conversations with Hossen Teimoorinia on machine learning applications in astronomy. Additionally, AFLB acknowledges several fruitful conversations on this research with Mirko Curti, Emma Curtis-Lake, Gareth Jones and James Trussler. We are also grateful to all contributors to the SDSS, MaNGA and CANDELS galaxy surveys, as well as to the LGalaxies semi-analytic model team. All of the analyses in this work were performed using {\small PYTHON}, and all machine learning applications were performed utilising {\small SCIKIT-LEARN}.

AFLB, RM, SB and JMP acknowledge ERC Advanced Grant 695671 QUENCH, and support from the UK Science and Technology Facilities Council (STFC). RM also acknowledges funding from a research professorship from the Royal Society. JMP also acknowledges funding from the MERAC Foundation. SLE acknowledges support from an NSERC Discovery Grant. 

{\bf Data Access:} All of the data used in this paper (both observational and simulated) are publicly available. For the SDSS see: \href{https://classic.sdss.org/dr7/}{https://classic.sdss.org/dr7/}; for MaNGA see: \href{https://www.sdss.org/surveys/manga/}{https://www.sdss.org/surveys/manga/}; and for CANDELS see: \href{http://arcoiris.ucolick.org/candels/}{http://arcoiris.ucolick.org/candels/}. More specifically, the bulge - disk decompositions used throughout this work are from \cite{Mendel2014} for the SDSS (also used in MaNGA); and from \cite{Dimauro2018} for CANDELS. To access these data products directly see - Mendel et al. : \href{https://vizier.u-strasbg.fr/viz-bin/VizieR?-source=J/ApJS/210/3}{https://vizier.u-strasbg.fr/viz-bin/VizieR?-source=J/ApJS/210/3}; Dimauro et al. : \href{https://vm-weblerma.obspm.fr/huertas/form\_CANDELS}{https://vm-weblerma.obspm.fr/huertas/form\_CANDELS}. The LGalaxies model data is available from \cite{Henriques2015}, and may be downloaded directly from: \href{http://galformod.mpa-garching.mpg.de/public/LGalaxies/downloads.php}{http://galformod.mpa-garching.mpg.de/public/LGalaxies/downloads.php}. We strongly recommend readers interested in working with these data to start by reading the following references - SDSS: \cite{Brinchmann2004, Abazajian2009, Simard2011, Mendel2014}; MaNGA: \cite{Bundy2015, Law2015, Bluck2020a}; CANDELS: \cite{Grogin2011, Dimauro2018}; LGalaxies: \cite{Henriques2013, Henriques2015}. Information regarding access to the data, explanation of the data products, and advice for using these data are provided in the above references.

This work makes use
of data from SDSS-I \& SDSS-IV. Funding for the SDSS has been provided by the Alfred P. Sloan Foundation, the Participating Institutions, the National Science Foundation, the U.S. Department of Energy, the National Aeronautics and Space Administration, the
Japanese Monbukagakusho, the Max Planck Society, and the
Higher Education Funding Council for England. Additional funding towards SDSS-IV has been
provided by the U.S. Department of Energy Office of Science. SDSS-IV acknowledges support and resources from
the Center for High-Performance Computing at the University of Utah. The SDSS website is: 
www.sdss.org

The SDSS is managed by the Astrophysical Research Consortium for the Participating Institutions of the SDSS Collaboration. For SDSS-IV this includes the Brazilian
Participation Group, the Carnegie Institution for Science,
Carnegie Mellon University, the Chilean Participation Group,
the French Participation Group, Harvard-Smithsonian Center
for Astrophysics, Instituto de Astrofisica de Canarias, The
Johns Hopkins University, Kavli Institute for the Physics
and Mathematics of the Universe (IPMU) / University of
Tokyo, Lawrence Berkeley National Laboratory, Leibniz Institut fur Astrophysik Potsdam (AIP), Max-Planck-Institut fur
Astronomie (MPIA Heidelberg), Max-Planck-Institut fur Astrophysik (MPA Garching), Max-Planck-Institut fur Extraterrestrische Physik (MPE), National Astronomical Observatory
of China, New Mexico State University, New York University,
University of Notre Dame, Observatario Nacional / MCTI,
The Ohio State University, Pennsylvania State University,
Shanghai Astronomical Observatory, United Kingdom Participation Group, Universidad Nacional Autonoma de Mexico, University of Arizona, University of Colorado Boulder,
University of Oxford, University of Portsmouth, University of
Utah, University of Virginia, University of Washington, University of Wisconsin, Vanderbilt University, and Yale University.

The MaNGA data used in this work is publicly available at: 
http://www.sdss.org/dr15/manga/manga-data/

The Participating
Institutions of SDSS-I \& II are the American Museum of Natural History, Astrophysical Institute Potsdam, University of Basel, University of Cambridge, Case Western Reserve University, University of Chicago, Drexel University, Fermilab, the Institute for Advanced Study, the Japan Participation Group, Johns
Hopkins University, the Joint Institute for Nuclear Astrophysics, the Kavli Institute for Particle Astrophysics and
Cosmology, the Korean Scientist Group, the Chinese Academy
of Sciences (LAMOST), Los Alamos National Laboratory,
the Max-Planck-Institute for Astronomy (MPIA), the Max-Planck-Institute for Astrophysics (MPA), New Mexico State
University, Ohio State University, University of Pittsburgh, University of Portsmouth, Princeton University, the United States Naval Observatory, and the University of Washington.


\bibliographystyle{aa}
\bibliography{Quench}


\appendix


\section{Bulge growth \& quenching in LGalaxies}

In this appendix we provide an overview of the relevant aspects of the LGalaxies galaxy evolution model with respect to star formation quenching and the growth of bulge structures. We also provide a novel parameterization of quenching in the model which is particularly instructive for explaining the observational results seen throughout this paper.

\subsection{The quenching model}

In the LGalaxies model, galaxies quench via radio-mode AGN feedback (see \citealt{Henriques2015}), with a relatively simple prescription based on \cite{Croton2006}. More specifically, galaxies quench when heating from hot-mode AGN accretion becomes greater than cooling from bremsstrahlung emission in the hot gas halo. Hence, the criterion for quenching in the model is:

\begin{equation}
\dot{E}_{\rm radio} \gtrsim \dot{E}_{\rm cool},
\end{equation}

\noindent where the heating power is given by

\begin{equation}
\dot{E}_{\rm radio} = \eta c^2 \dot{M}_{BH},
\end{equation}

\noindent with $\eta$ indicating the efficiency of energy release from accretion onto the supermassive black hole (see e.g. \citealt{Thorne1974, Elvis2004}). Note that dots over variables indicate derivatives with respect to time (as is conventional). Consequently, quenching is modelled as a function of hot-mode accretion, which is further parameterised as:

\begin{equation}
\dot{M}_{BH} = k_{\rm AGN} \bigg( \frac{M_{\rm Hot}}{10^{11}M_{\odot}} \bigg) \bigg( \frac{M_{BH}}{10^{8}M_{\odot}} \bigg) \propto \, M_{\rm vir} \, M_{BH} ,
\end{equation}

\noindent where $M_{BH}$ is the black hole mass, $M_{\rm Hot}$ is the hot halo gas mass, and $k_{\rm AGN}$ is a tuneable parameter in the model (see Henriques et al. 2015, 2019). We also assume that the hot gas mass is proportional to the virial mass of the halo (see e.g. \citealt{Guo2011}). Thus, the rate of energy input into the halo from radio-mode feedback is directly proportional to the black hole mass in this model. Indeed, one can show that a direct proportionality between the input energy from AGN accretion and the mass of the black hole must exist in the general class of energy-conserving models as well (see \citealt{Bluck2020a}, appendix B). Additionally, accretion onto the supermassive black hole is assumed to scale linearly with hot gas mass, and hence the virial mass of the halo. Note that this is a similar prescription to Bondi-Hoyle accretion, but avoids the need to accurately model the gas density at the very centre of the galaxy (see \citealt{Croton2006, Bower2006, Bower2008}).

The remaining piece of the puzzle is the gas cooling rate, which is modelled via the standard physics of the bremsstrahlung (free-free) interaction. That is, 

\begin{equation}
\dot{\mathcal{E}}_{\rm cool}(r)  = n_e n_I \Lambda(T_{\rm Hot}, Z_{\rm Hot})  \hspace{0.2cm}   \propto   \hspace{0.2cm}   \rho^{2}_{\rm Hot}(r) \, T_{\rm vir}^{1/2} \, Z^{2},
\end{equation}

\noindent where $\dot{\mathcal{E}}(r)$ is the cooling rate per unit volume at radius $r$; $\Lambda(T_{\rm Hot}, Z_{\rm Hot})$ is the equilibrium cooling function, assuming exclusively collisional processes (i.e. excluding radiative ionisation); $Z$ is the mean atomic number of ions in the plasma (which we hold constant); and $n_e$ \& $n_I$ indicate the number density of electrons and ions in the hot gas halo. The temperature of the hot halo ($T_{\rm Hot}$) gas is assumed to be at the virial temperature ($T_{\rm vir}$) of the dark matter halo. Consequently, the above functional form implies that the cooling rate per unit volume is proportional to the square of the density of plasma in the hot gas halo ($\rho_{\rm Hot}$), and square root of the virial temperature ($T_{\rm vir}$). 

The relationship between cooling power and density is obvious, given the functional form of eq. A4 (i.e. its dependence on the number density of electrons and ions). To see the reason for the relationship with temperature, it is instructive to view the formula for the virial temperature of a dark matter halo. This is given by (e.g. \citealt{Mo2010}):

\begin{equation}
T_{\rm vir} = \frac{\mu m_p}{2 k_B} V^2_{\rm vir} =  \frac{\mu m_p}{2k_{B}} (G M_{\rm vir})^{2/3} \bigg( \frac{\Delta_c(z) \Omega_{M}(z) H^2(z)}{2} \bigg)^{1/3} ,
\end{equation}

\begin{equation}
T_{200} = \frac{\mu m_p}{2k_{B}} \big( 100 \Omega_{M,0} H_0^2 \big)^{1/3}  (G M_{200})^{2/3} (1+z) ,
\end{equation}

\noindent where the root mean square velocity of particles in the hot gas halo (i.e. the velocity which goes into the fundamental equations of bremsstrahlung cooling: $\Gamma_{\rm brem} \sim n_e n_I \sigma_{T} v_{\rm rel}$) is taken to be equal to the virial velocity, $V_{\rm vir}$. Since the rate of free - free interactions is directly proportional to the relative velocity, this sets the $T^{1/2}_{\rm vir}$ dependence in eq. A4, given the above expressions. Additionally, $M_{\rm vir}$ is the dark matter halo mass contained within the virial radius ($R_{\rm vir}$), $H(z)$ is the Hubble parameter, $\Omega_M(z)$ is the mass density parameter, $\Delta_c(z)$ is the virial density parameter, $\mu$ indicates the mean atomic number of ions in the hot gas halo, and $m_p$ is the proton mass. In the second expression above (eq. A6) we set the virial density parameter $\Delta_c(z)$ = 200, as is done throughout the LGalaxies model, and make the redshift dependence on the Hubble and density parameters explicit.

At this point it is also useful to state the general forms of the virial velocity and virial radius, to aid in what follows. These are given by (see e.g. \citealt{Mo2010}):

\begin{equation}
V_{\rm vir} = (G M_{\rm vir})^{1/3} \bigg(\frac{\Delta_c(z) \Omega_{M}(z) H^2(z)}{2} \bigg)^{1/6},
\end{equation}

\begin{equation}
V_{200} =  \big(100 \Omega_{M,0} H^2_0 \big)^{1/6} (G M_{200})^{1/3} \, (1+z)^{1/2},
\end{equation}

\noindent and

\begin{equation}
R_{\rm vir} = (2G M_{\rm vir})^{1/3} \big( \Delta_c(z) \Omega_M(z) H^2(z) \big)^{-1/3},
\end{equation}

\begin{equation}
R_{200} = \big(200 \Omega_{M,0} H^2_0 \big)^{-1/3} (2G M_{200})^{1/3} \, (1+z)^{-1},
\end{equation}

\noindent where in the second line of each equation block we have set the virial density equal to 200, and extracted the explicit redshift dependence (as with with virial temperature above). In the above expressions, $M_{200}$ indicates the mass contained within $R_{200}$, which is defined as the radius at which the average density within that radius is 200 times the mass density of the Universe at redshift $z$.

In LGalaxies, the hot gas halo is modelled as an isothermal sphere, which is the simplest density distribution compatible with hydrostatic equilibrium. More specifically, the density of plasma at radius $r$ is given by:

\begin{equation}
\rho_{\rm Hot}(r) = \frac{M_{\rm Hot}}{4 \pi R_{200} \, r^2} = f_b \epsilon \frac{V^2_{200}}{4\pi G r^2} \propto \frac{M^{2/3}_{200} \, (1+z)}{r^2},
\end{equation}

\noindent where $f_b$ is the cosmic baryon fraction ($\equiv \Omega_b/\Omega_M$), and $\epsilon$ is the fraction of baryons residing in the hot gas halo, such that $M_{\rm Hot} = f_b \epsilon M_{200}$. Additionally, $V_{200}$ is the virial velocity and $R_{200}$ is the virial radius (defined above).

As a result of the above general expressions, as the dark matter halo increases in mass the rate of cooling per unit volume increases as a result of the increase in the density and the virial temperature (compare eqs. A4, A6 \& A11). As cooling progresses, the density of gas increases, which then leads to run-away cooling due to the squared dependence on density vs. the square root dependence on temperature (eq. A4). This process is inexorable, and hence has been dubbed the `cooling catastrophe' (e.g. \citealt{Ruszkowski2002, Fabian1994, Fabian1999, Fabian2012}). Without some form of stabilizing feedback, all hot gas haloes are expected to be thermodynamically unstable within (typically much less than) the Hubble time. These arguments are not strongly model dependent and indeed apply generally to any dark matter halo in which bremsstrahlung emission is the dominant hot gas cooling mode (approximately $M_{\rm Halo} \gtrsim 10^{12} M_{\odot}$). 

We now have all the ingredients we need to solve eq. A1 and hence determine whether a given halo will be star forming or quenched. However, in practice numerically solving the above equations is impractical, and moreover would ignore the dynamical time required for gas to condense into the halo (which can be very significant on these scales). As such, \cite{Henriques2015} reframe the problem in terms of gas mass accretion:

\begin{equation}
\dot{M}_{\rm cool, modified} = \dot{M}_{\rm cool} - 2E_{\rm radio}/V^2_{200} ,
\end{equation}

\noindent whereby the radio heating offsets a mass of cooling gas, determined by assuming the specific kinetic energy of the halo is constant (as required for an isothermal sphere). If the modified cooling mass becomes negative this is set to zero in the model. The rationale is that once cooling stops entirely, the black hole can no longer accrete gas. Explicitly, the condition for halo quenching becomes: 

\begin{equation}
\dot{E}_{\rm radio} \gtrsim \dot{E}_{\rm cool}   \hspace{0.3cm} \rightarrow \hspace{0.3cm}   \dot{E}_{\rm radio} \gtrsim \frac{1}{2} \dot{M}_{\rm cool} \, V^2_{200} .
\end{equation}

The entire problem is now reduced to inferring the gas cooling mass per unit time. To evaluate this, we first construct a general expression for the cooling time as a function of radius. The halo cooling time is defined by:

\begin{equation}
t_{\rm cool}(r) \equiv \frac{\mathcal{E}_{\rm Hot}(r)}{\dot{\mathcal{E}}_{\rm Hot}(r)} = \frac{3 \mu m_p k_{B} T_{200}}{2 \rho_{\rm Hot}(r) \Lambda(T_{200}, Z)}
\end{equation}

\begin{equation}
= \frac{6 \pi G \, \mu m_{p} \, k_{B} T_{200} \, r^2 }{f_b \epsilon \, V^2_{200} \, \Lambda(T_{200},Z)},
\end{equation}

\noindent where we have used the density - radius relation of eq. A11. Note also that there is a cancelling of one density term in the numerator and the denominator in the third term of the equation block. It is important to stress that the cooling time is dependent on radius. As such, one can define a cooling radius ($r_{\rm cool}$), which is the radius at which the cooling time is equal to the dynamical time ($t_{\rm dyn} \equiv R_{200} / V_{200}$). As such, the cooling radius is given by:

\begin{equation}
r_{\rm cool} = \bigg\{  \frac{t_{\rm dyn} M_{\rm Hot} \Lambda(T_{200}, Z)}  {6 \pi \mu m_p k_B T_{200} \, R_{200}}   \bigg\}^{1/2} \propto \bigg( \frac{M_{200}} {T^{1/2}_{200} V_{200}} \bigg)^{1/2}
\end{equation}

\begin{equation}
\propto M_{200}^{1/6} (1+z)^{-1/2},
\end{equation}

\noindent where we have used the standard virial relations for a dark matter halo (stated above); and assumed $\Lambda(T_{200}, Z) \propto T^{1/2}_{200}$ for constant $Z$. Hot gas within the cooling radius is unstable to collapse, but gas outside of the cooling radius is stable for at least one dynamical time and hence cannot collapse in that time window. Consequently, it is assumed in the model that the rate of gas accretion from the halo into the galaxy is given by:

\begin{equation}
\dot{M}_{\rm cool} = \frac{M_{\rm Hot}(<r_{\rm cool})}{t_{\rm dyn}} = \bigg( \frac{M_{\rm Hot}}{t_{\rm dyn}} \bigg)  \bigg( \frac{r_{\rm cool}}{R_{200}} \bigg)
\end{equation}

\begin{equation}
= \frac{f_b \epsilon M_{200} \, V_{200} \, r_{\rm cool}} {R^2_{200}}  \hspace{0.2cm}  \propto  \hspace{0.2cm}   M^{5/6}_{200} \, (1+z)^2,
\end{equation}

\noindent where $\Delta M \propto \Delta r$ due to the $1/r^2$ density dependence of the isothermal sphere (see eq. A11). Hence, in the region where the cooling time is less than the dynamical time, gas is assumed to condense from the hot halo into the galaxy in the dynamical time of the system. Of course, there is some ambiguity as to the correct definition of the dynamical time in this situation, but that does not significantly impact the general discussion to follow.

Utilising equations A2, A3, A13 \& A19, we may write the condition for intrinsic quenching in the LGalaxies model in a particularly illuminating (and novel) manner:

\begin{equation}
M_{BH} \gtrsim M_{BH,Q} (M_{200}, z),
\end{equation}

\noindent where,

\begin{equation}
M_{BH,Q} \, (M_{200}, z) \equiv \frac{V^3_{200} \, r_{\rm cool}}{2 \eta c^2 \, k_{\rm AGN} \, R^2_{200}}  \hspace{0.1cm}  \propto   \hspace{0.1cm}   M^{1/2}_{200} \, (1+z)^3,
\end{equation}

\noindent which we may write in a more convenient form as:

\begin{equation}
M_{BH,Q} \, (M_{\rm Halo}, z) = M_{BH,Q}\,(z=0) \, \bigg(  \frac{M_{\rm Halo}}{10^{12} M_{\odot}} \bigg)^{1/2} (1+z)^3,
\end{equation}

\noindent where $M_{BH,Q}\,(z=0)$ is the zero-point black hole mass quenching threshold in the model, defined as the black hole mass which just satisfies eq. A13 at redshift zero, for a halo mass of $10^{12}M_{\odot}$. Ultimately, the zero-point is a function of the tuneable parameter $k_{\rm AGN}$ in eq. A3, and also incorporates a variety of geometric, cosmological and dimensional constants, all of which may be ignored once a zero-point value is known. Note also that we now use the subscript `Halo' instead of `200', but it should be understood that this explicitly means the mass contained within $R_{200}$ (i.e. from now on we take $M_{\rm Halo} \equiv M_{200}$, in line with the rest of this paper).

It is important to appreciate that the critical black hole mass quenching threshold depends only on halo mass and redshift. The square root power law dependence of $M_{BH,Q}$ on $M_{\rm Halo}$ indicates that the threshold black hole quenching mass varies weakly with halo mass at a fixed epoch. Conversely, the cubic power-law exponent on the redshift term implies that there is strong redshift evolution in this threshold. More specifically, there is a (relatively weak) tendency for the critical black hole mass threshold to increase in more massive haloes (at a fixed redshift); and a strong tendency for the black hole mass threshold to increase with increasing redshift (for a fixed halo mass). 

Therefore, as a result of the above derivation, the key prediction of the LGalaxies model in terms of intrinsic galaxy quenching is that at any given epoch, there will be a threshold black hole mass at which galaxies quench. Thus, in any redshift bin, one expects to find that black hole mass is the key parameter driving quenching. Additionally, the black hole mass quenching threshold is expected to rise with redshift, as a result of dark matter haloes being more dense at earlier times, and hence intrinsically harder to stabilise against cooling and collapse (given the squared dependence on density in bremsstrahlung emission). Finally, there is a weak dependence on halo mass at any given redshift, whereby the critical black hole mass for quenching increases slightly with increasing halo mass. Hence, quenching in the model is ultimately a result of three processes: AGN heating (dependent on $M_{BH}$); the evolving structure of dark matter haloes (dependent on $z$); and cooling from the hot gas halo (ultimately dependent on $M_{\rm Halo}$).

The key insight of the above discussion is that entire haloes quench as a result of AGN radio-mode feedback. This occurs when cooling from bremsstrahlung emission is offset by heating from energy released by the supermassive black hole (assumed to occur via radio jets). The energetic condition may be reformulated as a black hole mass threshold, with an evolving dependence on redshift and halo mass. It is crucial to note that all components of the galaxy (i.e. bulge and disk) will quench as a result of halo cooling being shut down. As such, this type of starvation model naturally predicts a common quenching dependence of bulges and disks on the same parameters, as observed throughout this paper.

\subsection{Bulge \& black hole growth}

In the LGalaxies model, mergers are critical for both the growth of supermassive black holes (via quasar-mode accretion) and the growth of bulge structures. In a major merger, all of the stars and gas of both systems are placed into the bulge component of the descendent galaxy, i.e. major mergers create pure spheroids. However, gaseous disks may regenerate in the model as a result of cold gas accretion (along dark matter streams) and cooling of the hot gas halo (provided the threshold black hole quenching mass has not yet been reached). In a minor merger, the disk of the host galaxy remains intact, but all of the baryons and stars from the minor companion are placed into the central bulge. Due to the prevalence of minor mergers, this is the dominant route to bulge growth in the model (see \citealt{Henriques2015} and \citealt{Bluck2019} for further details).

Supermassive black holes impact star formation and halo cooling solely via radio-mode feedback in the model, as discussed in the previous sub-section. However, supermassive black hole growth is regulated almost entirely by quasar-mode accretion, associated primarily with mergers. During a merger event, the black hole mass increases by (\citealt{Henriques2015}):

\begin{equation}
\Delta M_{BH} = \frac{f_{BH} \, \mu_{\rm merg} \, M_{\rm cold}}{1 + V_{BH}/V_{200}},
\end{equation}

\noindent where $\mu_{\rm merg} = M_{\rm sat} / M_{\rm cen}$; and $f_{BH}$ \& $V_{BH}$ are tuneable parameters in the model. Additionally, $M_{\rm cold}$ indicates the total cold gas mass of both systems (roughly equivalent to the molecular plus atomic gas mass), and $V_{200}$ is the virial velocity of the central halo (as defined in eq. A8). Hence, some tuneable fraction of the cold gas content of the merger ends up being accreted into the black hole. Note that this accretion mode is suppressed for low virial masses, where it is assumed much of the merger gas may escape the system through quasar (and supernova) feedback. Again, this is a tuneable parameter in the model. Additionally, if the merging satellite galaxy contains a supermassive black hole, it is assumed to merge with the central's supermassive black hole after the Chandrasekhar time of the galaxy merger. This process also contributes significantly to the growth of the central galaxy's supermassive black hole over cosmic time.

As a result of the above brief discussion, it is clear that both bulges and supermassive black holes grow primarily in merger events in the model. Consequently, there is a natural expectation for a strong relationship between these two galactic components, as is also observed (see e.g. \citealt{Magorrian1998, Haring2004, Hopkins2007}). 

The key insight from this appendix is that in lieu of a measurement of black hole mass, bulge mass may act as a reasonable proxy for predicting quenching in the model. In Section 3.3 we explore this possibility quantitatively via a Random Forest classification analysis (see Fig. 4). As a result of these tests, we conclude that the excellent performance of bulge mass as a predictor of quenching in the observational data may ultimately be explained by its connection with black hole mass. In this hypothesis, galaxies will quench much as they do in LGalaxies, i.e. via preventative AGN feedback. This naturally leads to a common quenching of bulges and disks, and hence is also completely consistent with our observational evidence for both bulge and disk quenching depending primarily on the central most regions within galaxies (i.e. on bulge mass and/or central velocity dispersion; see Sections 4 \& 5).


\newpage

\section{The random forest approach}

\subsection{Mathematical details}

A Random Forest classifier offers one statistical machine learning approach to construct a mapping from a multi-dimensional input data set to a target class (here star forming or quenched). More specifically, a Random Forest consists of a set of decision trees, with enforced differences between them. The difference between individual trees in a Random Forest is ensured by bootstrapped random sampling of the input data set (with return), and sometimes additionally by partial random sampling of the features. Hence, there will be subtle differences between each tree, which are averaged over, yielding a more accurate final result than a single decision tree could achieve (see \citealt{Bluck2020a} for a discussion). 

Each tree in the Random Forest is structured as a series of decision forks, with a binary criterion at each node. To generate the tree, in the training step, the Random Forest classifier selects from the available features the most effective variable (and threshold) in order to split the data such that the reduction in impurity is maximised at each node. The impurity of the data arriving in a given node is measured via the Gini coefficient (see \citealt{Pedregosa2011}), defined as:

\begin{equation}
G(n) \equiv 1 - \sum_{i=1}^{c} \bigg( P_i(n)^2 \bigg),
\end{equation}

\noindent where $P_i(n)$ is the probability of randomly selecting objects of a given class from the sample arriving at each node ($n$). The Gini impurity measures the probability of misclassifying the data, if one selects based on the distribution. To appreciate the logic of this definition, it is helpful to consider a two class problem (i.e. c = 2). In this case the Gini coefficient is given by:

\begin{equation}
G(n) = 1 - \sum_{i=1}^{2} \bigg( P_i(n)^2 \bigg) = 1 - P_1^2 - P_2^2 
\end{equation}

\begin{equation}
= (P_1 + P_2) - P_1^2 - P_2^2 = P_1(1-P_1) +  P_2(1-P_2) ,
\end{equation}

\noindent where the subscripts 1 \& 2 refer here to the two arbitrary classes. Looking at the final expression above, to understand the functional form, note that $P_1$ is the probability of randomly selecting class-1, which is the likelihood of `guessing' class-1 based on the distribution of classes (at the node in question). Now, $(1-P_1)$ is the probability of being wrong about that guess. Hence, the product $P_1(1-P_1)$ is the probability of guessing class-1 and being incorrect. Therefore, the probability of misclassifying the data based on the distribution is the sum of the two possible options for being incorrect (i.e. eq. B3, right hand expression). The equivalence with the standard expression (eqs. B1 \& B2) follows from the fact that the probability of being any class is unity (i.e. in the 2-class problem, $P_1 + P_2 = 1$). 

The Gini coefficient has a peak value of 0.5 (for an even distribution of classes in a given node, which is equivalent to the maximum entropy state) and a minimum value of zero (for only one class in a given node, which is equivalent to the minimum entropy state). As such, one seeks to minimise the Gini coefficient, in order to solve the classification problem. This is why it is often referred to as the  `Gini impurity'.

The above description is only exact for the `parent' node (i.e. the node which is to be split into two at the next level in the decision tree). At the subsequent level, the `daughter' nodes, there are two branches which must be considered. The needed generalisation is constructed by simply summing over the two Gini coefficients (of the left and right branches), weighted by the probability of arriving in each branch. That is,

\begin{equation}
G_D(n+1) = P_L(n+1)G_L(n+1) + P_R(n+1)G_R(n+1),
\end{equation}

\noindent where, for example, 

\begin{equation}
P_L(n+1) = \frac{N_L(n+1)}{N_L(n+1) + N_R(n+1)} = \frac{N_L(n+1)}{N(n)},
\end{equation}

\noindent with $N_L(n)$ and $N_R(n)$ indicating the number of data points arriving in the left and right branches of level $(n+1)$, respectively. The probability of residing in the right branch is defined in exact analogy to the left branch, shown above (i.e. switch the subscripts $L \leftrightarrow R$). The Gini coefficients for the right and left nodes in eq. B4 are each computed by eq. B1. The subscript, $D$, in eq. B4 just makes it explicit that we are considering both of the daughter nodes.

In order to select which feature is used by the Random Forest classifier at any given decision fork, in training `truth' labels are made available to the classifier. Hence, the classifier can compare the efficacy of each possible threshold on each feature at each node (by measuring the reduction in Gini coefficient between node $n$ and its daughter nodes for each choice). Explicitly, this is defined by:

\begin{equation}
\Delta G(n) \equiv G(n) - G_D(n+1).
\end{equation}

\noindent That is $\Delta G(n)$ quantifies the change in the Gini coefficient between the parent node (at level $n$) and its daughter nodes (at level $n+1$) in the decision tree, with $G(n)$ defined in eq. B1 and $G_D(n+1)$ defined in eq. B4. Note that due to the ordering in the definition above, minimising the daughter nodes' combined Gini coefficient corresponds to maximising $\Delta G(n)$.

Conceptually, it is straightforward to identify the optimal parameter and threshold for each decision fork once the data, truth labels, and impurity metric are specified. The {\small SCIKIT-LEARN} python package offers a particularly fast and efficient algorithm to achieve this (see the online documentation\footnote{https://scikit-learn.org}). As a pedagogical example, we sketch a simple (though inefficient) algorithm which would solve this data science problem to arbitrary precision: For each feature in the sample, construct trial decision boundaries by a coarse-grained discretisation of the parameter space from its minimum to its maximum value, with $N_{\rm init}$ steps. Evaluate $\Delta G(n)$ for each trial decision boundary in each feature. Systematically increase the resolution on the coarse-graining (e.g. to $r \times N_{\rm init}$, where $r$ is a resolution parameter, increased in integer steps from unity) until no further change in $\Delta G(n)$ is found for any threshold choice in any feature. Select the feature and threshold which maximises $\Delta G(n)$ to use as the criterion for the decision fork at node, $n$. Proceed to level $n+1$. Repeat the prior steps until either no further increase in $\Delta G(n)$ is possible, or else a pre-defined limit is reached (used to avoid over-fitting, see Section 3.2).

The class prediction from the trained Random Forest for an object (i.e. a galaxy or spaxel in our application) with given features (e.g. bulge mass, disk mass and so forth) is given by the mode of the truth class labels in the  `leaf-node' (i.e. the final node in the decision branch) in which that particular object ends up. In the case of fully developed trees the mode is simply equal to the class of the unique training object residing in each leaf-node, whereas in general there will be several (not necessarily identical) truth values in each leaf-node. As such, once a Random Forest is trained to classify certain data into given categories, one can use it on novel data to ascertain a prediction for the class of the object based on its features, without the need to determine the object class directly. This is the primary use of Random Forest classification in astronomy. 

However, the relatively simple architecture of the Random Forest (in comparison to other machine learning techniques) enables another deeply valuable usage, which we fully exploit in this work. Namely, one can determine precisely how much reduction in impurity is generated by each feature in each decision tree, and then average over the features' individual tree importances throughout the Random Forest, yielding a final average importance for each variable of interest. Hence, in our application, through Random Forest classification we establish precisely how informative any given variable is for the process of galaxy quenching.

Explicitly, the relative importance of an arbitrary feature, $k$, in a given tree is defined by:

\begin{equation}
I_{R, \, \mathrm{tree}}(k) \equiv \frac{I_{k, \, \mathrm{tree}}}{\sum_j I_{j, \, \mathrm{tree}}}  =  \frac{\sum_{n_k} N(n_k) \, \Delta G(n_k)} {\sum_{n} N(n) \, \Delta G(n)},
\end{equation}

\noindent where the $j$-summation is performed over all features; the $n_k$-summation is performed over nodes which utilise feature $k$ to separate the data; and the $n$-summation is performed over all nodes in the decision tree. It is important to appreciate that the summation in the numerator in the third expression above is evaluated just for the nodes which utilise feature $k$, whereas the summation in the denominator is evaluated for all nodes in the decision tree. The change in Gini coefficient ($\Delta G(n)$, see eq. B6) is weighted by the number of objects which reach each parent node ($N(n)$). This results in features utilised at decision forks which induce a large change to the data being up-weighted over those which have a smaller impact, i.e. impact a smaller fraction of the data. 

Finally, the relative importance from each tree is averaged throughout the entire Random Forest to yield the final importance statistic on each variable ($k$). Explicitly, this is evaluated as:

\begin{equation}
I_R(k) = \frac{1}{N_{\mathrm{trees}}} \, \sum \limits_{\mathrm{trees}}  \bigg\{ I_{R, \, \mathrm{tree}} (k) \bigg\}   =  \frac{1}{N_{\mathrm{trees}}} \, \sum \limits_{\mathrm{trees}}  \Bigg\{      \frac{\sum_{n_k} N(n_k) \, \Delta G(n_k)} {\sum_{n} N(n) \, \Delta G(n)}      \Bigg\} ,
\end{equation}

\noindent where $I_{R, \, \mathrm{tree}}(k)$ is defined as in eq. B7. This is the key statistic we use to rank parameters, and extract insights from the observational data. In the main body of the paper we refer to this statistic most often as the `relative quenching importance' (due to our specific scientific application).


\begin{figure*}
\includegraphics[width=0.33\textwidth]{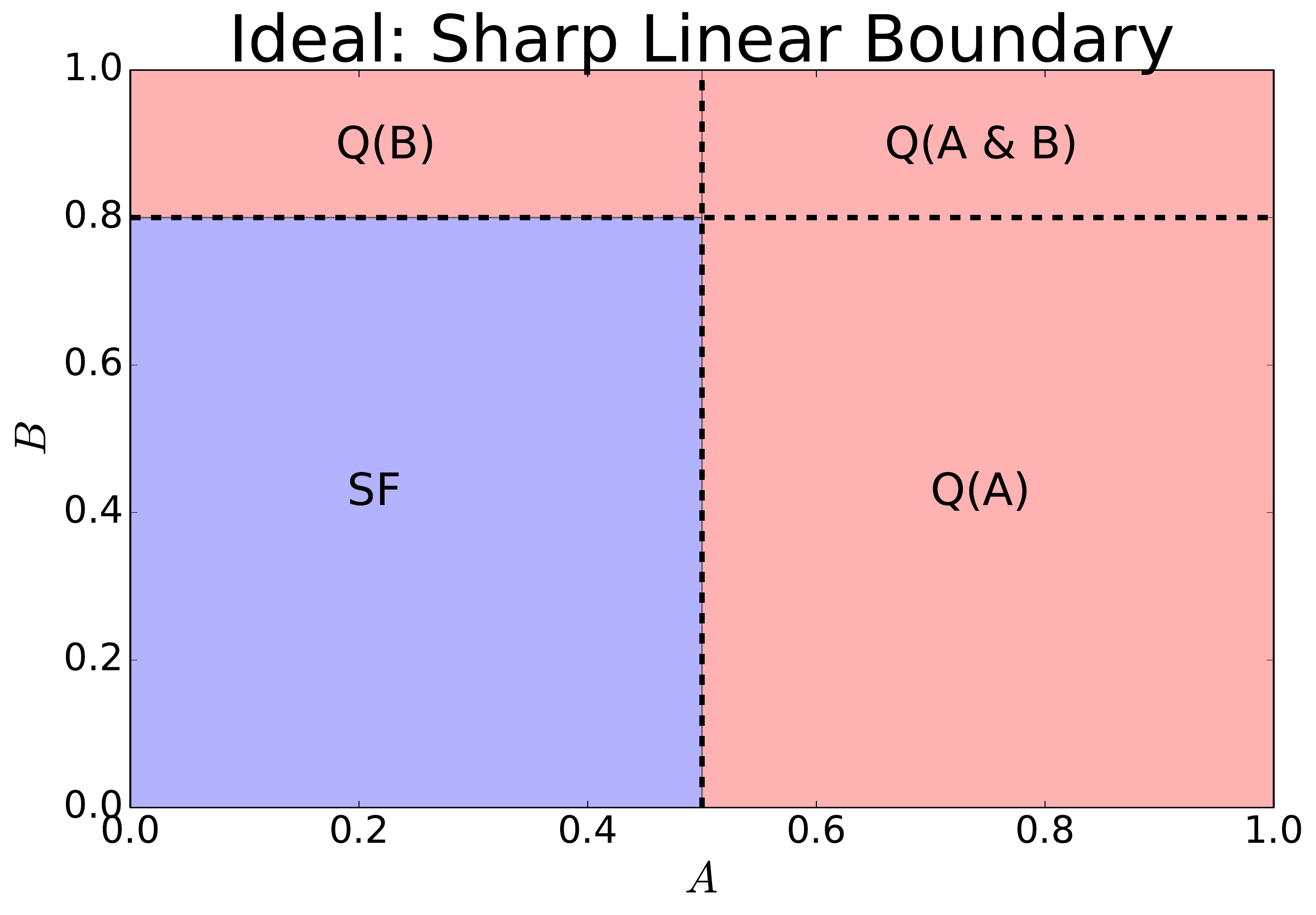}
\includegraphics[width=0.33\textwidth]{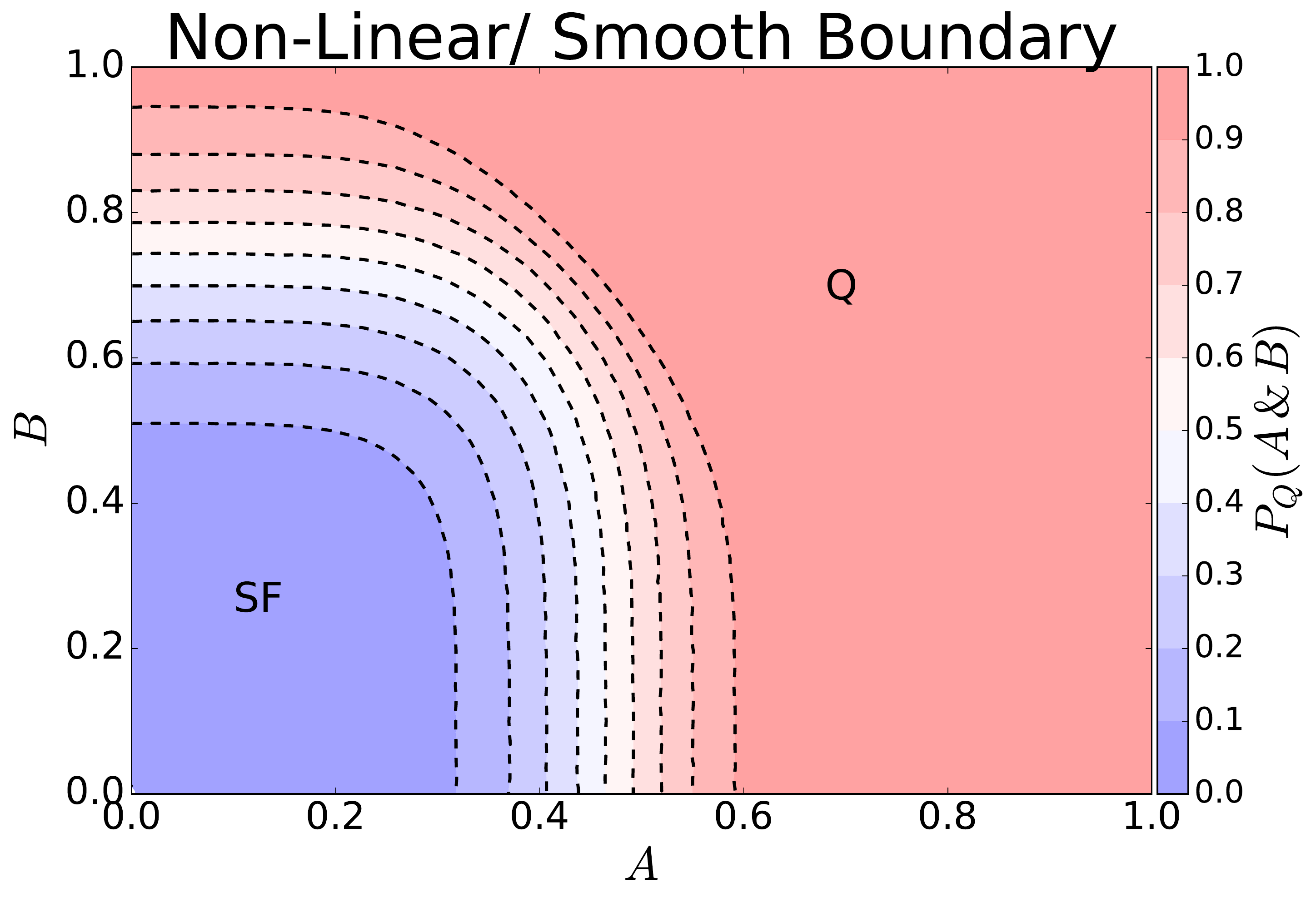}
\includegraphics[width=0.33\textwidth]{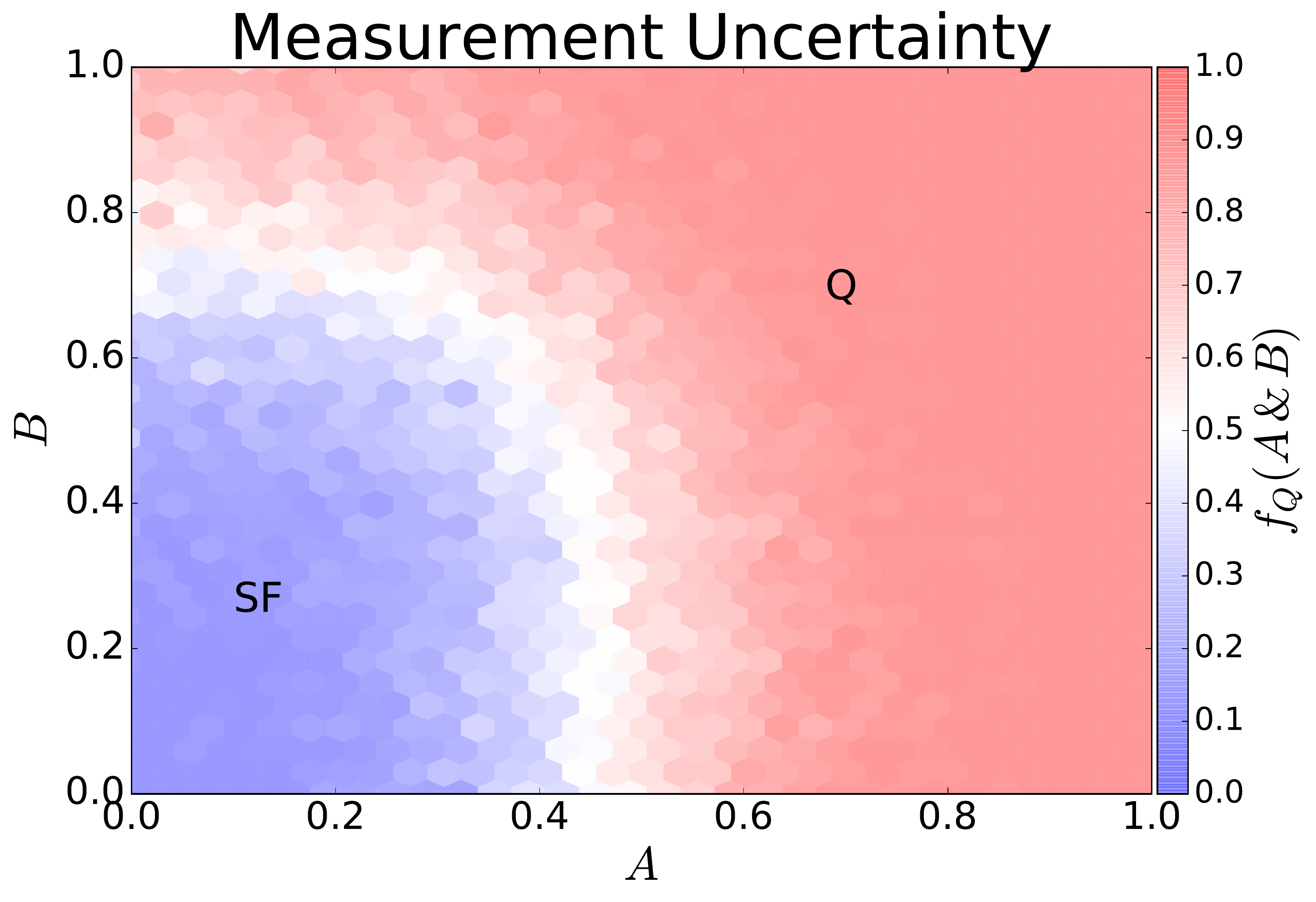}
\includegraphics[width=0.33\textwidth]{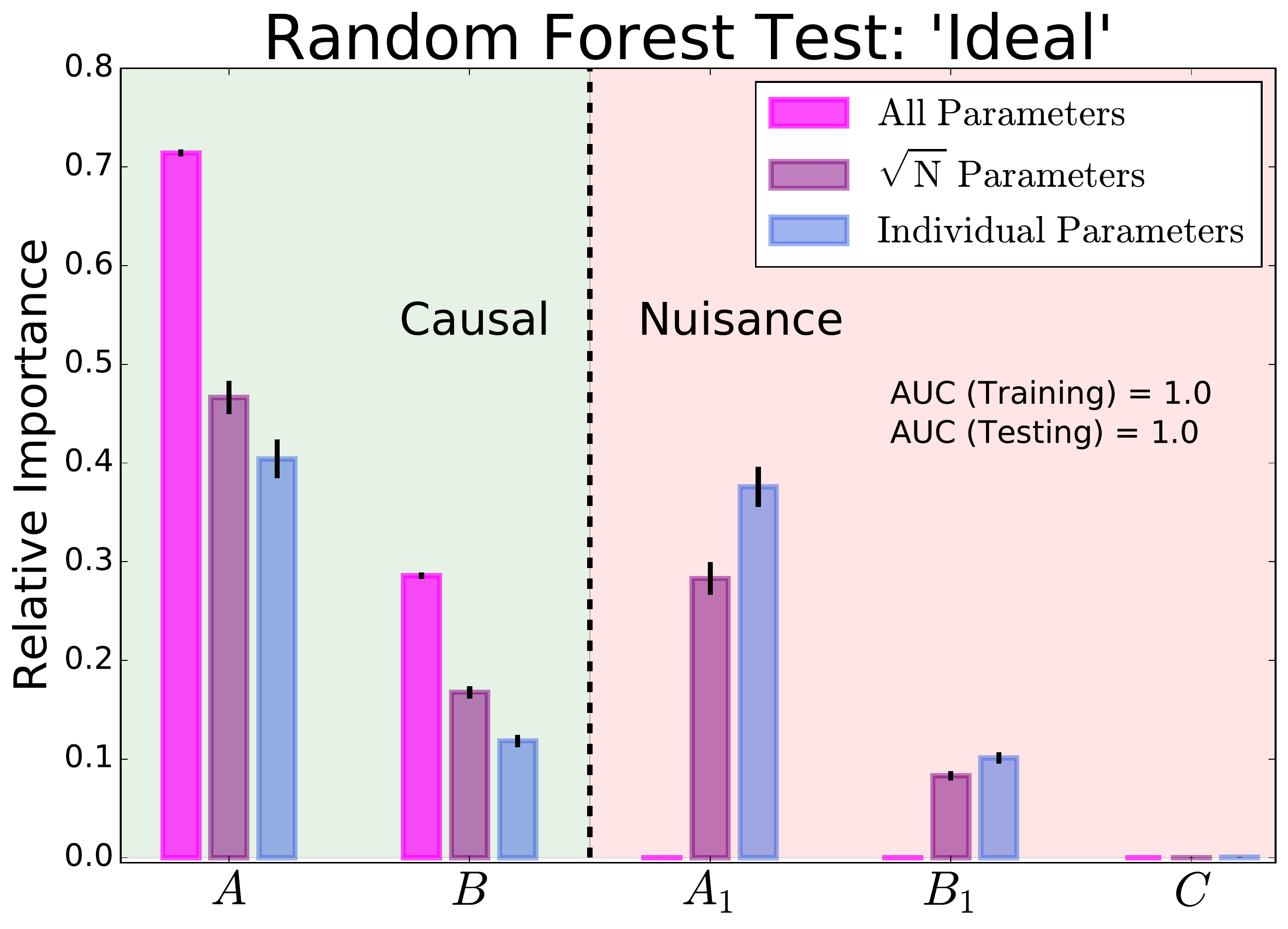}
\includegraphics[width=0.33\textwidth]{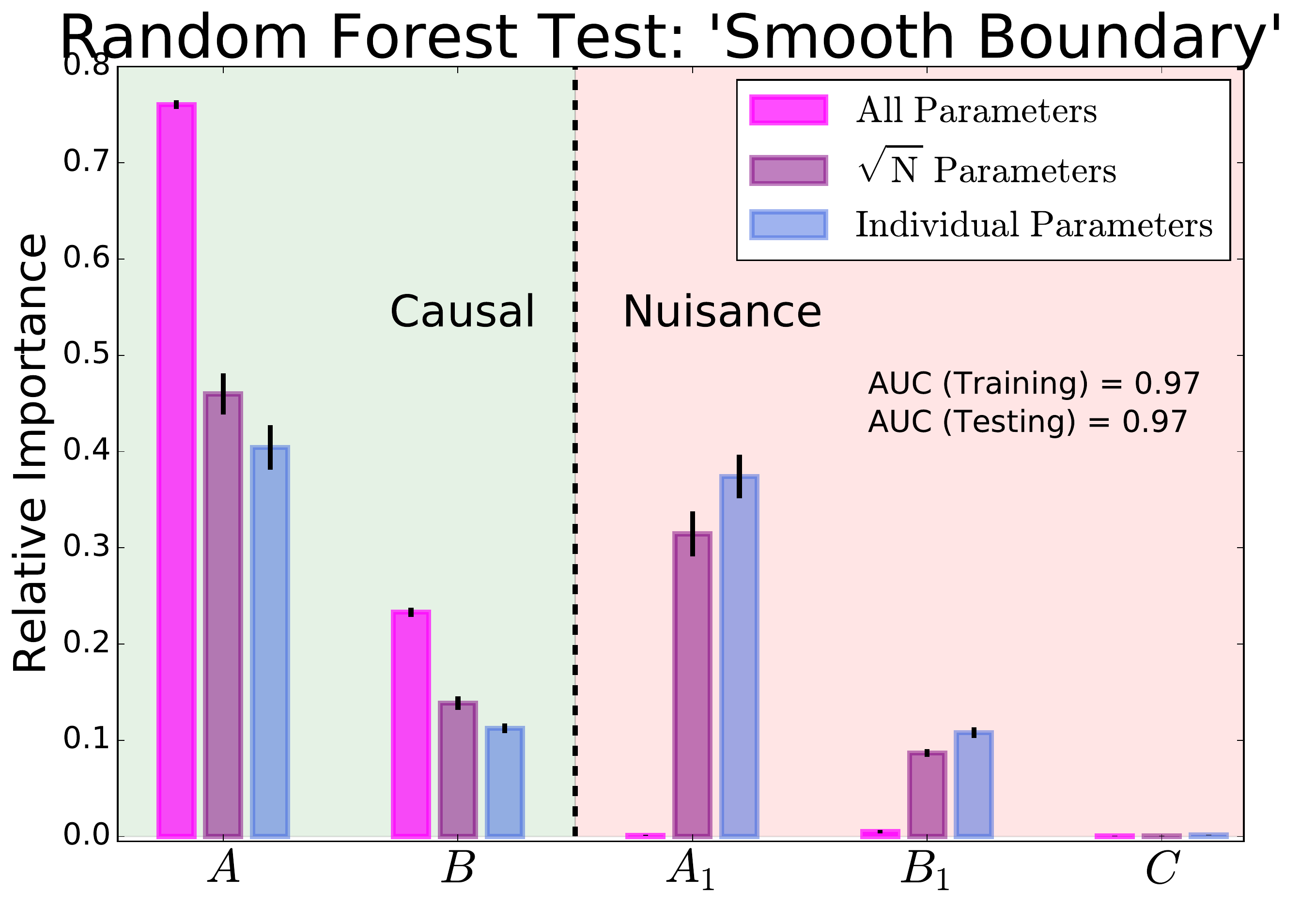}
\includegraphics[width=0.33\textwidth]{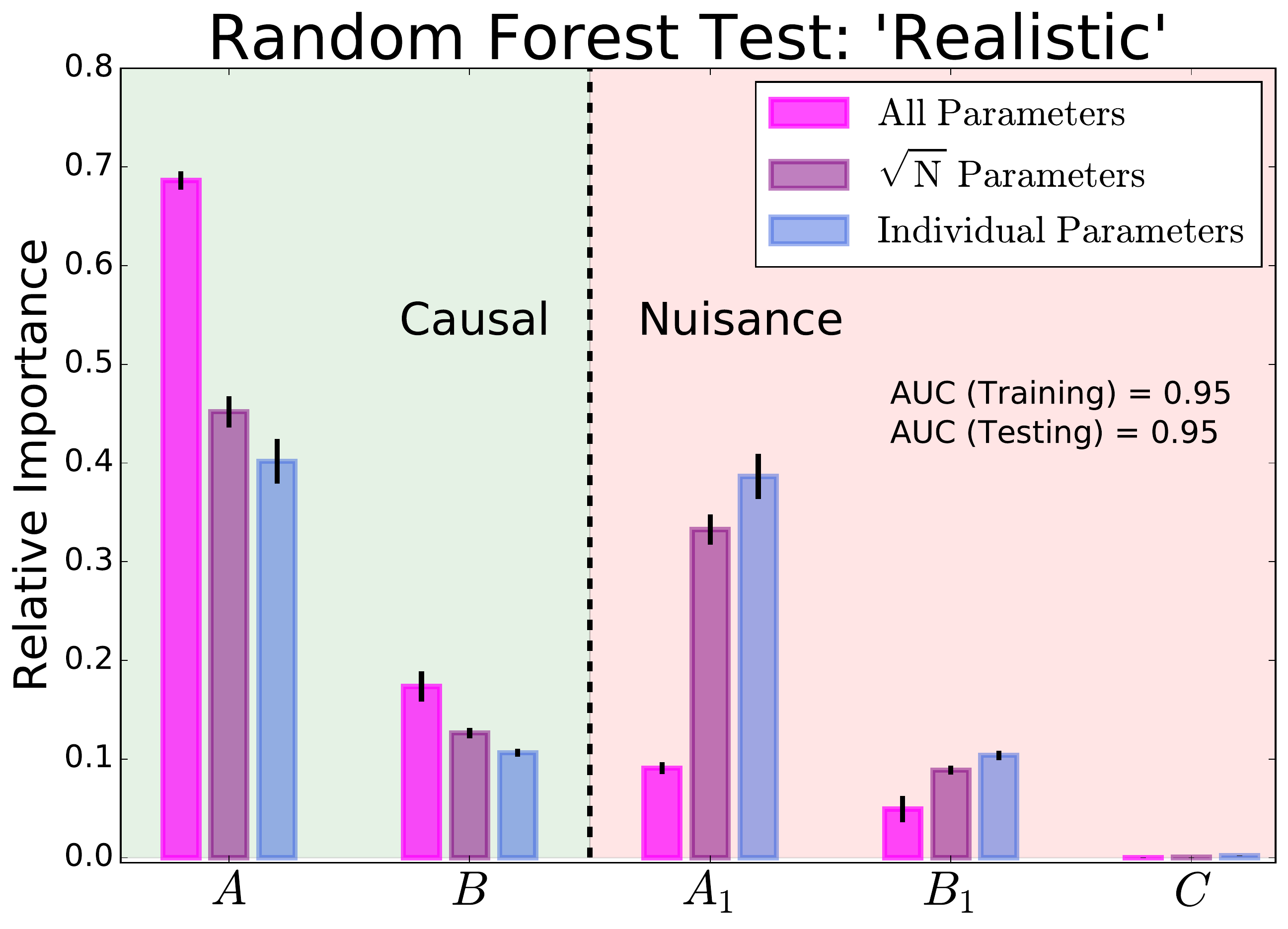}
\caption{Tests on the Random Forest approach. {\it Top panels: } Simple 2-dimensional classification problems, with increasing complexity from left-to-right. {\it Top-left panel: } `Quenching' occurs as a result of the following criteria: $(A > 0.5) \, \cup \, (B > 0.8)$. {\it Top-middle panel: } `Quenching' occurs probabilistically as a result of a smooth non-linear function of the $A$ \& $B$ parameters (see eqs. B10 \& B11). {\it Top-right panel: } Same as the previous panel but with added Gaussian random noise. {\it Bottom panels: } Results from Random Forest classification analyses to ascertain the relative importance of the following parameters: $A$ (the most important causal parameter), $B$ (the secondary causal parameter), $A_1$ (an a causal `nuisance' parameter, highly correlated with $A$), $B_1$ (an a causal `nuisance' parameter, highly correlated with $B$), and $C$ (an a causal parameter uncorrelated with either $A$ or $B$). The Random Forest classification is run in three modes: i) considering all parameters at each fork in the decision tree; ii) considering $\sqrt{N}$ parameters at each fork (randomly selected); iii) considering only one parameter at each fork (randomly selected). The run utilising all variables at every decision fork correctly assigns the highest importance to the causal parameters at all levels of complexity.} 
\end{figure*}

\subsection{Tests on the random forest approach}

Correlation does not imply causation, as is well known. Nonetheless, it is the search for causation (not correlation) which is the hallmark of modern science. To bridge the epistemic gulf between correlation and causation is a notoriously difficult problem, fraught with many fundamental challenges. Modern machine learning techniques offer one technique to aid in disentangling the true causes of relationships between complex data. In this appendix we illustrate how the Random Forest technique can be used in a very specific mode to correctly isolate causal from a causal `nuisance'  parameters in successively more realistic classification problems. 

In Fig. B1 (top-left panel) we show the set-up for a very simple classification problem. In this idealised example, data may be classed as `star forming' (shown in blue) or `quenched' (shown in red), although for now these are just arbitrary labels (we could equally well have called these `cats' and `dogs'). We start by constructing two arbitrary variables: $A$ and $B$. The class of each datum is then set by a very simple criteria:

\begin{equation}
\mathrm{if} \,\, (A > 0.5) \,\, \cup \,\, (B > 0.8)  \,  \rightarrow \, Q.
\end{equation}

\noindent That is, if the above criteria is met, then the object is `quenched', otherwise it is `star forming'. This is just an example of a simple decision boundary incorporating two variables. The $A$ and $B$ variables are constructed to be independent and are both modelled as a uniform distribution on the space \{0,1\}, with 100,000 realisations of each. As a result of the above decision boundary, there is, in a sense, a causal relationship between `quenching' in the simple model and both the $A$ \& $B$ parameters. Due to the different chosen thresholds, the $A$ parameter will dominate `quenching' in the sample, but there will be a significant secondary dependence on the $B$ parameter. 

In order to demonstrate the value of the Random Forest technique, we construct three additional parameters: $A_1$, which is designed to have an extremely high correlation of $\rho = 0.99$ with $A$ but have no direct role in `quenching'; $B_1$, which is designed to have an extremely high correlation of $\rho = 0.99$ with $B$ but have no direct role in `quenching'; and $C$ which is uncorrelated with either $A$ or $B$ and has no causal connection to `quenching'. It is important to appreciate that in terms of correlation, $\{A, A_1\}$ and $\{B, B_1\}$ are virtually indistinguishable, yet $C$ is clearly distinct from the other parameters.

In Fig. B1 (bottom-left panel) we show the results from a Random Forest analysis to solve this simple classification problem. First, the parameters are separated into causal (shown with a light green background) and a causal `nuisance' (shown with a light red background) sets. The $y$-axis shows the relative importance of each variable (i.e. how useful it is for reduction in Gini impurity throughout the Random Forest, see the previous sub-section). We consider three modes of operation for the Random Forest classifier: i) only one, randomly selected, parameter considered at each decision fork (`Individual Parameters'); ii) the square root of the number of available features, randomly selected, considered at each decision fork (`$\sqrt{N}$ Parameters', here rounded down to two); iii) all features considered at each decision fork (`All Parameters').

For the Individual Parameters run, $C$ is clearly found to be unimportant for `quenching', as expected. However, the relative importance of both $\{A, A_1\}$ and $\{B, B_1\}$ are indistinguishable within their 1$\sigma$ uncertainties. Thus, the Individual Parameter run fails to identify the causal structure of the problem. This is a direct result of the extremely high inter-correlation between the variables, and indeed the Individual Parameter run may be thought of as the natural generalisation of a correlation to a classification problem. Ultimately, this issue is the fundamental origin of the truth behind the statement, `correlation does not imply causation'.

For the $\sqrt{N}$ Parameters run\footnote{We note that this is the default Random Forest classification method in the SciKit-Learn package.}, there is a marked improvement. Parameter $A$ is found to be significantly more important than $A_1$, and the same is true for $B$ and $B_1$. Hence, by using a Random Forest in this mode one can identify the superiority of causal parameters over a causal parameters, even to a level of inter-correlation of $\rho = 0.99$ (see also \citealt{Bluck2020a, Bluck2020b, Piotrowska2021}). This feat is achieved through the competitive nature of the Random Forest in this mode, where on occasion $A$ and $A_1$ (or $B$ and $B_1$) are directly compared in their potential to reduce impurity. The slight improvement of the causal parameter is enough for the Random Forest to favour this variable. For example, if one controls for $A$ when assessing the importance of $A_1$, the latter is revealed as being a causal (there is no value to adding $A_1$ over $A$ alone). Alternatively, when considering these parameters the other way around, there remains information in $A$, even at a fixed $A_1$. Thus, $A$ is used first, claiming the bulk of importance. However, there remains a problem in this mode. The parameter $A_1$ is found to have more importance than $B$, even though the latter is causal and the former is not. This is because $A$ is more important than $B$, and $A_1$ is an excellent substitute for $A$, when it is not randomly selected at a given decision fork. Hence, the solution must be to always consider every available parameter (as we do next).

Finally, for the All Parameter run, where every feature is considered at every decision fork, the Random Forest cleanly separates the causal from the a causal `nuisance' parameters. More specifically, $A$ followed by $B$ are found to be the most important parameters governing `quenching', and $A_1$, $B_1$, $C$ are all found to have no importance at all. This exactly solves this simple classification set up. In so doing, it illustrates how the Random Forest is a useful tool to separate causal from a causal `nuisance' parameters in a simple classification problem. Note also that the classifier achieves an overall performance of AUC = 1.0, i.e. a perfect classification (as displayed on Fig. B1 bottom-left panel), for this simple problem. 

It should be stressed that this is not `magic', and nor is it in any way a `black box'. The RF solves the classification problem by controlling for all nuisance parameters in the assessment of relative importance. Ultimately, this is the only leverage we truly have in science to move from correlation to causation. The Random Forest just offers a fast, efficient, and robust method to carefully control for potentially a causal parameters before offering a verdict on the relative importance of each variable (i.e. relative to the other variables made available to the classifier).

At this point one might rightly wonder how well the Random Forest classifier can perform at isolating the causal structure in a classification problem with a greater level of complexity. To address this, in Fig. B1 (top-middle panel) we show a more complex classification problem based on the same causal parameters: $A$ \& $B$. Here we replace the sharp linear boundaries of the above simple problem with non-linear smooth boundaries. As such, there is a substantial region of the parameter space in which both `star forming' and `quenched' objects may reside. More specifically, the previous deterministic quenching criteria is replaced with a probabilistic quenching criterion as follows:

\begin{equation}
P_Q = \bigg(1 - \exp\bigg\{-(A/A_{\mathrm{crit}})^\alpha\bigg\}\bigg) + \bigg(1 - \exp\bigg\{-(B/B_{\mathrm{crit}})^\beta \bigg\} \bigg)
\end{equation}

\noindent and

\begin{equation}
\mathrm{if } \,\, P_Q > 1  \,  \rightarrow \, P_Q = 1,
\end{equation}

\noindent where $P_Q$ is the probability that a given datum will be `quenched', given its values of $A$ and $B$. The above functional form is motivated by the modelling in \cite{Peng2010}, where it was used to parameterise mass and environment quenching. Here we set $A_{\rm crit} = 0.5$, $B_{\rm crit} = 0.8$ and $\alpha$ = $\beta$ = 5. Note that the secondary criterion is required to prevent the probabilities extending beyond unity, whilst preserving the independence of the $A$ and $B$ `quenching' channels. 

In Fig. B1 (bottom-middle panel) we present the results from a Random Forest analysis to solve this more complex classification problem. The parameters $A_1$, $B_1$ and $C$ are created in the same way as before and given to the Random Forest along with $A$ and $B$. The results are very similar to the simple classification problem, although the overall performance is lowered to AUC = 0.97 (as expected). Most importantly, the All Parameter run is still able to accurately identify $A$ and $B$ as the causal parameters governing `quenching' in the more complex model, yielding essentially zero importance to all of the nuisance parameters (despite their extremely high inter-correlation with the causal parameters). This result clearly establishes that a Random Forest applied in the All Parameter mode is capable of identifying causality among extremely highly inter-correlated parameters for a non-linear multi-parameter classification problem, even when the classes are only probabilistically defined.

Finally, in Fig. B1 (top-right panel) we add Gaussian random noise to the smooth-boundary classification problem to mimic a realistic astronomical data set. We add noise at the level of 10\% of the full range of each parameter (which is comparable to the measurement uncertainty on the observational parameters we investigate in this work). As such, we now have a noisy, probabilistic, non-linear classification problem. As before, we add in the nuisance variables, and test how well the Random Forest classifier can extract the underlying causality in various modes. Once again, the Random Forest classifier fails to identify the input causality of the classification problem in both the Individual Parameters and $\sqrt{N}$ Parameters modes. Yet, in the All Parameters mode, the Random Forest classifier correctly gives the highest importance to the dominant causal parameter ($A$), followed by the secondary causal parameter ($B$), exactly as in the simpler classification problems. 

It is important to highlight that $A$ is found to be far more important than $A_1$, and $B$ is found to be significantly more important than $B_1$, despite both of these pairings being inter-correlated at the level of $\rho = 0.99$. However, unlike in the simpler cases, some importance is given by the Random Forest classifier to the nuisance parameters $A_1$ and $B_1$, albeit at a much low level than in the other modes. This is an important result as it cautions against over-interpreting the significance of very low relative importances in correlated data with significant measurement uncertainty (i.e. they may be spurious). Note also that as we add complexity from left-to-right in Fig. B1, the overall performance of the classification, as measured by AUC (presented on each panel), decreases as expected. Nevertheless, the Random Forest classifier is still capable of identifying the causal parameters as the most important parameters at all levels of complexity studied here. Clearly then, the RF classification is advantageous over other simpler (correlation-based) analysis methods, when one seeks to identify causation in inter-correlated data.

There are two important caveats which we outline in closing this appendix. First, in order for the Random Forest to identify the causal variable(s), they must be present in the data. Whilst obvious, this is a major issue in any realistic astronomical application, which must be carefully thought about. Second, whilst the Random Forest is remarkably stable to global measurement uncertainty (applied equally to all features), it is not stable to differential measurement uncertainty. The Random Forest classifier does not require a perfect measurement of any parameter in order to function successfully, but differences in the accuracy with which two highly correlated variables are measured can and will affect the results. For example, in the case of variables $A$ and $A_1$ (which are extremely highly correlated), only a modest amount of Gaussian random noise added only to $A$, would yield $A_1$ as the superior parameter. This is intuitive since $A_1$ becomes a better estimate of intrinsic-$A$ than the measured-$A$ parameter in this scenario.

Ultimately, differential measurement uncertainty becomes more important the greater the level of inter-correlation between variables. In this work, variables are inter-correlated at a level less than (often much less than) $\rho \sim 0.9$. As such, small deviations in the accuracy of measurements are tolerable. Nonetheless, we rigorously test whether differential measurement errors could mimic any of the observational results in this work, and this is never a serious problem for any of our main science conclusions.

As a result of the above tests, we find that a Random Forest analysis in the All Parameter mode is capable of extracting the causal structure in realistic classification problems, where standard statistical techniques would fail. In this paper we apply this technique to observed and simulated galaxy catalogues to investigate the dependence of quenching in galaxies, bulges and disks. 

It is important to acknowledge that in our early work applying Random Forest classification to quenching in MaNGA (see \citealt{Bluck2020a, Bluck2020b}), we utilised the default $\sqrt{N}$ Parameter method (as recommended in the S{\small CIKIT}-L{\small EARN} documentation). As such, our prior RF results were not optimal. Indeed, we were aware of the limitations of the technique in that mode, such that variables correlated with the causal parameter(s) would have spurious importance (and we carefully tested for this in our prior papers). However, following the new results in this appendix, it is now clear that by switching to the All Parameter mode it is possible to remove the issue of inter-correlated variables almost entirely in Random Forest classification. Consequently, the results of this appendix provide further motivation to revisit the MaNGA data set, as well as the SDSS, CANDELS and LGalaxies (which have never been analysed using these promising techniques).

\end{document}